\documentclass[a4paper,fleqn,authoryear]{cas-dc}

\usepackage[numbers]{natbib}
\usepackage{aas_macros}
\usepackage[left,pagewise]{lineno}

\usepackage{glossaries}

\makeglossaries

\newglossaryentry{planetesimals}{
name={planetesimals},
description={$\sim$1-100\,km-size bodies that are a hypothetical intermediate step in planet formation and could be related to comets or the parent bodies of asteroids}
}

\newglossaryentry{Rayleigh-Taylor instability}{
name={Rayleigh-Taylor instability},
description={An instability of the interface between two fluids when a less dense fluid is accelerated into a denser fluid, e.g. by buoyancy}
}

\newglossaryentry{mare}{
name={mare},
description={The lunar mare are dark, volcanic plains, predominantly on the near side, that are $\sim$1 Gyr younger than the surrounding crust}
}

\newglossaryentry{cation-anion polyhedra}{
name={cation-anion polyhedra},
description={The geometric arrangement of atoms in which a cation (e.g. Si$^{+4}$) is surrounded by and bonds with anions (e.g., O$^{-2}$) either in regular (in minerals) or transient (in liquids) arrangements}
}

\newglossaryentry{equilibrium crystallization}{
name={equilibrium crystallization},
description={Crystallization of a silicate melt (magma) when the melt remains in thermodynamic equilibrium with all resulting solids}
}

\newglossaryentry{fractional crystallization}{
name={fractional crystallization},
description={Crystallization of silicate melt (magma) when the solids are continuously removed from the melt, e.g. by negative buoyancy, and the chemical composition of the system evolves as a result}
}

\newglossaryentry{plastic yield strength}{
name={plastic yield strength},
description={An applied stress above which a material will become ductile or flow}
}

\newglossaryentry{superliquidus}{
name={superliquidus},
description={Temperature and pressure conditions where a given silicate composition is completely molten}
}

\newglossaryentry{dissipation quality factor}{
name={dissipation quality factor},
description={A dimensionless factor $Q$ that is the inverse of the fractional energy dissipated during one cycle of a periodic system}
}

\newglossaryentry{Rossiter-McLaughlin effect}{
name={Rossiter-McLaughlin effect},
description={A time-dependent shift in a star's apparent radial velocity during the transit of a stellar or planet companion due to the occultation of different regions of the rotating star}
}

\newglossaryentry{homopause}{
name={homopause},
description={The altitude in an atmosphere above which molecular diffusion dominates over turbulent (eddy) diffusion, and individual gases are not well-mixed and decrease with altitude according to molecular weight}
}

\newglossaryentry{eddy diffusivity}{
name={eddy diffusivity},
description={A measure of turbulent transport by eddy motion in a fluid, in contrast to, but with the same units of, molecular diffusivity}
}

\newglossaryentry{T Tauri star}{
name={T Tauri star},
description={A young (typically less than a few Myr old) approximately solar-mass star with a disk of gas and dust in which planets are expected to form}
}

\newglossaryentry{Laplace resonance}{
name={Laplace resonance},
description={An orbital resonance in which three bodies have orbital periods in the ratio 1:2:4}
}

\newglossaryentry{peridotite}{
name={peridotite},
description={An iron-magnesium-rich rock, primarily composed of \gls{olivine} and \gls{pyroxene} minerals, that makes up most of Earth's upper mantle}
}

\newglossaryentry{smooth particle hydrodynamics}{
name={smooth particle hydrodynamics},
description={A method of numerical simulation of continuum mechanics in which a fluid is simulated by discrete Lagrangian particles and the properties at any particle location are calculated by summing over neighboring particles weighted by a kernel function}
}

\newglossaryentry{coronal mass ejection}{
name={coronal mass ejection},
description={The ejection of plasma and any magnetic field that accompanies it from the corona of the Sun or another star into the solar (stellar) wind}
}

\newglossaryentry{olivine}{
name={olivine},
description={An iron-magnesium silicate mineral that is abundant in Earth's crust and upper mantle}
}

\newglossaryentry{volatile}{
name={volatile},
description={A volatile element is in gaseous form in most planetary settings and condenses as a liquid or solid only at temperatures below a few hundred K, depending on ambient chemistry}
}

\newglossaryentry{refractory}{
name={refractory},
description={A refractory element typically condenses as a solid above 1500\,K, depending on ambient chemistry}
}

\newglossaryentry{phase angle}{
name={phase angle},
description={The angle subtended by lines between the illumination source (star), the reflecting/emitting object (planet) and the observer}
}

\newglossaryentry{redox buffer}{
name={redox buffer},
description={An assemblage of compounds/minerals that are in chemical equilibrium at a certain \gls{oxygen fugacity} and thus tend to constrain fO$_2$}
}

\newglossaryentry{pyroxene}{
name={pyroxene},
description={A series of (usually) Mg- and Fe-rich silicate minerals common in igneous rocks and the crusts and mantles of planetary bodies}
}

\newglossaryentry{plagioclase}{
name={plagioclase},
description={A series of Na-, Ca-, and Al-rich silicate minerals in the feldspar group and a major component of planetary crusts.  Anorthite is a plagioclase}
}

\newglossaryentry{partial melting}{
name={partial melting},
description={Melting in which only some solid phases melt, which can occur when rising material in a silicate mantle undergoing \gls{subsolidus convection} experiences lower pressures and ambient conditions that cross the \gls{solidus} in a pressure-temperature diagram}
}

\newglossaryentry{differentiation}{
name={differentiation},
description={Gravitational segregation of silicate and metal phases into the mantle and core of a planetary body during its formation}
}

\newglossaryentry{primitive melt}{
name={primitive melt},
description={Silicate melt that has not experienced the removal or segregation of any its liquid or crystalline components}
}

\newglossaryentry{diapirs}{
name={diapirs},
description={Regions of upwelling of buoyant material in geologic settings}
}

\newglossaryentry{phonolitic}{
name={phonolitic},
description={Describing magmas or volcanic rocks which are both more silicon-rich and alkali element-rich than basalts}
}

\newglossaryentry{andesitic}{
name={andesitic},
description={Describing magmas or volcanic rocks which are more silicon-rich than basalts; andesitic magmas are also more viscous}
}

\newglossaryentry{geostrophic flow}{
name={geostrophic flow},
description={Fluid flow in which the Coriolis force is balanced by a pressure gradient due to buoyancy differences along a temperature gradient}
}

\newglossaryentry{lock-up temperature}{
name={lock-up temperature},
description={Rheology-based temperature below which a magma transitions from a suspension of individual crystals to chains of crystals and the viscosity increases markedly}
}

\newglossaryentry{Love number}{
name={Love number},
description={One of a set of three dimensionless numbers that describe the response of a body to an imposed tidal potential, in this case $k$, which parameterizes the change in the potential due the tide}
}

\newglossaryentry{intermediate- and high-mass stars}{
name={intermediate- and high-mass stars},
description={Stars with masses of 3-8 solar masses and $>8$ solar masses, respectively.  Intermediate mass stars rapidly in red giants and lose their nucleosynthetic element-enriched envelopes as winds into space.  The cores of massive stars collapse, ejecting the envelope in a supernova event}
}

\newglossaryentry{pallasites}{
name={pallasites},
description={Stony-iron meteorites consisting of large olivine crystals in iron-nickel matrix thought to originate from a disrupted,differentiated parent body}
}

\newglossaryentry{achondrites}{
name={achondrites},
description={Stony meteorites that have recrystallized and/or melted due heating within a parent body}
}
\newglossaryentry{solidus}{
name={solidus},
description={Temperature at the onset of melting in a geologic material at a given pressure}
}

\newglossaryentry{liquidus}{
name={liquidus},
description={Temperature at which a geologic material becomes completely molten at a given pressure}
}

\newglossaryentry{coordination number}{
name={coordination number},
description={The number of atoms or ions that immediately surround a given ion or atom}
}

\newglossaryentry{bridgmanite}{
name={bridgmanite},
description={A high-pressure phase of magnesian silicate (MgSiO$_3$), thought to the first mineral to crystallize from Earth's magma ocean}
}

\newglossaryentry{silicate perovskite}{
name={silicate perovskite},
description={High pressure phases of silicates with the formula (Fe,Mg)SiO$_3$ or CaSiO$_3$ and a common crystal structure, thought to be a dominant constituent of Earth's deep mantle}
}

\newglossaryentry{oxygen fugacity}{
name={oxygen fugacity},
description={Chemical potential of oxygen in an environment, or the tendency of substances in chemical equilibrium to be oxidized or reduced, expressed as a partial pressure, although often not representing an actual gas phase}
}

\newglossaryentry{chondritic}{
name={chondritic},
description={Referring either to relatively unaltered groups of meteorites that are accumulations of smaller condensates or melted silicates (chondrules), or to elemental abundances that reflect the initial bulk composition of condensible matter in the Solar System}
}

\newglossaryentry{subsolidus convection}{
name={subsolidus convection},
description={Convection in a solid material, particularly a silicate mantle, due to ductile deformation}
}

\newglossaryentry{spectral type}{
name={spectral type},
description={A system of stellar classification by temperature-sensitive absorption lines in a stellar spectrum, and a proxy for radiative temperature and, for stars fusing hydrogen in their cores, mass}
}

\newglossaryentry{M dwarf star}{
name={M dwarf star},
description={A star with a radiative temperature $<$3900K, a mass $\lesssim$0.5 of the Sun, and a luminosity $<0.1$ of the Sun.  M dwarf stars are the most numerous type of star and important targets for exoplanet searches}
}

\newglossaryentry{cumulate}
{
    name=cumulate,
    description={Igneous rock formed by accumulation of minerals separating from a crystallizing magma}
}
\newglossaryentry{anorthosite}
{
    name=anorthosite,
    description={A relatively low density igneous rock made of $\>90\%$ by volume Na-, Ca-, and Al-rich \gls{plagioclase}}
}
\newglossaryentry{compatible}
{
    name=compatible,
    description={A compatible element prefers the solid phase (as opposed to the liquid) in a crystallizing melt}
}
\newglossaryentry{Goldschmidt classification}
{
    name=Goldschmidt classificatio,
    description={Geochemical classification of elements based on bonding affinities. Includes \gls{siderophile}s, \gls{chalcophile}s, \gls{lithophile}s, and \gls{atmophile}s}
}

\newglossaryentry{siderophile}
{
    name=siderophile,
    description={Elements that have an equilibrium affinity for the metallic (iron) phase (as opposed to silicate phases) and which tend to be more abundant in the metallic core of a differentiated body}
}
\newglossaryentry{chalcophile}
{
    name=chalcophile,
    description={\Gls{Goldschmidt classification} for elements that ten to bond with chalcogens other than oxygen, most commonly sulphur. They tend to be located near the surface, in the upper mantle and crust}
}
\newglossaryentry{lithophile}
{
    name=lithophile,
    description={Elements that have an equilibrium affinity for the silicate phases (as opposed to metallic phase) and which tend to be more abundant in the silicate mantle of a differentiated body}
}
\newglossaryentry{atmophile}
{
    name=atmophile,
    description={\Gls{Goldschmidt classification} for elements that show affinity for the gas phase, tending to primarily exist in the atmosphere}
}
\newglossaryentry{SNC meteorites}
{
    name={SNC meteorites},
    description={Shergotites, nahklites, and chassignites, a group of basaltic meteorites demonstrated to be sourced from Mars}
}

\newglossaryentry{light curve}
{
    name={light curve},
    description={A plot of the flux of a celestial object versus time}
}

\newglossaryentry{equilibrium temperature}
{
    name={equilibrium temperature},
    description={The temperature of a planet if it were radiating as a uniform blackbody}
}

\newglossaryentry{exosphere}
{
    name=exosphere,
    description={The outermost part of a planet's atmosphere the density is low enough that the mean free path between collisions is larger than the atmospheric scale height or planetary radius}
}

\newcommand{\altwosix}{$^{26}$Al}
\newcommand{\water}{H$_2$O}
\newcommand{\jwst}{\emph{JWST}}
\newcommand{\kepler}{\emph{Kepler}}
\newcommand{\tess}{\emph{TESS}}
\newcommand{\wpersqm}{W\,m$^{-2}$}
\newcommand{\mwpersqm}{mW\,m$^{-2}$}
\newcommand{\fotwo}{fO$_2$}
\newcommand{\mearth}{$M_{\oplus}$}
\newcommand{\rearth}{$R_{\oplus}$}
\newcommand{\cnce}{55\,Cnc\,e}

\begin{document}
\let\WriteBookmarks\relax
\def\floatpagepagefraction{1}
\def\textpagefraction{.001}
\shorttitle{Lava Worlds}
\shortauthors{}

\title [mode = title]{Lava Worlds: From Early Earth to Exoplanets}                      


\author[1]{Keng-Hsien Chao}[orcid=0000-0002-0329-7511]
\fnmark[1]
\credit{Writing - Original Draft, Visualization}

\address[1]{Department of Earth Sciences, University of Hawai'i at M\={a}noa, Honolulu, HI 96822 USA}

\author[1]{Rebecca deGraffenried}[orcid=0000-0003-2885-1870]
\fnmark[1]
\credit{Writing - Original Draft, Visualization}

\author[1]{Mackenzie Lach}[orcid=0000-0002-4778-9219]
\fnmark[1]
\credit{Writing - Original Draft, Visualization}

\author[1]{William Nelson}[orcid=0000-0001-7296-697X]
\fnmark[1]
\credit{Writing - Original Draft, Visualization}

\author[1]{Kelly Truax}[orcid=0000-0002-1283-3064]
\fnmark[1]
\credit{Writing - Original Draft, Visualization}

\author[1]{Eric Gaidos}[orcid=0000-0002-5258-6846]
\cormark[1]
\ead{gaidos@hawaii.edu}
\credit{Conceptualization, Supervision, Writing - Original Draft, Visualization}

\cortext[cor1]{Corresponding author}
\fntext[fn1]{These authors contributed equally to this work.}


\begin{abstract}
The magma ocean concept was first conceived to explain the geology of the Moon, but hemispherical or global oceans of silicate melt could be a widespread "lava world" phase of rocky planet accretion, and could persist on planets on short-period orbits around other stars.  The formation and crystallization of magma oceans could be a defining stage in the assembly of a core, origin of a crust, initiation of tectonics, and formation of an atmosphere.  The last decade has seen significant advances in our understanding of this phenomenon through analysis of terrestrial and extraterrestrial samples, planetary missions, and astronomical observations of exoplanets.  This review describes the energetic basis of magma oceans and lava worlds and the lava lake analogs available for study on Earth and Io.  It provides an overview of evidence for magma oceans throughout the Solar System and considers the factors that control the rocks these magma oceans leave behind.  It describes research on theoretical and observed exoplanets that could host extant magma oceans and summarizes efforts to detect and characterize them.  It reviews modeling of the evolution of magma oceans as a result of crystallization and evaporation, the interaction with the underlying solid mantle, and the effects of planetary rotation.  The review also considers theoretical investigations on the formation of an atmosphere in concert with the magma ocean and in response to irradiation from the host star, and possible end-states.  Finally, it  describes needs and gaps in our knowledge and points to future opportunities with new planetary missions and space telescopes to identify and better characterize lava worlds around nearby stars.
\end{abstract}





\begin{keywords}
magma oceans \sep Moon \sep Io \sep exoplanets \sep planet formation \sep planetary atmospheres
\end{keywords}

\maketitle

\tableofcontents

\section{Introduction}

At Earth's surface, silicates form minerals and rocks, but at depth in volcanic areas where the temperature exceeds the \gls{solidus}, which is about 1300\,K at 1 bar for an Earth-like composition \citep{Sarafian2017}, silicates are partially molten and readily flow.  These conditions are not presently widespread at or near the surface of any Solar System body: Venus and the subsolar point of Mercury are at 735\,K and 800\,K, respectively.  It is only in active volcanic regions on Earth and the Jovian satellite Io that silicate melts appear, and only transiently.  Yet, geochemical studies of Earth and returned samples, remote sensing by spacecraft of the Moon and other planets, and models of planetary accretion from smaller planetesimals all suggest that the silicate mantles of many Solar System bodies began in a partially or completely molten state \citep{ElkinsTanton2012}.  Recent surveys of exoplanets, i.e. planets orbiting other stars, have discovered a class of rocky planets similar in size to Earth on very close-in orbits around their host stars which experience irradiation orders of magnitude higher than the terrestrial value \citep{Winn2018}.  These have no analogs in the Solar System, and their \gls{equilibrium temperature} could reach or exceed 2000\,K, at which point silicates are completely molten and begin evaporating.  Close-in planets on eccentric orbits or with non-synchronous rotation rates will experience strong tides from the central star, and dissipation within the planets could lead to widespread melting and global volcanism \citep{Jackson2008} analogous to but on a larger scale than Io \citep{Park2019}.   Young stars may harbor newly accreted, magma ocean-hosting planets at the present time.

Planets with molten silicates at their surfaces have  been  termed ``lava planets" \citep{Leger2011} and the liquid melt bodies ``magma oceans', ``lava oceans", or ``lava ponds".  The geologic definition of lava is simply magma that is erupted to the surface, a local phenomenon.  Because this review is concerned with hemispherical or global bodies of melt not specifically related to volcanism, we will refer to these melt volumes as magma pools, seas, or oceans, although for terrestrial examples in craters and calderas they will continue to be termed lava lakes.  We reserve the term ``lava planet" or ``lava world" for rocky planets with surface temperatures reaching the solidus either globally or at least over much of a hemisphere \emph{or} where the global average heat flow is at least $\sim$1 \wpersqm, typical of terrestrial volcanic regions, sufficient to maintain a subsurface temperature gradient of 700 K km$^{-1}$, and permissive of melt within $\sim$2\,km of the surface regardless of surface temperature.  Planets, including early Earth, Mars, and present-day Venus, could have "basal" magma oceans at their core-mantle boundaries that are well insulated from the surface; although these are discussed in this review, we do not term these lava worlds.  Likewise, larger Neptune-like planets with magma oceans under thick, low-molecular weight envelopes, could be widespread \citep{Kite2019}, but are not considered here.  Additional discussion of terminology can be found in \citet{Solomatov2015}.  

The presence or absence of a magma ocean could have been a deterministic branch point in the evolution of the Solar System's rocky planets, governing their geologic history and, by influencing the distribution of the planets' \gls{volatile} elements, atmospheric composition and evolution, climate, and habitability \citep{ElkinsTanton2012,Hamano2013,Armstrong2019}. This presumably also holds for rocky exoplanets, which are expected to be more diverse in terms of mass, formation history, and composition.  Exoplanets that are very close to their host star are, by current principal methods, less difficult to detect and characterize (i.e., measurements of mass, temperature, and atmosphere). They will be among the high priority targets for future space missions such as the \emph{James Webb Space Telescope} \citep[\jwst,][]{Beichman2018}.  ``Evaporating" planets with tails of escaping or recondensing dust are an opportunity to probe interior composition \citep{Bodman2018b}, complementary to other measurements such as comparing mass and radius to interior models.

This review is motivated by: (a) The successful completion of the prime mission of the \emph{Transiting Exoplanet Survey Satellite}  \citep[\tess,][]{Ricker2014}.  \tess\ is designed to discover ``super-Earth"-size planets on close-in orbits around bright, nearby stars; in the case of the most luminous host stars the planets could be lava worlds.   These planets are amenable to mass measurements using the Doppler radial velocity method with a new generation of precision spectrographs.  (b) The planned launch of \jwst\ in 2021, which will usher in a new era of spectroscopy at infrared wavelengths where the emission from lava worlds peaks.  With \jwst\ it will be possible to measure the temperature of a planet and infer the re-distribution of heat by any atmosphere, as well as identify some atmospheric constituents such as CO$_2$.  And (c) a suite of new laboratory instruments to analyze meteorites and returned samples, and recent and future missions to Mercury (\emph{Messenger}, \emph{Bepi-Columbo}) and Mars (\emph{Curiosity}, \emph{InSight}, \emph{Perseverance}), planets which may have had magma oceans, as well as proposed missions to Venus and Io, which could have interior magma oceans at the present time.

We begin this review with the fundamentals, i.e., an overview of sources and transport of heat on and in lava worlds (Sec. \ref{sec:energetics}).  In Section \ref{sec:lavalakes} we discuss lava lakes of Earth and Io as ``natural laboratories" for the diverse behavior of magma near or at the surface, the parameters that may govern that behavior, and the scaling laws relevant to applying lessons to other (exo)planets.   Next we summarize what we have learned about surface and interior magma oceans in the  inner Solar System beginning with the first-established and best-studied example, the Moon (Sec. \ref{sec:solarsystem}).  We review the evidence for magma oceans on Earth, other planets, and asteroids, and their potential physical and chemical diversity.  We move beyond the Solar System in Section \ref{sec:exoplanets} to review the contributions of the field of exoplanets to the study of lava worlds, beginning with a summary of relevant observational techniques followed by a discussion of the population of short-period lava world candidates that these have uncovered and characterized. CoRoT-7b, Kepler-78b, and 55 Cancri e are presented as comparatively well-studied examples. The theoretical diversity of lava world exoplanets is addressed, with attention given to several scenarios where planets or satellites with magma oceans might in the future be detected on wider orbits.  In Sec. \ref{sec:dynamics} we discuss the physical and dynamical principles that determine the dynamics and stability of magma as controlled by the relative density and buoyancy of melts and solids, as well as the effects of rotation on magma ocean circulation.  In Sec. \ref{sec:atmospheres} we discuss the role of atmospheres on lava worlds, the likely initial conditions that may have accompanied primordial magma oceans, the processes that governed the evolution of those atmospheres, and their Mercury-like and Venus-like end-states.  Finally, we close with a synthesis of our current understanding, highlight outstanding questions and gaps in our knowledge, and describe promising future directions of research.  

\begin{figure*}
	\centering
		\includegraphics[width=\textwidth]{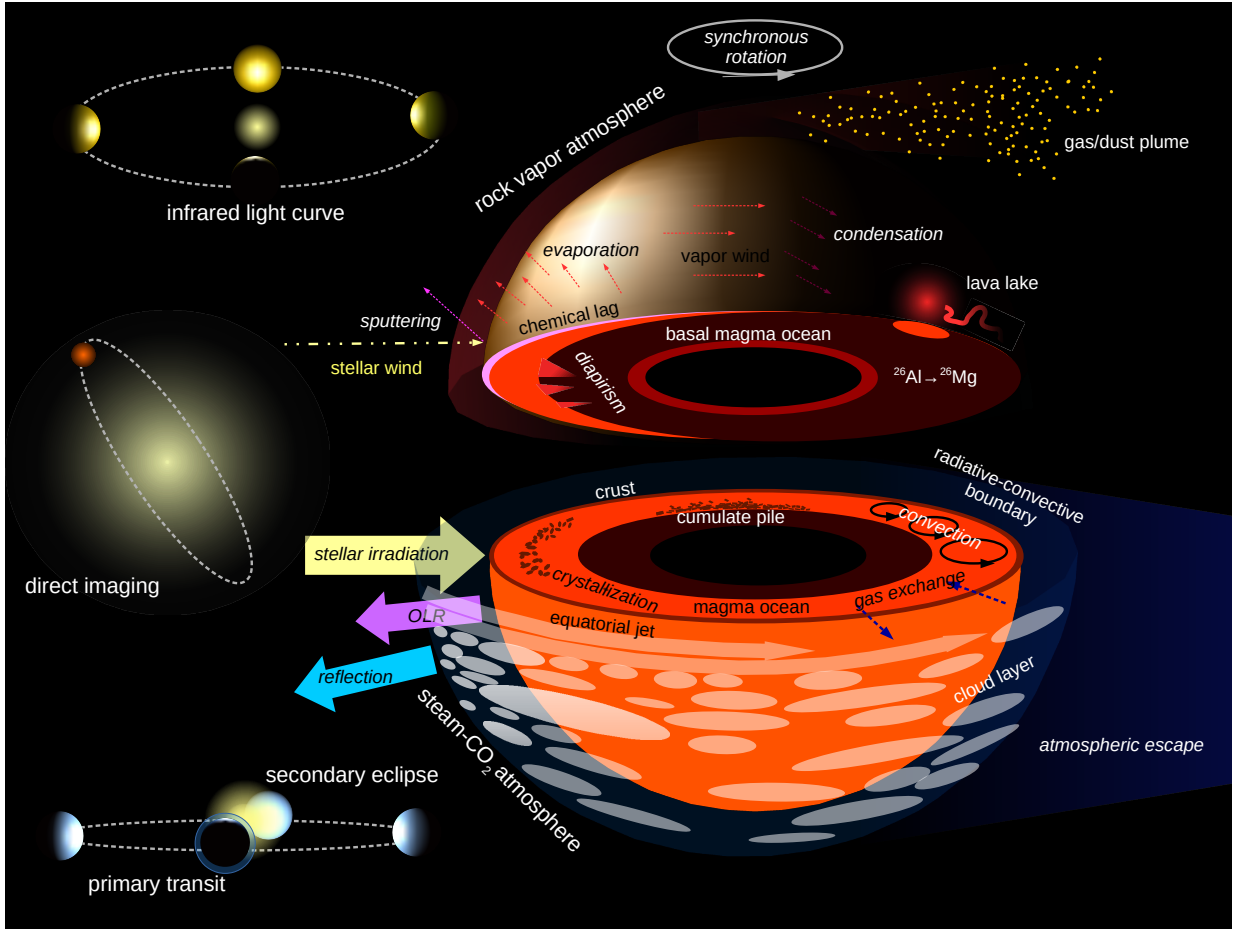}
	\caption{Schematic illustration of two ``lava world" scenarios: a tidally-locked, synchronously rotating planet with one irradiated hemisphere, a hemispherical magma sea, and a thin rock vapor atmosphere (top); and a planet with a thick steam-CO$_2$ atmosphere and a global magma ocean (bottom).  Components (normal font) and processes (italic labels) discussed in this review are represented, but not to scale or indicating the relative prominence.  The left side figures illustrate methods of exoplanet detection/characterization discussed in Sec. \ref{sec:methods}.  OLR = outgoing long-wavelength (infrared) radiation.}
	\label{FIG:lavaworldcartoon}
\end{figure*}

\section{Energetics of Lava Worlds}
\label{sec:energetics}

\subsection{Sources of Heat}
\label{sec:heat_sources}

Different sources of heat can form and maintain a magma ocean or pond, depending on a planet's size, age, and distance from the star (Fig. \ref{fig:sources}). A magma ocean will be maintained indefinitely on a mantle with an Earth-like (\gls{peridotite}) composition if the surface temperature exceeds $\approx$1300,K.  In the absence of an atmosphere, this requires an energy flux of $1.5 \times 10^5$ \wpersqm, or $>$300 times the present global mean insolation of Earth, to balance radiative loss.  For a Venus-like CO$_2$ "greenhouse" atmosphere ), the energy requirement falls to $\approx$2300\,\wpersqm, and for an optically thick, steam proto-atmosphere like that which presumably accompanied accretion, it is only $\approx$300\,\wpersqm \citep{Hamano2013}; 1.3 times current terrestrial globally-averaged insolation or $8 \times 10^3$ times Earth's current internal heat flow (see Section \ref{sec:atmospheres}.  

\subsubsection{Gravitational Energy}
\label{sec:accretion}

Transient magma oceans can be formed during planetary accretion as gravitational energy gained by infalling planetesimals is converted into heat during impact.  As a protoplanet grows, the energy liberated per unit mass becomes comparable to the heat of melting of silicates.  The energy flux per unit area, averaged over the time scale over which many accretionary impacts occur, is:
\begin{equation}
\label{eqn:accretion_energy}
    q_{\rm accretion} = \left[\frac{GM_p}{R_p}+\frac{1}{2}v_{\infty}^2\right]\frac{\dot{M_p}}{4\pi R_p^2},
\end{equation}
where $G$ is the gravitational constant, $M_p$ and $R_p$ are the instantaneous mass and radius of the planet, and $v_{\infty}$ is the approach velocity of the impacting bodies.  For a rocky planet $R_p \propto M_p^{\epsilon}$, with $\epsilon = 0.25-0.33$ \citep{Grasset2009}, and the rate of energy release is at least $2.3 \times 10^4 M_p^{2-3\epsilon} \tau_{\rm Myr}^{-1}$ \wpersqm\ where $M_p$ is in Earth masses and $\tau_{\rm Myr}$ is the planet's accretion time in Myr.  The corresponding blackbody equilibrium temperature is $T_{\rm eq} = 800M_p^{(2-3\epsilon)/4} \tau_{\rm Myr}^{-1/4}$K.  Since accretion is likely to form a substantial atmosphere of volatiles released by impacts, the temperature at the boundary between radiative and convective heat transport (Section \ref{sec:heat_transfer}), rather than the surface temperature, is near $T_{\rm eq}$; the surface temperature will be much higher.  The temperature at the surface of an Earth-sized planet with a thick steam atmosphere can be maintained above the solidus if the energy flux exceeds 300 W\,m$^{-2}$ \citep{Hamano2013}, the case if accretion is continuous and happens in less than 100\,Myr, which also happens to be about the inferred time for Earth's accretion \citep{Jacobson2014}.  

In the canonical scenario, the late stage of rocky planet accretion is not continuous but consists of a small number of "giant" impacts by Moon- to Mars-size objects \citep{Morbidelli2012} which will release enough energy to melt a fraction or most of the silicate mantle, which could then re-crystallize, depending on the depth of melting and the presence or absence of an atmosphere \citep{Reese2011,deVries2016}.  The largest of these impacts onto the proto-Earth is the leading scenario for the Moon's formation, which would have also generated magma oceans on both bodies (Sec. \ref{sec:solarsystem}).   Importantly, impacts deposit energy at depth, driving vigorous convection and mixing in the magma ocean as heat escapes to the surface.  Additional energy will be released if metals (particularly Fe) segregate from silicates and sink to a growing core \citep{Rubie2007}.

\subsubsection{Stellar Irradiation}
\label{sec:irradiation}

Once accretion is complete, the principal energy source will be radiation from the host star.  If the energy is uniformly distributed over the planet's surface by rotation and advection (by an ocean or atmosphere, see Sec. \ref{sec:heat_transfer}), the time-averaged energy flux over a nearly circular orbit is \citep{Williams2002}:
\begin{equation}
\label{eqn:stellar_energy}
    q_{\rm stellar} = \frac{L_* (1-A)}{16\pi a^2} = \frac{L_*}{4} \left(\frac{\sqrt{\pi}}{4 G M_* P_K^2}\right)^{2/3},
\end{equation}
where $L_*$ and $M_*$ are the stellar luminosity and mass, respectively, $a$ is the orbital semi-major axis, $A$ is the planet's Bond (phase- and spectral-averaged) albedo, and $P_K$ is the Keplerian orbital period.  If $L_*$ and $M_*$ are expressed in solar units and $P_K$ is in days, then $q_{\rm stellar} \approx 8.9 \times 10^5 (1-A) L_* M_*^{-2/3} P_K^{-4/3}$ \wpersqm.  If there is no re-distribution of heat, i.e. no atmosphere, and the planet is locked by tides to synchronous rotation, then the local incident flux is maximum at the substellar point where it is four times that in Eqn. \ref{eqn:stellar_energy}, and it varies with the cosine of the angle $\theta$ from that point to zero at the terminator (see Sec. \ref{sec:heat_transfer}).  If $A$ is low, the emissivity is near unity, and a stellar mass-luminosity relation for low-mass hydrogen-burning stars $L_* \propto M_*^{4.84}$ \citep{Eker2015} is adopted, then $T_{\rm eq} \approx 2000M_*^{1.04} P_K^{-1/3}$\,K for complete heat redistribution, and $T_{\rm eq} \approx 2800M_*^{1.04} P_K^{-1/3} (\cos \theta)^{1/4}$\,K for no heat redistribution and synchronous rotation.  The sensitivity to stellar mass means that, for a given orbital period, planets around stars with $M_* \gtrsim M_{\odot}$ can more readily host magma oceans, bearing in mind that the distribution of planets with orbital period also depends on stellar mass \citep[e.g.,][]{Mulders2015}.   

Besides local stellar irradiance (time-averaged if the planet is non-synchronously rotating), the albedo and the efficacy of heat redistribution govern the temperature distribution on an irradiated planet.  The albedo of Solar System bodies varies from a few percent for primitive comets to nearly 80\% for cloudy Venus (see Table \ref{tbl:albedo} and \citealt{Madden2018}).  On an airless rocky body, the albedo will be determined by surface particle size distribution and composition, primarily the abundance of iron, which has a rich spectrum of transitions in the optical, and energetic particle bombardment or ``space weathering" \citep{Pieters2016}.   Carbon in the form of pure graphite has a moderate ($\approx$0.25) reflectivity \citep{Papoular2014} but its actual reflectance in a planetary setting will depend on impurities.  Since albedo is averaged over the spectrum of the star, it can vary significantly with the host star's \gls{spectral type}.  For instance, aerosols scatter less at longer wavelengths and thus a planet with a hazy atmosphere would have a lower albedo around an M dwarf star.

\begin{table*}
\caption{Bond Albedo of Planetary Surfaces}\label{tbl:albedo}
\begin{tabular}{l|l|l}
\toprule
\textbf{body or surface} & \textbf{value(s)} & \textbf{references}\\
\midrule
Moon & 0.08-0.11 & \cite{Buratti1996,Helfenstein1997} \\
Mercury & 0.09-0.12 & \cite{Mallama2017} \\
Venus & 0.76 & \cite{Haus2016} \\
Io & 0.52 & \cite{Simonelli2001} \\
Fresh basalt & $\sim$0.04 & \cite{Spinetti2009} \\
H$_2$O-CO$_2$ atmosphere & 0.55-0.85 & \cite{Pluriel2019} (model)\\
\bottomrule
\end{tabular}
\end{table*}

\subsubsection{Aluminum-26}
\label{sec:al26}

The decay of short-lived radionuclides (SLRs)\footnote{To be distinguished from long-lived radionuclides (i.e., $^{40}$K, $^{232}$Th, $^{235}$U, $^{238}$U) with mean lives of Gyr and total contributions of 10s to 100s of mW m$^{-2}$ to the heat flow from rocky Earth-like planets.}, unstable isotopes with mean lives of kyrs to Myrs, in particular \altwosix{} (mean life of 1\,Myr), was probably the principal internal heat source during the first few Myr of the Solar System.  The presence of \altwosix{} is evidenced by a fraction of $^{26}$Mg (the daughter isotope of \altwosix) in primitive meteorites that correlates with $^{27}$Al.  Regions of the inner Solar System from which we have samples had an initial (``time zero") abundance of  \altwosix{} relative to $^{27}$Al of about $4.5 \times 10^{-5}$ \citep{Dauphas2011}.  Its decay in a body of primitive \gls{chondritic} composition would have released $6.7 \times 10^6$ J kg$^{-1}$ \citep{Moskovitz2011}, significantly larger than the heat of fusion of the bulk rock ($\approx 2 \times 10^6$ J kg$^{-1}$).  In bodies larger than a few km, diffusion of this heat to the surface is slower than its release and melting would have occurred.  This is thought to be the cause of the melting and \gls{differentiation} of the parent bodies of the iron meteorites \citep{Qin2008}, \gls{pallasites}, and \gls{achondrites}, and, by extension, any body that accreted within $\sim$2\,Myr of time zero.  Magnesium isotopes indicate that the  accretion of the parent body of three related groups of meteorites -- most likely from Vesta -- occurred within 0.3 Myr of time zero, and that heat from from the decay of \altwosix{} would have formed a magma ocean \citep{Schiller2011,Schiller2017} (see Sec. \ref{sec:vesta}).  There is evidence for other SLRs in the early Solar System, but their contribution to the heat budget would have been very minor \citep{Huss2009}.

\altwosix{} was definitely present in the early Solar System, but neither its source nor its distribution within the protoplanetary disk are well established; both bear directly on expectations for the abundance of \altwosix{} in other planetary systems.   \altwosix{} is endogenously produced by spallation reactions of high energy protons from the Sun with Mg atoms \citep{Duprat2007}, and exogenously produced in \gls{intermediate- and high-mass stars} and released into the interstellar medium by winds and supernova explosions and could have been incorporated into the Sun's parent molecular cloud \citep{Huss2009}.  High energy protons are produced by flares and shocks during \gls{coronal mass ejection}s; these are universal manifestations of magnetic activity on solar-type stars which is especially elevated when a star is younger and more rapidly rotating \citep{Wright2011,Davenport2019}.  On the other hand, an exogenous source of \altwosix{} in massive stars requires its formation, transport, and introduction into a protoplanetary disk within a few mean lifes and the efficiency of this process could vary markedly from star to star \citep{Gaidos2009b,Lichtenberg2019}.  While the planets of the inner Solar System seem to have accreted in 10-100 Myr \citep{kleine2002}, after the essentially complete decay of \altwosix, in other systems formation could have occurred more rapidly, i.e. close to the host star \citep{Chiang2013,Chatterjee2014}, especially around M dwarf stars where many planets on short-period orbits are found \citep{Dressing2015,Mulders2015,Gaidos2016}.  In these objects, intense heating by residual \altwosix{} could have readily sustained magma oceans for Myr (Fig. \ref{fig:sources}).

\subsubsection{Tides}
\label{sec:tides}

The dissipation of tidal deformation can be a much longer-lived internal heat source in planets.  Tidal deformation can arise from non-synchronous rotation, non-zero orbital eccentricity, or non-zero obliquity \citep{Driscoll2015}.  Heat production by these effects increases markedly with decreasing semi-major axis.  For a synchronously rotating, homogeneous planet on a slightly eccentric orbit,
\begin{equation}
\label{eqn:tides}
q_{\rm tide} = \frac{84\pi^4k_2R_p^3e^2}{QGP_K^4},
\end{equation}
where $k_2$ is the \gls{Love number} and $Q$ is the \gls{dissipation quality factor} \citep{Peale1978}.  For rocky planets, the modified quality factor $Q' \equiv 3Q/(2k_2)$ is taken to be $\approx$100 \citep{Clausen2015}, but is expected to be weakly frequency dependent \citep{Ray2012}, and could differ substantially for planet with magma oceans.  For an Earth-size planet, $q_{\rm tide} \sim 10$ \wpersqm $(e/0.1)^2 (P_K/10 {\rm\,day})^{-5}$.  Thus, tidal dissipation in planets on moderately eccentric, close-in ($P<10$\,day) orbits  can drive widespread melting of their interiors (Fig. \ref{fig:sources}).  One caveat is that this same dissipation circularizes the orbit, unless the eccentricity is continuously excited by a perturbing planet or stellar companion.  Another is that $Q$ depends on mantle rheology and will decrease as interior temperature approaches the solidus, e.g., Fig. 3 in \citet{Driscoll2015}.  Tidal heating could be maximum at a certain distance from the star and be lower on interior/exterior orbits where the planets are hotter or cooler and thus less or more dissipative.

\subsubsection{Induction}
\label{sec:induction}

Induction heating, the ohmic dissipitation of currents induced by the motion of a conducting body through a magnetic field, could be an important long-term heat source in a planetesimal or planet.  The magnetic field is provided by the star, the motion can be a combination of rotation, orbital motion, or a stellar wind, and the conductor is either the metal core or a silicate mantle which becomes more conducting at high pressure and/or temperature \citep{Soubiran2018}.  Unipolar induction can take place if there is a complete electrical circuit that includes the planet, as is the case for Io's ionosphere and Jupiter \citep{Goldreich1969}.  If the planet is electrically isolated, induction heating can occur as dissipation of alternating eddy currents driven by oscillatory components of the magnetic field in the rest frame of the body due to, e.g., orbital motion through an inclined field or an eccentric orbit in an axisymmetric field.  AC induction heating was first considered for the parent bodies of meteorites \citep{Sonett1968}, the Moon \citep{Sonett1975} and Io \citep{Colburn1980}.  The strength of a magnetic dipole field scales as $a^{-3}$, motivating models of this as a heat source in close-in planets around white dwarfs \citep{Veras2019}, and M dwarf stars like TRAPPIST-1 \citep{Kislyakova2017,Kislyakova2018}, as well as planetesimals close to young, magnetically active stars \citep{Bromley2019}.  

An important aspect of this phenomenon is the skin depth $\delta$, the characteristic scale over which eddy currents induced by a periodic field (here assumed to have an angular frequency equal to $2\pi/P_K)$ cancel the imposed magnetic field:
\begin{equation}
    \label{eqn:skin}
    \delta = \sqrt{\frac{P_K}{\pi \mu_0 \sigma}},
\end{equation}
where $\mu_0$ is the magnetic permeability (close to the free space value) and $\sigma$ is the electrical conductivity.  Adopting $\sigma = 0.5\, \Omega^{-1}$\,m$^{-1}$,  appropriate for basaltic melts at low pressure \citep{Gaillard2005,Pommier2010}, then $\delta \approx 200 P_K^{1/2}$\,km, where $P_K$ is in days.  Thus dissipation could occur throughout most or all of a surface magma ocean, but at high pressures and temperatures, especially within more massive planets, electrical conductivity increases by several orders of magnitude \citep{Soubiran2018} and the skin depth will be a few km.

\begin{figure}
	\centering
		\includegraphics[width=0.48\textwidth]{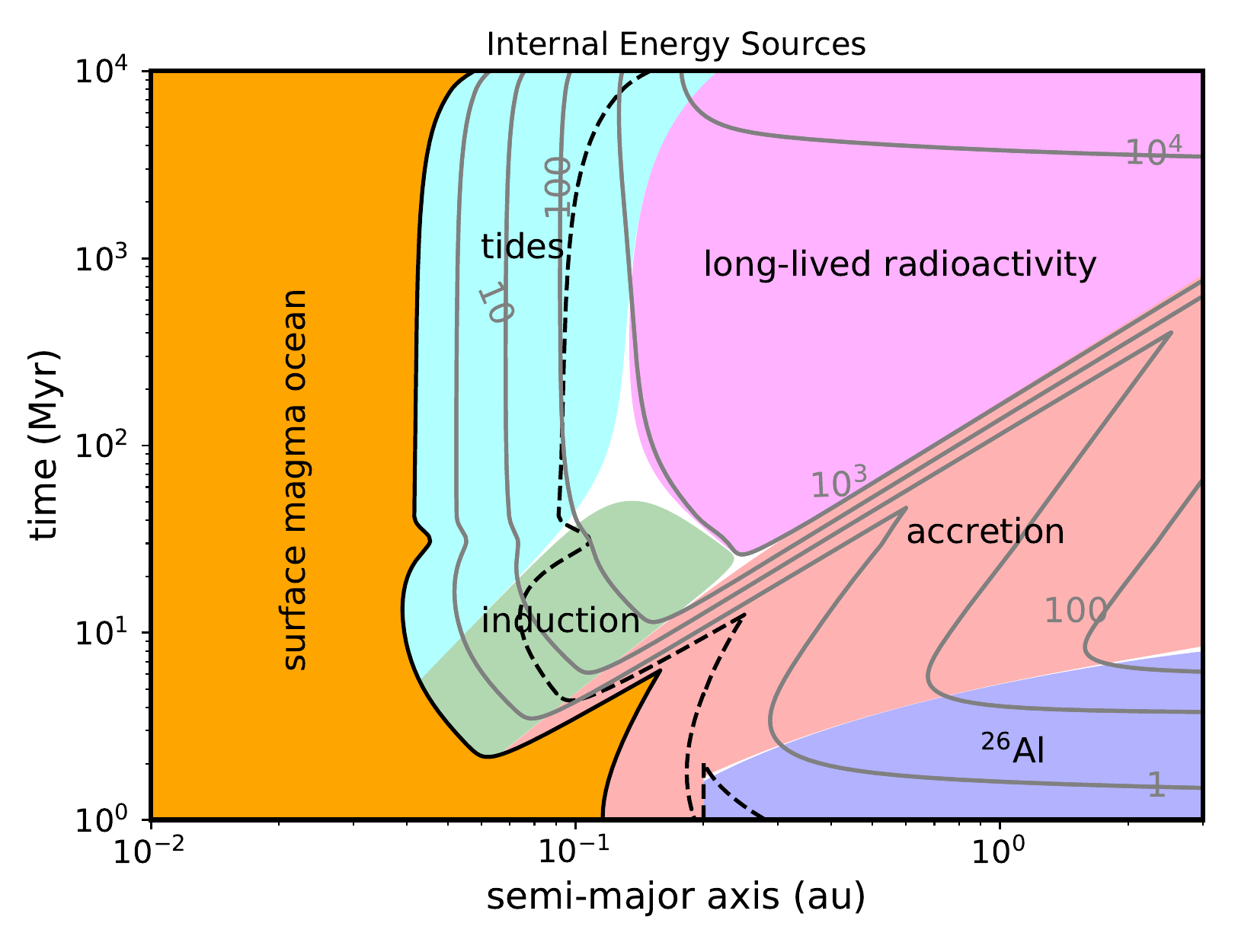}
	\caption{Approximate domains of dominant internal heat sources in an Earth-size planet orbiting a Sun-like star as a function of orbital semi-major axis and age.  The orange zone is the area where the crust is $<$10 cm, i.e. a surface magma ocean, assuming efficient heat transport by an atmosphere but no greenhouse effect; the dashed line is the same boundary for a Venus-like 500\,K greenhouse effect.  In other parts of parameter space, gray contours indicate the thermal equilibrium thickness in meters of a crust.  Since the crust can continuously founder, a surface magma could effectively exist over all but the upper right-hand region of this parameter space. Irradiation calculations use a model track from \citet{Dotter2008} and assume $A = 0.2$.  Tidal heating assumes that orbital eccentricity is maintained at 0.05.  A Solar System canonical initial \altwosix\ abundance is assumed.  Long-lived radioactivity assumes an Earth-like inventory with an average mean life of 2.8 Gyr.  Induction heating adopts the relationship between rotation period and magnetic field strength of \citet{Vidotto2014} and a solar-like \citet{Skumanich1972} rotation history.  Accretion assumes linear growth over 50 Myr at 1 AU and exponential decay of the growth rate as the protoplanetary disk is cleared with a 5 Myr time constant at 1 AU, and that the time constants scale with orbital period.  In the absence of tidal heating that zone would be dominated either by induction heating (in younger planets) or long-lived radionuclides (in older planets).}
	\label{fig:sources}
\end{figure}

The energy flux for eddy current dissipation in a uniform alternating field that induces a current density $J$ is given by the volume integral:
\begin{equation}
    q_{\rm ind} = \frac{1}{4\pi R_p^2}\int_V \frac{\vec{J} \cdot \vec{J}}{\sigma} dV.
\end{equation}
If $\delta \ll R_p$, the dissipation can be approximated as occurring within a thin spherical shell of radius $R_p$ \citep{Nagel2018,Manser2019}. For a planet on an orbit that is moderately inclined ($i$) to a stellar dipole magnetic field with surface intensity $B_*$ , 
\begin{equation}
    q_{\rm ind} = \frac{27 \pi^{3/2} B_*^2 \cos^2 i}{4 \mu_0^{5/2}G^2\rho_*^2 P_K^{9/2}}
\end{equation}
If $P_K$ is in days and $B$ is in Gauss, $q_{\rm ind} \approx 3 \times 10^{-6} B^2 P_K^{-9/2}$\,\wpersqm\ and thus is negligible around slowly-rotating solar-type stars like the Sun ($B \sim 1$ Ga).  It could, however, be appreciable for stars with stronger magnetic fields, i.e. close to rapidly rotating M dwarfs, young stars \citep[$B \sim 500$ Ga,][]{Vidotto2014,Kislyakova2018,Bromley2019}, or white dwarfs \citep[$B \sim10^4$ Ga][]{Manser2019}.  Given the extreme period dependence, only planets with $P_K \lesssim 1$ day would have significant heat flow (Fig. \ref{fig:sources}).  A higher-order field would produce an even more extreme period dependence.  Remaining uncertainties in this scenario are whether this process can ``bootstrap" from small amounts of existing melt, and the degree to which the planet's surface and interior would be shielded by any electrically conducting ionosphere.

\subsection{Transport of heat}
\label{sec:heat_transfer}

Vertical heat transport regulates the temperature profile and hence depth of a magma ocean, and lateral (meridional or longitudinal) heat transport regulates its extent.  Both are usually manifested as internal circulation driven by buoyancy differences.  Heat transport occurs by conduction, convection, advection, or radiation; convection is common in optically-thick regions of atmospheres (or above an irradiated surface), while in regions above the surface that are optically thin, net radiative transport from warmer lower layers upwards will flatten the temperature profile and suppress convection.

A dichotomy in the nature of heat transport will exist between ``Mercury-like" planets with thin or nonexistent atmospheres and ``Venus-like" planets with thick greenhouse atmospheres (Fig. \ref{FIG:lavaworldcartoon}).  On the former, surface temperatures will be chiefly governed by local radiative balance\footnote{Condensates, e.g. dust, in such an atmosphere can nevertheless influence this radiative balance.}, and large gradients will exist between the sub-stellar point and terminator:  $T_s \approx T_0 (\cos \theta)^{1/4}$ for synchronous rotation, where $T_0$ is the sub-stellar point temperature and $\theta$ is the angle from that point, or between the equator and poles ($\bar{T_s} \approx T_0 ((\cos \theta)/\pi)^{1/4}$ for rapid asynchronous rotation, where $\theta$ is then the latitude).  Associated density differences and buoyancy forces will govern the circulation and temperature structure in a magma ocean, and heat transport can be locally outward or inward.\footnote{The consequences of strongly asymmetric surface temperature profiles for \gls{subsolidus convection} in the planet's silicate mantle have been explored by \citet{vanSummeren2011}.}  

For "Venus-like" planets with an optically-thick atmosphere containing greenhouse gases like H$_2$O and CO$_2$ (Sec. \ref{sec:atmospheres})\footnote{In atmospheres hotter than 1000\,K or 1500\,K, alkali metals or Fe, respectively, will provide substantial opacity in the visible.}, radiative balance occurs near the radiative-convective boundary in the atmosphere, typically at a pressure altitude of $\sim$0.1\,bar \citep{Robinson2014}, where the optical depth to space falls below unity.  Above this altitude, radiation dominates and flattens the vertical temperature profile and the \gls{eddy diffusivity} decreases with altitude, eventually reaching the molecular diffusion value at the \gls{homopause}. Below this altitude, temperatures follow an adiabatic profile, and atmospheric circulation will impose a nearly uniform temperature distribution around the planet.   In this case, the interior heat flow is everywhere outwards and set by internal sources of heat and cooling.   

\subsubsection{Magma Ocean Heat Transport}
\label{sec:magmaheattrans}

For the case of a hemispherical magma ocean of scale $R_p$ on a tidally-locked, synchronously rotating ($P_{\rm rot} = P_K$) planet, the boundary between the "Mercury-like" and "Venus-like" regimes of magma ocean heat transport is where $q_{\rm circ} = q_{\rm int}$, which for a hemispherical ocean of radius $R_p$ on a tidally-locked planet is:
\begin{equation}
\label{eqn:qcirc1}
    q_{\rm circ} = \frac{\rho c_p \delta_T \left(T_{0} - T_{l}\right)}{\tau_T} 
\end{equation}
where $\delta_T$ and $\tau$ are the thermal boundary layer thickness and turnover time, respectively, and $T_l$ is the \gls{lock-up temperature}, where elevated viscosity prevents circulation, respectively \citep{Kite2016}.  Assuming \gls{geostrophic flow} over a hemispherical ocean of radius $\approx R_p$, then to a factor that is approximately unity:
\begin{equation}
\label{eqn:qcirc2}
\begin{split}
    q_{\rm circ} \approx c_p \rho \alpha^{1/3}(\Delta T)^{4/3}\kappa_T^{2/3}g^{1/3}P_K^{1/3}L^{-2/3} \\
    \approx 20\,\mathrm{W m}^{-2} \left(\Delta T/100\,K\right)^{4/3} M_P^{1/3}R_P^{-4/3} P_K^{-4/3},
\end{split}
\end{equation}
where $M_p$ and $R_p$ are in Earth units, $P_K$ is in days, and $\alpha$ is the thermal expansion coefficient.  Thus $q_{\rm circ}$ usually dominates over $q_{\rm int}$ (see Sec. \ref{sec:heat_sources}), even for modest surface temperature gradients $\Delta T /R_p$.  At least on airless planets, ocean circulation will be governed by the surface temperature gradient, but this heat transport is itself negligible compared to radiative terms ($>10^5$\,\wpersqm) and will not influence the surface temperature \citep{Kite2016}.  

In the absence of surface temperature-driven circulation (i.e., near-uniform temperatures under a thick atmosphere), magma oceans will experience vigorous convection driven by the escape of interior heat, and the temperature profile will follow an adiabat:
\begin{equation}
\label{eqn:adiabat}
    \frac{\partial T}{\partial z} = \frac{\alpha' g T}{c'_p},
\end{equation}
where $\alpha'$ and $c'_p$ are the thermal expansion coefficient and heat capacity, respectively, each modified to account for the volume change $\rho/\Delta \rho$ and latent heat $L$ release associated with crystallization in a magma ocean, or condensation of solids in an atmosphere \citep{Solomatov2015}.  The heat flow by conduction along such a gradient is $q_{\rm crit} = k \alpha'gT/c'_p$, where $k$  is the thermal conductivity.  If $q > q_{\rm crit}$ then convection occurs.  For completely melted silicates with a basalt- or peridotite-like composition, $k \approx 2$\,W\,m$^{-1}$K$^{-1}$, $\alpha \sim 10^{-4}$\,K$^{-1}$ and $c_p \approx 2 \times 10^3$ J\,kg$^{-1}$\,K\,$^{-1}$ \citep{Lesher2015} and with an Earth-like gravity, $q_{\rm crit} \sim 1$\,mW\,m$^{-2}$.  In a crystallizing magma ocean, $\alpha'/c'_p \approx \Delta \rho / (\rho L)$ \citep{Solomatov2015} and again $q_{\rm crit} \sim 1$\,mW\,m$^{-2}$.  Thus $q$ is usually $\gg q_{\rm crit}$ (see Sec. \ref{sec:heat_sources}) and, in the absence of other effects such as a temperature inversion or compositional gradient, convection will occur.  The \emph{style} of convection is characterized by the Grashof number, the ratio of convective to diffusive transport \citep{Turcotte2002}:  
\begin{equation}
Gr \approx \frac{g \rho \Delta \rho D^3}{\eta^2},
\end{equation}
where $\Delta \rho$ is the density change due to temperature differences or crystallization, $D$ is the magma ocean depth, $\eta \sim 1$\,Pa-s is the dynamic viscosity, $\rho = 2800$\,kg\,m$^{-1}$ \citep{Lesher2015} differences in mass density of a few percent, and $D$ is in km, Gr $\sim 10^{15} D^3$.  \footnote{Alternatively, the flow can be characterized by the Rayleigh number Ra = Gr\,Pr, where the Prandtl number Pr $\gtrsim 10^3$.}    This regime of ``hard" turbulence is characterized by large-scale circulation consisting of distinct hotter, ascending and cooler, descending plumes \citep{Solomatov2015}.  It is only when the temperature is near the lock-up temperature that the crystallization fraction approaches the maximum packing crystal fraction (experimentally estimated to be $\approx$60\%, \citealp{Solomatov2015}), and the viscosity soars does convection transition to disorganized ``soft" turbulence.  

On a rotating planet, flows can be rapid enough and the scales large enough that the Coriolis effects can be important, particularly since ultra-hot planets are likely to be synchronously rotating ($P \sim 1$ day).  Two other dimensionless parameters capture this effect: the Ekman number, which is the ratio of viscous forces to Coriolis acceleration:
\begin{equation}
    \label{eqn:Ekman}
    Ek = \frac{\eta P_{\rm rot}}{4\pi \rho D^2 \sin \theta},
\end{equation}
and the local Rossby number\footnote{To be distinguished from the global Rossby number in which $\sin \theta$ is omitted.} at latitude $\theta$, which is the ratio of the inertial forces to Coriolis acceleration:
\begin{equation}
\label{eqn:rossby}
    Ro = \frac{P_{\rm rot}}{4 \pi \tau \sin \theta}.
\end{equation}
Here, $\tau$ is a characteristic circulation time.  For molten silicates, $\eta \sim 1$ Pa-s and $Ek \ll 1$ signifying that inertial/pressure forces balance centrifugal terms (the condition for geostrophic flow) and viscous forces are significant only in a thin ``Ekman layer".  When $Ro \gg 1$ rotational effects are negligible.  For rapidly rotating planets where $R_o < 1$ meridional flows are suppressed, and convective flow becomes more two-dimensional and less efficient (discussed further in Sec. \ref{sec:dynamics}).

\subsubsection{Atmospheric Heat Transport}
\label{sec:atmosheattrans}

A substantial atmosphere can redistribute heat around the planet, i.e. from the illuminated substellar point to the poles and the night side.  In astronomical systems  where little is known about the details of the planet other than the assumption of synchronous rotation, heat transport is sometimes described as a dimensionless parameter $C$ in terms that modify the radiation balance equation:
\begin{equation}
\label{eqn:rad_balance}
    \sigma T^4 = \left(1-A\right)q_{\rm stellar}(\theta = 0) \left[(1-C) \cos \theta \mathcal{H}\left(\pi/2-\theta\right)  +  \frac{C}{4}\right],
\end{equation}
where $\mathcal{H}$ is the Heaviside step function.  $C=0$ represents no heat transport and $C=1$ represents efficient heat transport and a uniform temperature distribution \citep[e.,g.][]{Kreidberg2016}.  Heat transport by the atmosphere across a temperature difference $\Delta T$ will scale as
\begin{equation}
\label{eqn:atm1}
    q_{\rm atm} \approx \frac{c_p \rho H v \Delta T }{R_p},
\end{equation}
where $H$ is the atmosphere scale height and $v$ is a characteristic meridional velocity, which, again assuming geostrophic flow, such that $v = P_{\rm rot}\nabla p /(4 \pi \rho \sin \theta)$, and ideal gas behavior,
\begin{equation}
\label{eqn:atm2}
\begin{split}
    q_{\rm atm} \approx \frac{\gamma p P_{\rm rot}}{2 \pi (\gamma-1)g \sin \theta}\left(\frac{R\Delta T}{\mu R_p}\right)^2 \\
    \sim 10^4\,{\rm W m}^{-2}\,p_{\rm bar} P_K \Delta T^2 R_p^{-2},
\end{split}
\end{equation}
where $\gamma$ is the ratio of specific heats, $R$ is the gas constant, and $\mu$ is the molecular weight, the numerical expression being for a CO$_2$ atmosphere.  Comparing Eqn. \ref{eqn:atm2} to Eqn. \ref{eqn:stellar_energy} shows that temperature gradients of $<100$\,K are sufficient to redistribute most heat around a magma ocean world.  On a rapidly rotating planet, i.e. one that is tidally-locked to star on a short-period orbit (Sec. \ref{sec:exoplanets}), with a low Rossby number, transport may take the form of eastward equatorial jets (``superrotation") \citep{Showman2011}.  

\subsection{Putting it together: Magma ocean lifetime}

The characteristic cooling timescale of a magma ocean or pond will be the total latent heat of crystallization divided by the rate at which that energy can be lost to space:
\begin{equation}
    \tau_{\rm MO} = \frac{\rho D \mathcal{L}}{q},
\end{equation}
where $\rho$ is the magma density, $D$ the ocean depth, $\mathcal{L}$ the latent heat of crystallization, and $q$ the energy flux.  There are three limiting cases; in the first, a stable crust or ``lid" of thickness $L$ forms and heat transfer is limited by heat transport through the lid.  That time scale is $D L \rho \mathcal{L}/(k \Delta T \sim 2 {\rm Gyr} (L/100 {\rm km})(D/1000 {\rm km})$, thus deep magma oceans can be stable for a significant fraction of a planetary system's age \citep{orourke2020}.  (Lids thicker than $\sim$100 km will experience subsolidus convection and not prolong crystallization.)  In the second case, no stable lid forms but heat transfer is limited by a thick steam atmosphere to $q \approx$ 400 \wpersqm\ (Sec. \ref{sec:atmospheres}) and the timescale will be $\sim$ 3 ($D$/100 km) kyr.  The third case is a planet with no or a thin atmosphere, with an exposed magma ocean radiating its energy to space at $\approx$1300\,K, i.e. $q = 1.6 \times 10^5$ \wpersqm.   Unless balanced by a heat source of equal magnitude (Sec. \ref{sec:heat_sources}), a lava ocean cannot persist under such circumstance and crystallization occurs in decades. These timescales are also the relevant timescales for the entirely of a lava ocean to respond to changes in heat input, although surface boundary layers can respond far more rapidly \citep{Kite2016}. Under such conditions, heat loss is limited by heat transport within the lava ocean, by convection and/or crystal settling.  Cooling and crystallization at the surface of a magma ocean will release latent heat and stabilize the magma column against convection.   If individual crystals cannot settle quickly enough a surface mush layer will develop until the density difference causes overturn via a Rayleigh-Taylor instability \citep{Michioka2005,Culha2020}.  If crystals are more buoyant than the melt then a stable crust will develop, as happened on the Moon (Sec. \ref{sec:solarsystem}).

\section{Lava lakes as Analogs in the Solar System}
\label{sec:lavalakes}

\subsection{Why Study Lava Lakes?}
\label{sec:why_lakes}

Magmas are complex mixtures of liquids, crystalline solids, and gases, and their properties, especially on large scales, are challenging to investigate in the laboratory or simulate numerically.  With the possible exceptions of Io (Sec. \ref{sec:io_lakes}) and Venus (Sec. \ref{sec:venus}) there are no longer magma oceans in the Solar System and the largest known bodies of magma at the surface of any Solar System body are lava lakes on Earth and (probably) Io.  Lava lakes, particularly those on Earth (Table \ref{tbl:lavalakes}), have been studied extensively, and are useful analogs to understand fundamental processes such as convection, degassing, crystallization, and crust formation \citep{Lev2019} that could be relevant to magma oceans.  However, issues of scaling from lava lakes to lava worlds must be considered, and there are numerous caveats.  An obvious difference between lava lakes and magma oceans is size; all terrestrial lava lakes are $<$500\,m across \citep{Lev2019}, and the largest lava lake on Io is 200\,km in size \citep{Matson2006}.  Another is the geometry of the heat sources, i.e. advective heat transfer from below vs. stellar irradiation from above.  Thus, it is important to consider how processes scale through dimensional analysis, i.e. dimensionless numbers that typically relate competing effects.  It is also important to remember that the range of compositions and temperatures that are spanned by the lakes of the Solar System is probably limited compared to what is proposed or inferred for exoplanets.  The general framework to begin tackling such scaling problems is beginning to be developed, but tying together multi-scale and multi-phase flow is at the forefront of modeling \citep[e.g.,][]{Keller2019}.  

\begin{table*}
\caption{Extant or Recent Active Lava Lakes ($\dagger$ = drained or solidified in 2018)}\label{tbl:lavalakes}
\begin{tabular}{l|l|l|l|l|l|l}
\hline
\textbf{volcano} & \textbf{size [m]} & \textbf{temp [K]} & \textbf{$\eta$ [Pa-s]} & \textbf{\% crust} & \textbf{chemistry} & \textbf{references} \\ 
\hline
Ambrym$^\dagger$ & 20/50 & 1350--1400  & $10^3$ & 20--30 & basaltic & \cite{Firth2016,Shreve2019}\\
Erebus & 30 & 1250 & 5000 & 50 & phonolitic & \cite{Birnbaum2020}\\
Erta\,'Ale & 34-100 & 1413 & 30 & 95 & basaltic & \cite{Harris2005,Spampinato2008} \\
Kilauea$^\dagger$ & 250 & 1443 & $10^2$ & 50--95 & basaltic & \cite{Patrick2019}\\
Masaya & 50 & 1500 & 630 & 30 & basaltic & \cite{Aiuppa2018,Pering2019} \\
Mt. Michael & 110 & 1262--1552 & ? &?  & ? & \cite{Gray2019} \\
Nyamuragira$^\dagger$ & 20--200 & ? & ? & ? & nephelenitic & \cite{Campion2014,Coppola2016} \\
Nyiragongo & 50--240 & 1370-1400  & 60-150  & 90 & nephelenitic & \cite{Valade2018} \\
Villarica & 30 & 1408 & 850 & 20--30 & basalt/andesitic & \cite{Witter2004} \\

\hline
\end{tabular}
\end{table*}

\subsection{Lava Lakes on Earth}
\label{sec:earth_lakes}

Terrestrial lava lakes are our only opportunity at present to directly study the behavior of large bodies of magma.  Lava lakes are categorized as either "active" lakes that form on top of a volcanic conduit and are constantly fed fresh magma (including gas) from depth, or "passive" lakes that form when lava travels from its initial eruption site and collects in a topographic depression then cools passively \citep[e.g.,][]{tilling1987}.   To compare with magma oceans which are expected to have some circulation (Sec. \ref{sec:heat_transfer}), we will only consider active lakes.  In these, hotter, gas-rich magma erupts from a conduit, then cools, de-gasses and descends back to the conduit.  Most long-lived lava lakes on Earth have been subjects of extensive study (Table \ref{tbl:lavalakes}).  Most of these lakes are basaltic or \gls{andesitic}-basaltic in composition, though other compositions are represented. 

Investigations of terrestrial lava lakes have measured gas emission to determine magma exchange rate \citep{Lev2019} and the stability of the lake against draining away \citep{Witham2006}. Many lava lake studies have described heat flow \citep{Harris1999,Cipar2012,Coppola2016} and surface velocity \citep{Peters2014,Valade2018,Pering2019}. A few studies have also used analog models \citep{witham2006analogue}. These large and varied data on lake behavior have permitted the development and testing of sophisticated models of lake dynamics.  Observations of surface motion have constrained models of convection \citep[e.g.,][]{Harris2008,Birnbaum2020}. \citet{Harris2008} used observations of surface motion and heat flux at Erta 'Ale to develop a model of lava lake convection, including density changes due to both crystallization and temperature (but not vesiculation, see below).  This model shows that convection can be driven by fractional density differences as small 0.2\%.

Vesiculation (formation of gas bubbles) influences the dynamics of lava lakes via its lowering of bulk density.  The effect of vesiculation on convection in Ray Lake (Erebus) was quantitatively addressed by \citet{Birnbaum2020}.  Their 2-d model, which combines the effects of cooling, crystallization, and vesiculation on density, exhibits two regimes of circulation: \gls{diapirs} occur when there is a large density contrast between freshly supplied and recirculated lava.  Pulsating plumes occur when the density difference is small between the older lake lava and newly supplied lava.  \citet{Birnbaum2020} find that that the relative rate of gas loss to gas influx govern the density difference and hence the circulation regime.  However, the Erebus lava lake is atypical in terms of its \gls{phonolitic} composition and higher viscosity (10\textsuperscript{4} Pa-sec, as compared to the typical 10\textsuperscript{2} Pa-sec for basalts; \citep{sweeney2008}), and it remains to be shown how this model applies to other lakes.

A numerical study by \citet{Witham2006} demonstrates that the long-term stability of lava lakes is strongly dependent on the relative rates of gas supply and loss. Lakes that have large exsolved gas fractions are typically unstable and will collapse and drain away with system perturbations (e.g., the Pu'u O'o and Kilauea lava lakes; \citep{Witham2006}). \citet{Lev2019} synthesized data available on the main active lava lakes and found a strong relationship between lake surface velocity and the ratio of total gas flux to lava lake area. Additionally, they categorized lava lakes into two types of surface behavior - chaotic and organized. Surface behavior controls the relative rates of gas loss. Chaotic lakes tend to lose gas more efficiently, and organized lava lakes tend to develop crusts that trap gas, thus reducing gas loss.

\subsection{Lava Lakes on Io}
\label{sec:io_lakes}

Jupiter's satellite Io was first recognized as volcanically active from data collected by the \emph{Voyager 1} spacecraft \citep{Morabito1979}.  Partial melting of the interior is caused by dissipation of tides on Io's orbit, the eccentricity of which is maintained by a \gls{Laplace resonance} with the other Galilean satellites.  Subsequent observations by \emph{Galileo} showed that eruptions occur primarily from \emph{paterae}, volcanic-tectonic depressions similar to calderas on Earth \citep[e.g.,][]{Davies2001,Radebaugh2001}. Resurfacing by lava is largely confined to these paterae, which are thought to host active lava lakes, with the rest of the satellite's surface being mostly covered by plume deposits \citep{Lopes2004}. 

The largest and most well-studied of these paterae is Loki Patera. Despite accounting for only 0.07\% of Io's surface, it accounts for 10-20\% of Io's total heat output \citep{Veeder1994,Matson2006}.   Loki was first hypothesized to contain a lava lake by \citet{Rathbun2002}, but whether the patera actually contains a lake or is only resurfaced by lava flows is still debated \citep[e.g.,][]{davies2003,dePater2004,Matson2006,Gregg2008,DeKleer2017,deKleer2017b,DePater2017}. Recent high-resolution ground-based observations in the infrared support the lava lake hypothesis \citep{Conrad2015,DePater2017}.  This illustrates the difficulty of interpreting the limited information provided by remote sensing (temperatures, fluxes, and morphology) and using it to test models, a problem that is vastly more exacerbated in the case of exoplanets, where at best, only fluxes and spectra from the unresolved planet are available (Sec. \ref{sec:exoplanets}).  The current model of Loki lava lake dynamics is one of foundering of denser crust, producing one or multiple resurfacing wave(s) of lava that propagate(s) across the patera \citep{Rathbun2002,Matson2006,DeKleer2017}.  This is analogous to multiple locations of down-welling in many terrestrial lava lakes \citep[e.g.,][]{Patrick2017}.  The densification and foundering of the crust due to cooling and degassing might be accelerated by by repeated extension and compression of the crust by the tidal stresses experience by Io on its eccentric orbit around Jupiter \citep{Matson2006}. 

Io may distinguish itself as the only body in the Solar System for which there is at least indirect evidence -- induction by Jupiter's field in the satellite's interior  -- of an interior magma ocean \citep{Khurana2011}.  \citet{deKleer2019} also describe orbital periodicity in volcanic activity that could be a result of changes in magma conduits from tidal stresses.  The magma ocean interpretation remains tentative, with \citet{Bloecker2018} showing that the changes in local magnetic field observed by the \emph{Galileo} spacecraft could also be explained by plasma interactions with Io's atmosphere.

\subsection{Scaling from Lava Lakes to Lava Worlds}
\label{sec:scaling}

\subsubsection{Laminar vs. turbulent flow regimes}
\label{sec:laminar_turbulent}

One to two orders of magnitude in scale separate magma oceans from the lava lakes of Io; another three orders of magnitude separate Io from the much more accessible terrestrial counterparts.  This size difference translates into a difference in flow regime, as characterized by the Reynolds number
\begin{equation}
\label{eqn:reynolds}
    Re = \frac{\rho v L}{\eta}
\end{equation}
where $L$ and $v$ are the characteristic size and velocity, respectively, of a convective eddy, and the transition from laminar to turbulent flow occurs at $Re \approx 200$ \citep{Harris2008}.   Eddy sizes are typically $\lesssim$10\,m and for basaltic lava viscosities ($\sim10^2$ Pa-sec) and $v \sim0.1-1$ m\,sec$^{-1}$, convection in most lava lakes is laminar.  However, eddy sizes will scale with the depth of magma oceans (many km) and Re values will fall well inside the turbulent regime \citep{Solomatov2015}.  Higher temperatures, \gls{superliquidus} or silica-poor (i.e, low polymerization) melts on lava worlds will have substantially lower viscosity (see Sec. \ref{sec:heat_transfer}), with yet higher Reynolds numbers and stronger turbulence (see Sec. \ref{sec:heat_transfer}).  Turbulence affects heat transport but also mixing, i.e. mixing of crystallizing solids vs. gravitational settling of those solids into a \gls{cumulate} pile \citep{Solomatov2015}.  Convection in terrestrial lava lakes is clearly not representative of these processes in magma oceans.  Investigations of the lava lakes of Io, which are substantially larger, and potentially much hotter ($>$1600\,K) than basaltic magmas \citep{Davies2001} and thus presumably less viscous, could partly bridge this gap.     

\subsubsection{Crystallization}

Along with vesiculation, crystallization produces differences in density that drive circulation within magma bodies.  In a magma with a crystalline component, the latent heat absorbed or released will buffer temperatures between the solidus and liquidus and reduce that variable's contribution to density.   Crystals and as well as bubbles dramatically increase the viscosity of magmas; viscosity increases with crystal fraction and depends on crystal aspect ratio \citep{Mader2013}.  These effects will persist in turbulent, well-mixed magmas \citep{Costa2009}.  In low Re, crystal-rich magmas, crystals will settle collectively in down-welling plumes, initiating a process known as crystal-driven convection \citep{Michioka2005,Culha2020}.

Crystallization will not be a factor in magma oceans that are at superliquidus temperatures, e.g., at the substellar point of highly-irradiated, tidally-locked planets on very short-period orbits (see Sec. \ref{sec:exoplanets}).  They will apply, however, to cooling, crystallizing magma oceans and to the regions of tidally planets close to the terminator, on the ``shores" of a magma ocean where temperatures are well below the liquidus, crystal fractions reach the ``lock up" value of $\approx$60\% \citep{Leger2011,Kite2016}, a finite \gls{plastic yield strength} becomes important \citep{Pinkerton1992} and ultimately there is a transition to subsolidus-style convection.     

\subsubsection{Degassing and vesicularity}
\label{sec:degassing}

Bubbles will form and grow (vesiculation) in ascending magma as the ambient pressure falls below the saturation pressure for any dissolved volatile, e.g., H$_2$O, CO$_2$, and SO$_2$.  Bubbles lower the bulk density and increase the buoyancy of the magma, potentially accelerating its ascent and further vesiculation.  The degree to which the acceleration happens depends on the viscosity of the magma, which governs how strongly bubbles are retained; in high-viscosity rhyolitic melts bubbles are strongly coupled to the melt, and and magma ascent and bubble growth accelerates \citep{Cashman2013}. Bubbles more readily escape from low-viscosity basaltic melts and acceleration by vesiculation is less pronounced.  The two scenarios will lead to vigorous upwelling through the magma column vs. a more passive style of convection, respectively.  The dynamic of vesiculation in and gas escape from a magma column is predicted to produce a wide range in lava lake behavior \citep{Witham2006}.  

While vesiculation can be profoundly important for active lava lakes, it is unclear if it has any role in convecting magma oceans.  Without in-gassing, the magma ocean will devolatilize and equilibrate with any atmosphere in a few convective turnover times.  In-gassing could occur during melting of the underlying mantle in response to the evaporation of the ocean and advection of more volatile constituents to cooler regions of the planet \citep{Kite2016}.  It could also occur if the planet is actively accreting gas from the protoplanetary disk \citep{Olson_2019}.  Or volatiles might be trapped as bubbles in a crust and then recycled into the interior when the crust eventually founders (see Sec. \ref{sec:surface} below).  A very thick, Venus-like atmosphere (see Sec. \ref{sec:venus-like}) can suppress bubble formation, regardless of whether a source of volatiles is maintained.  

\subsubsection{Surface dynamics and crust formation}
\label{sec:surface}

Wherever the equilibrium surface temperature is below the solidus, crystallization can occur in a thermal boundary layer at the surface, and these crystals can aggregate into a crust which will usually founder because of the higher density of solids.  In some circumstances the crystals are less dense than the melt, such as the flotation crust of the Moon (see Sec. \ref{sec:moon}).  

Otherwise, crust formation depends on the rate of radiative cooling of the surface and the convective turnover time (which together regulate the thickness of the boundary layer) and the (negative) buoyancy of the crust (which regulates foundering).  In terrestrial lava lakes, the process is thought to occur as hotter, relatively buoyant lava erupts through the crust and floods it \citep{Stovall2009}. Bubbles can be trapped in the crust, increasing its buoyancy and hence lifetime.  On the other hand, vigorous vesiculation, that accelerates upwelling and, on larger scales, tides \citep{Matson2006}, can disrupt the boundary layer and inhibit crust formation.

\begin{figure*}
	\centering
		\includegraphics[width=\textwidth]{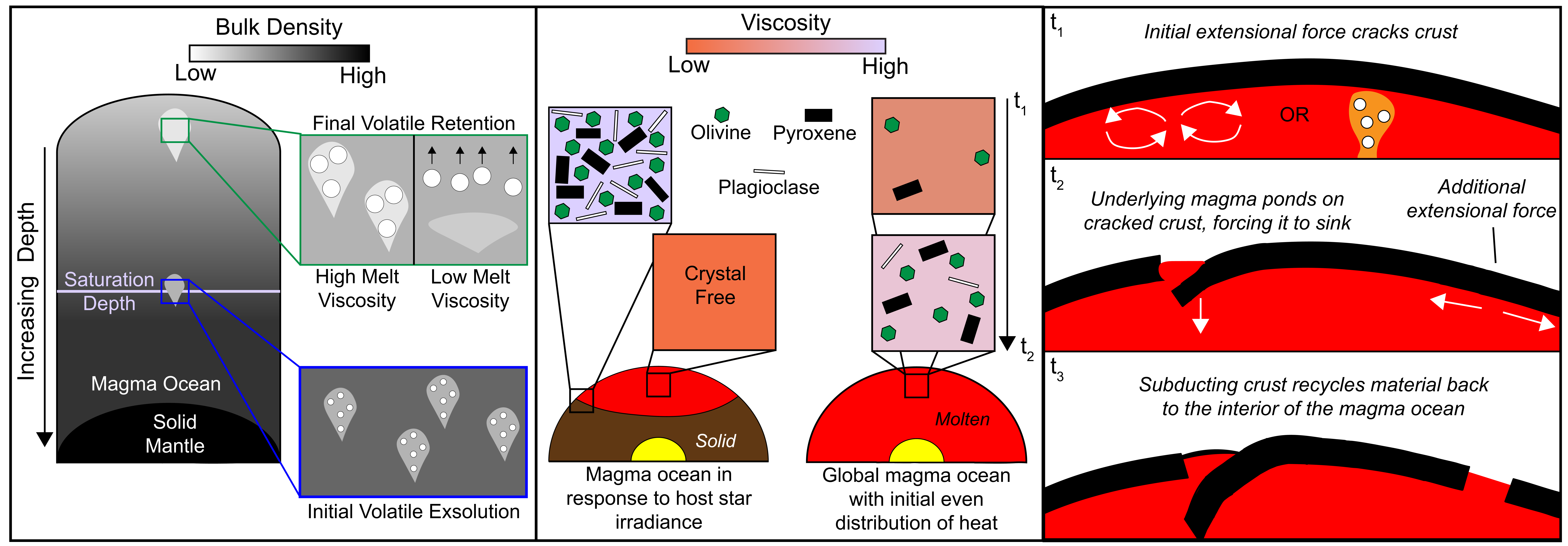}
	\caption{Schematic representation of processes in lava lakes and their potential role in the dynamics of magma oceans/seas.  Left: Effect of vesiculation (bubble formation) and melt viscosity on the rise of a plume.  Bubbles will initially accelerate the rise of a volatile-rich plume, but the outcome is also determined by whether the melt is viscous enough to retain bubbles. The gray scale indicates bulk density.  Middle: Variations in bulk viscosity of a magma ocean depend on thermal evolution and crystallization.  In a magma ocean maintained by stellar irradiance on an airless planet (left), crystallization will be minimal except at "shores" near the terminator. Crystallization in a cooling global magma ocean (right) will proceed through time, increasing bulk viscosity.  Background colors in the boxes indicate bulk viscosity; other colors are used merely to distinguish components.  Right: Foundering of the crust of a lava lake, as described by \citet{Stovall2009}. Tension is applied to a dense, mature crust, e.g., by underlying convection currents or a rapidly ascending plume.  The crust fails, and lava floods the broken crust, causing it to sink.  This foundering can impose additional extensional stress in the crust and cause failure at other locations.}
	\label{FIG:oceancartoon}
\end{figure*}

\section{Lava Worlds in the Early Solar System}
\label{sec:solarsystem}

As a consequence of energy liberated by accretion and short-lived radionuclide decay during planet formation (Sec. \ref{sec:energetics}), magma oceans could have been widespread in the early Solar System (Table \ref{tbl:bodies}).  The concept of a magma ocean was originally introduced to explain lunar geology as revealed by samples returned by the Apollo 11 mission \citep{Wood1970}.  The Moon is both the best preserved and most accessible ancient surface, and the Lunar Magma Ocean (LMO) remains a touchstone against which other bodies are compared and models are tested \citep{Warren1985}, but the LMO is unlikely to be representative of diverse magma oceans in the past inner Solar System.     

\subsection{The Moon}
\label{sec:moon}

While the generally accepted theory of lunar formation -- one or more giant impacts \citep{hartmann1975,cameron1976} -- followed the idea of a LMO, the former is necessary for any LMO model because the exact formation scenario determines the initial extent of melting of the Moon.  A scenario with a single giant impact of a Mars-sized impactor (dubbed "Theia") successfully explains the Moon's bulk composition, small iron core, and paucity of volatiles \citep{halliday2000}.  However, in its simplest form this scenario fails to explain the extreme chemical and isotopic similarity, particularly oxygen isotopes, between Earth and the Moon\citep{wiechert2001,Young2016,Greenwood2018}.   This is because models predict that the Moon should contain proportionally more material from Theia and be chemically distinct from Earth \citep{Asphaug2014}.  One explanation is that Theia somehow accreted on the same orbit and from the same reservoir as Earth \citep{Belbruno2005}.  Alternatively, the Moon could have accreted from the cumulative debris of many smaller impacts to which the impactors contributed much less \citep{rufu2017}.  At the other extreme, a single large impactor could have completely homogenized the two bodies \citep{Canup2012}. Or, a large, high angle impact completely vaporized the proto-Earth, forming a rapidly rotating ellipsoid of silicate gas ("synestia") from which condensed Earth and smaller bodies that formed the Moon \citep{lock2018}.  Increasing numbers and precision of measurements might eventually resolve the question:  \citet{Cano2020} report small but significant variation in lunar oxygen isotopes that suggest the Moon's \emph{interior} is different from Earth.

The original evidence for a LMO, i.e. a crystallizing global melt, are the petrological characteristics of the different suites of lunar crustal rocks, chiefly widespread \gls{anorthosite} in the lunar highlands.  Anothosites are low-density volcanic rocks that are uncommon on Earth and typically found only in large igneous intrusions. However, they are widespread in the lunar crust, having been identified in lunar samples, lunar meteorites \citep{Nagaoka2014}, and by remote sensing \citep{Yamamoto2012,Cheek2013}.  Shortly after the Apollo 11 mission, \citet{Smith1970} and \citet{Wood1970} proposed that lunar anorthosites were produced by accumulation of crystals into a buoyant crust on a LMO.   For separation to occur, buoyancy forces must overcome viscous forces and this is favored in an anhydrous magma ocean that crystallizes at higher temperature and a lower viscosity.  A volatile-depleted Moon is the expected outcome of the giant impact scenario, but this consistency with observations has been upset by accumulating evidence for lunar water \citep{McCubbin2013,Hui2013} and has prompted more sophisticated approaches to reconciliation \citep{Lin2017b}.  

The oldest suite of rocks on the Moon, the ferrous iron-containing (ferroan) anorthosite suite and the magnesium-rich (magnesian) suite are characterized by a mixture of minerals that crystallized first or early from a \gls{primitive melt} \citep{Dymek1975,Papike1997}.  Younger rocks such as the KREEP basalts (see below) and the alkali suite anorthosites \citep{SNYDER1995} contain quartz and alkali feldspars implying that they crystallized from more evolved melts \citep{Dowty1976,SNYDER1995}.  This favours a progressively evolving source, as one would expect in a crystallizing magma ocean.

Progressive segregation of crystals from the melt would have driven chemical evolution in the LMO and ``fingerprinted" lunar rocks. One such fingerprint is enrichment of potassium, rare earth elements, and phosphorous (KREEP) in younger lunar rocks such as \gls{mare} basalts, the alkali suite anorthosites, and KREEP basalts \citep{Rapp2018}.   The most evolved lunar rocks form from the final residual liquid before complete crystallization, the "ur-KREEP" melt, \citep{warren1979}.  Elements such as europium and strontium that are \gls{compatible} in plagioclase are depleted from the melt  \citep{Walker1975, Papike1996}.  This is reflected in negative anomalies (depletion) of Eu and Sr in all lunar rocks that formed from melts that are younger than the anorthosite crust \citep{Papike1996}:  

\subsection{Earth}
\label{sec:earth}

The energy released during the giant impact(s) that formed the Moon is also expected to have formed a terrestrial magma ocean (TMO).  As in the case of the Moon, the extent of melting and degree of mixing between the bodies depends on the impact scenario \citep{Nakajima2015}.  Unlike the Moon, plagioclase is not expected to crystallize early at higher pressures in the TMO and form a floatation crust \citep{Elkins-Tanton2012}.  Also unlike the Moon, plate tectonics and crustal recycling have destroyed the terrestrial rock record older than 4 Gyr, erasing any petrologic evidence for a TMO.  Evidence for the TMO is therefore strictly geochemical.  The mantle abundance of \gls{siderophile} elements are elevated with respect to that  predicted by chemical equilibrium between silicates and core-forming metallic iron at the surface \citep{ringwood1966}.  These elements become less siderophilic and there is better agreement with mantle abundances if equilibration happened at high pressure, i.e. at the base of a deep magma ocean where metallic iron may have temporarily pooled \citep{li1996,righter1997}.  Some (but not all) of this excess, particularly in the highly siderophilic elements, could also be due to a late ``veneer" of siderophile-rich, chondritic material that never equilibrated with the core \citep{Kruijer2015,Creech2016}.  This also must be reconciled with the isotopic evolution of Earth \citep{Fischer-Godde2020} and the isotopic similarity between Earth and Moon (see Sec. \ref{sec:moon}).

Another line of evidence for a TMO is the disparity in neodymium (Nd) isotopes between Earth and primitive chondritic meteorites.  \citet{boyet2005} reported that chondritic meteorites have a $^{142}$Nd/$^{144}$Nd that is 20 ppm lower than the majority of terrestrial rocks. (See also \citet{Cipriani2011}.) This requires either that Earth accreted from material with a non-chondritic $^{142}$Nd/$^{144}$Nd that is not represented in meteorite collections, or that the isotopes were fractionated between reservoirs within the mantle, e.g. by global melting.  The corollary is that there must be an unsampled reservoir in Earth's interior with $^{142}$Nd/$^{144}$Nd that is sub-chondritic.  Age constraints using multiple radiometric isotope systems indicate that this global differentiation event took place within 200 Myr of "time zero" of Solar System formation \citep{Caro2005,caro2006}, bracketing estimates of the time of lunar formation. 

\subsection{Mars}
\label{sec:mars}

Evidence for a Martian magma ocean and crustal differentiation has accumulated from analyses of \gls{SNC meteorites} \citep{mcsween1985}.  Radiometric dates point to a global differentiation event (and by inference, a magma ocean) within the first 50 Myr of ``time zero" \citep{wood1982, Borg1997}.  Radiometric Nd ages suggest that the magma ocean had a protracted lifetime, perhaps due to a thick proto-atmosphere \citep{Debaille2007}.   However, \citet{Bouvier2018} and \citet{Kruijer2020} used the U-Pb and the Mn-Cr radioisotope systems, respectively, to infer that crystallization of any magma ocean was complete by about 20 Myr.   

Recently, the idea of a magma ocean at the base of the Martian mantle has been suggested. \citet{Zeff2019} predicted that the residual liquids of melts produced by impacts on Mars would be denser than the surrounding mantle and would sink and accumulate at the core-mantle boundary. This  explains the lack of an internally-generated magnetic field, as this hot layer would limit the escape of heat from the core that would otherwise drive a magnetodynamo.  Melts become negatively buoyant because of the comparatively high Fe content of the martian mantle and the tendency of heavier Fe to remain in a melt \citep{taylor2013}; this process might not operate in other, more iron-poor mantles in the inner Solar System.  

\subsection{Mercury}
\label{sec:mercury}

Evidence for a past Mercurian magma ocean is based on remote sensing, particularly by the recent \emph{Messenger} mission \citep{Solomon2007}, as no meteorites have been tied to Mercury.   Mercury has many unusual characteristics that point to a magma ocean formed under comparatively reducing (low \gls{oxygen fugacity} or \fotwo) conditions \citep{Malavergne2010,Cartier2019}.  The crust of Mercury is iron-poor but sulphur-rich relative to other planets \citep{Nittler2011,evans2012}, which could reflect S-rich but Fe-poor magmas that would erupt during the late stages of crystallization of a low \fotwo\ magma ocean \citep{Zolotov2013}.  A more volatile-rich crust would help explain Mercury's substantial phase of late-stage volcanism \citep{Thomas2014}.   Under reducing conditions, carbon becomes more of a lithophile than a siderophile, and graphite can 
saturate in a magma ocean:  Regions of unusually low albedo could represent exposures of a global graphite-rich flotation crust \citep{peplowski2016}, and such a crust could have allowed a magma ocean to survive for longer than otherwise.  In contrast, silicon becomes a siderophile at low \fotwo, and dissolution of significant Si into the metallic core could partly explain Mercury's proportionally larger core \citep{Cartier2019}.

\subsection{Venus}
\label{sec:venus}

Despite its importance as Earth's "sister" planet, we have only cursory knowledge of the composition and geologic history of Venus due to the absence of samples and the presence of a thick atmosphere which blocks most remote sensing of its surface. The surface appears very young ($\lesssim$1\,Gyr) as a result of some global resurfacing event that destroyed most or all of any older crust \citep{phillips1992}.   Despite these gaps in our understanding of Venusian geology, the similarities in its interior structure to that of Earth \citep{aitta2012} and by inference, accretion history, suggest an early magma ocean phase.  Of course, Venus has no Moon.  It is also possible that a basal magma ocean  has survived to the present time.   \citet{orourke2020} shows that, should a basal magma ocean have formed, the absence of plate tectonics and reduced heat flow from the mantle would have allowed such a magma ocean to persist to the present.  The thermal buffering provided by its ongoing crystallization would suppress heat flow from the core and prevent the operation of a magnetodynamo, explaining why Venus currently lacks a magnetic field. 

\subsection{Vesta}
\label{sec:vesta}

The large main belt asteroid Vesta is the only sub-planetary body widely agreed to have had a magma ocean.   Vesta is thought to be the source of the Howardite-eucrite-diogenite (HED) suite of meteorites that are grouped by their similar oxygen isotopes and linked to the asteroid by spectral similarities \citep{McCord1970, Binzel1993}, and the existence of both an impact crater and a family of smaller, related steroids (``vestoids") that could be debris from the impact and the intermediary source of HED meteorites \citep{Binzel1993}.  This connection has been strengthened with observations by the \emph{Dawn} mission \citep{McSween2013}.  There is direct, but model-based evidence for a magma ocean, i.e.., petrogenesis of the HED suite by \gls{equilibrium crystallization} followed by \gls{fractional crystallization} of a chondritic parent melt \citep{righter1997}.  There is also indirect evidence, i.e., the existence of an iron core as supported by the depletion of moderately siderophile elements in the HEDs \citep{righter1997} and a paleomagnetic signatures of a magnetic field during crystallization \citep{Fu2012}.

Though the existence of a Vestan magma ocean is widely accepted, the extent of melting of the body is debated. \citet{Mandler2013} argue that Vesta's primary plausible heat source -- the decay of \altwosix\ (see Sec. \ref{sec:heat_sources}) was efficiently retained by a conductive lid and as a consequence the interior was entirely molten.  In contrast, \citet{Neumann2014} argue that \altwosix\ will partition (along with its stable counterpart $^{27}$Al) into silicate melt and migrate toward the surface faster than the mean life of the isotope, thus reducing the effectiveness of heating and maintaining a shallow ocean for about 150 Myr (see also \citet{Moskovitz2011}).  They invoke another radioisotope -- $^{60}$Fe -- to explain the convection in the core needed for a magnetodynamo, but the abundance of this isotope is now considered to have been much lower than previously thought \citep{Tang2012}, and there are alternative mechanisms to drive convection \citep{Nimmo2009}.  Likewise, there is a range of potential scenarios for the crystallization of the magma ocean: \citet{Kawabata2017} showed that the chemistry (i.e., Mg abundance) of the cumulate crust that is the source of the HED meteorites can be used to constrain the thickness of the conductive crust and the size and hence settling time of crystals.   

\subsection{Other Small Bodies}
\label{sec:othersmall}

The canonical theory of planet formation includes a phase with many small bodies and it seems implausible that Vesta was the only such body with a magma ocean, although it might be the only one that has survived intact \citep[c.f.,][]{Wilson2017}.  The meteorite collections contain hints that there may have been others.  Several related achondritic meteorites, Northwest Africa (NWA) 7325, 8014 and 8486, contain abundant plagioclase, have enrichments in Eu and Sr like those seen in lunar anorthosites, and have a bulk chemistry consistent with that derived from melting of a \gls{pyroxene}-rich anorthosite \citep{Frossard2019}, all suggesting they derive from a parent body which had a lunar-like anorthosite crust.  However, the bulk element abundances rule out a lunar origin, and these meteorite are unrelated to any known meteorite groups.  Potential counterparts to these enigmatic meteorites could be searched for among the Main Asteroid Belt \citep{Reddy2015}.

\begin{table*}[width=1.8\linewidth,cols=3,pos=h]
\caption{Solar System Bodies with Suspected Past or Present Magma Oceans.}\label{tbl:bodies}
\begin{tabular*}{\tblwidth}{l|l|l|l}
\toprule
\textbf{object} & \textbf{interior structure}$^a$ & \textbf{evidence} & \textbf{reference(s)}\\
\midrule
\multirow{4}{*}{Moon}& \multirow{4}{*}{\includegraphics[scale=0.1,angle=270]{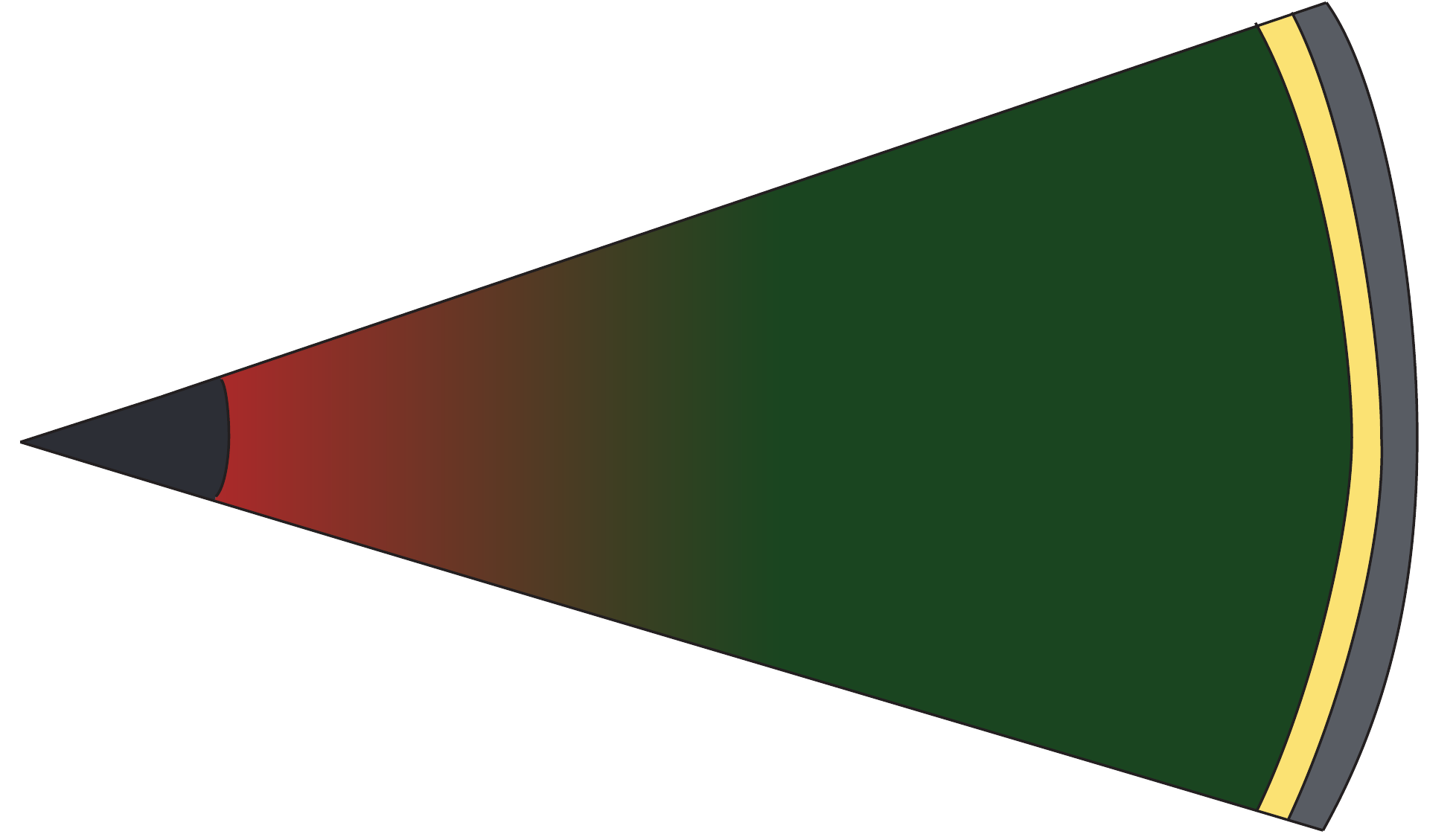}} & anorthosite flotation crust & \cite{Wood1970}\\
&& Fe-rich lavas & \cite{Borg2019}\\
&&KREEP enrichments&\cite{warren1979}\\
&&Oxygen isotopes&\cite{Cano2020}\\
\midrule
\multirow{4}{*}{Earth}&\multirow{4}{*}{\includegraphics[scale=0.1,angle=270]{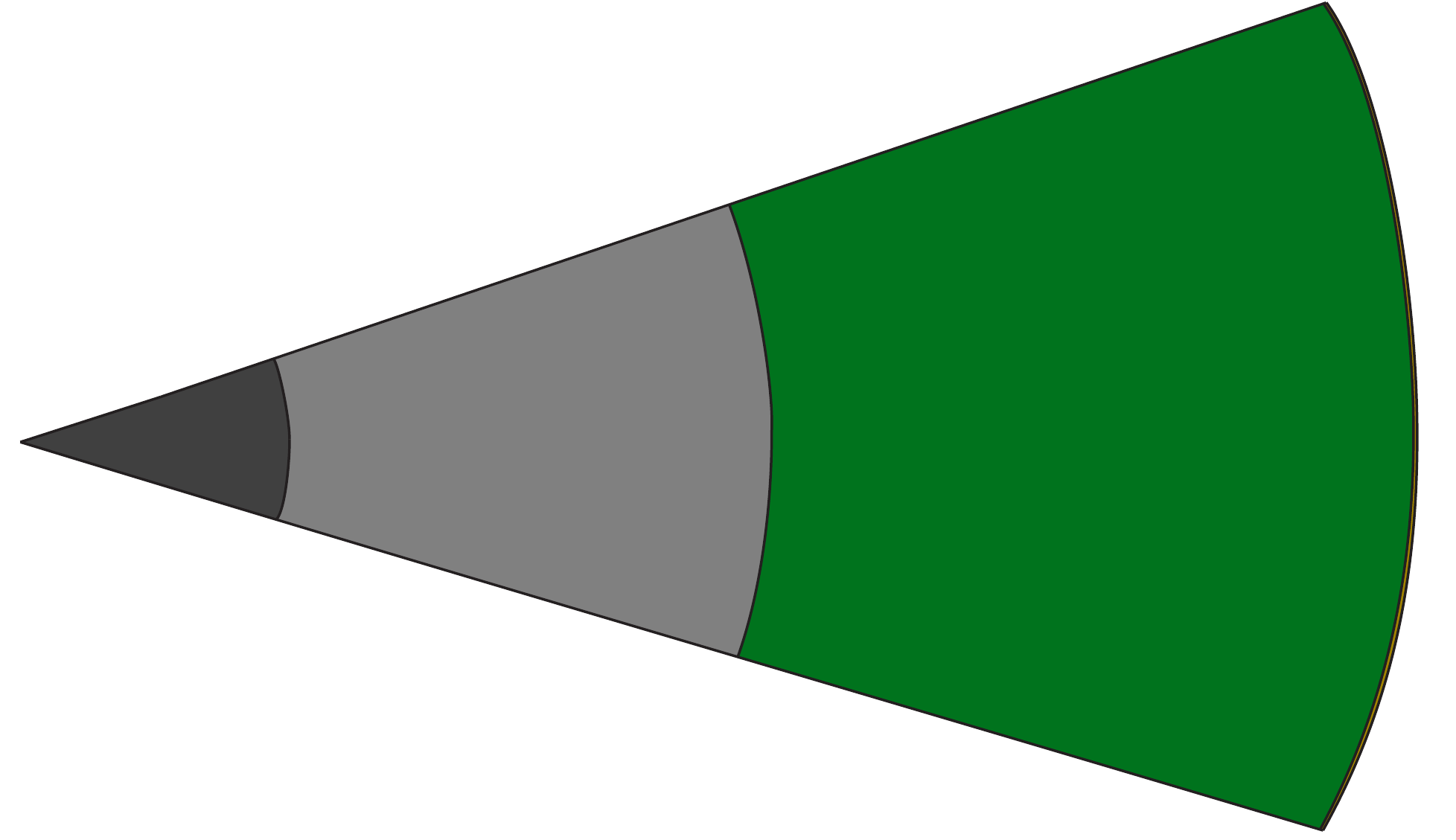}} & Nd isotopes & \cite{boyet2005,Cipriani2011} \\
& &siderophile abundances in mantle & \cite{li1996,righter1997}\\
& & & \\
& & & \\
\midrule
\multirow{4}{*}{Mercury}&\multirow{4}{*}{\includegraphics[scale=0.1,angle=270]{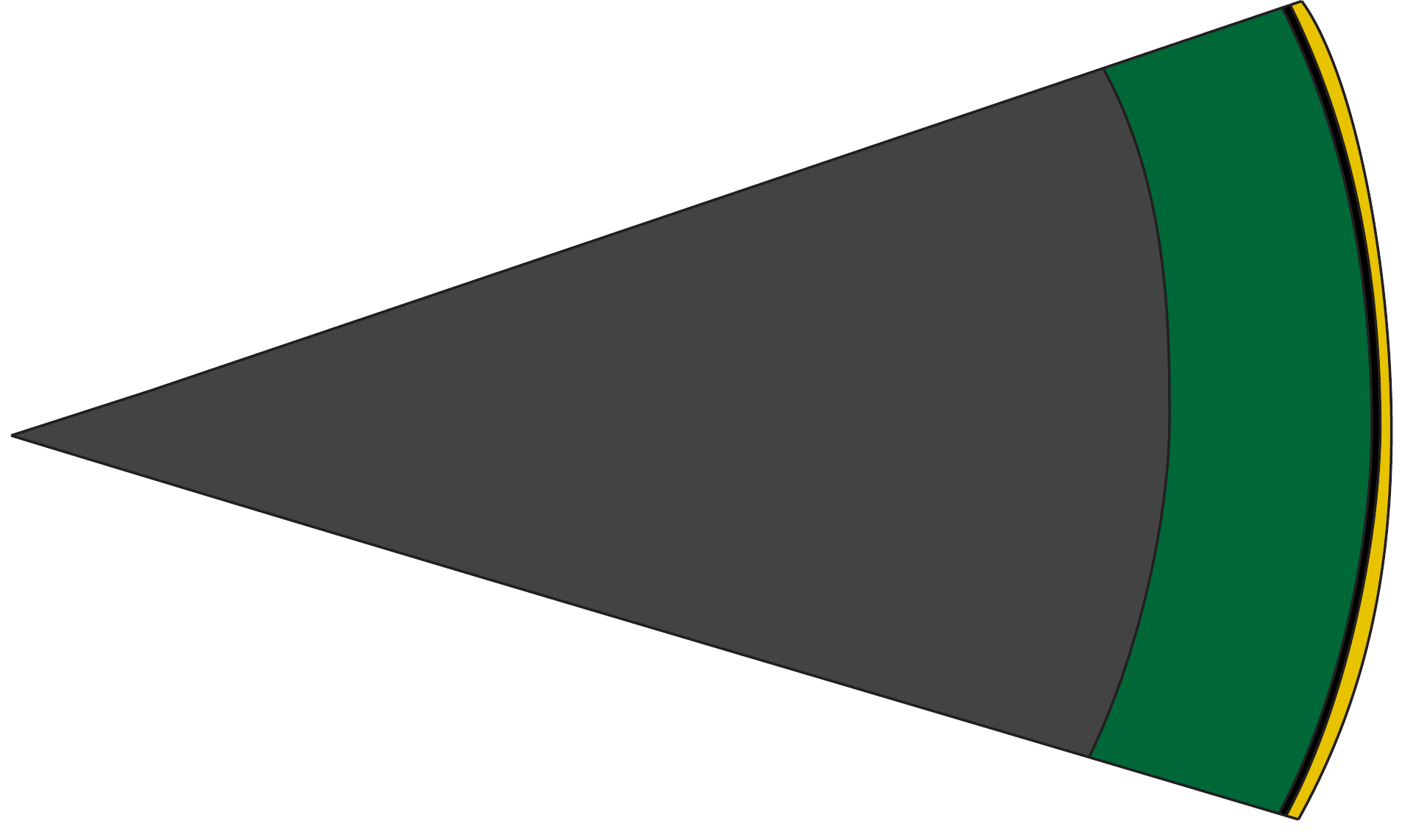}}& carbon-rich crust & \cite{peplowski2016,klima2018} \\
 & & sulfur rich basalts & \cite{Cartier2019}\\
 & & & \\
 & & & \\
\midrule
\multirow{4}{*}{Venus} & \multirow{4}{*}{\includegraphics[scale=0.1,angle=270]{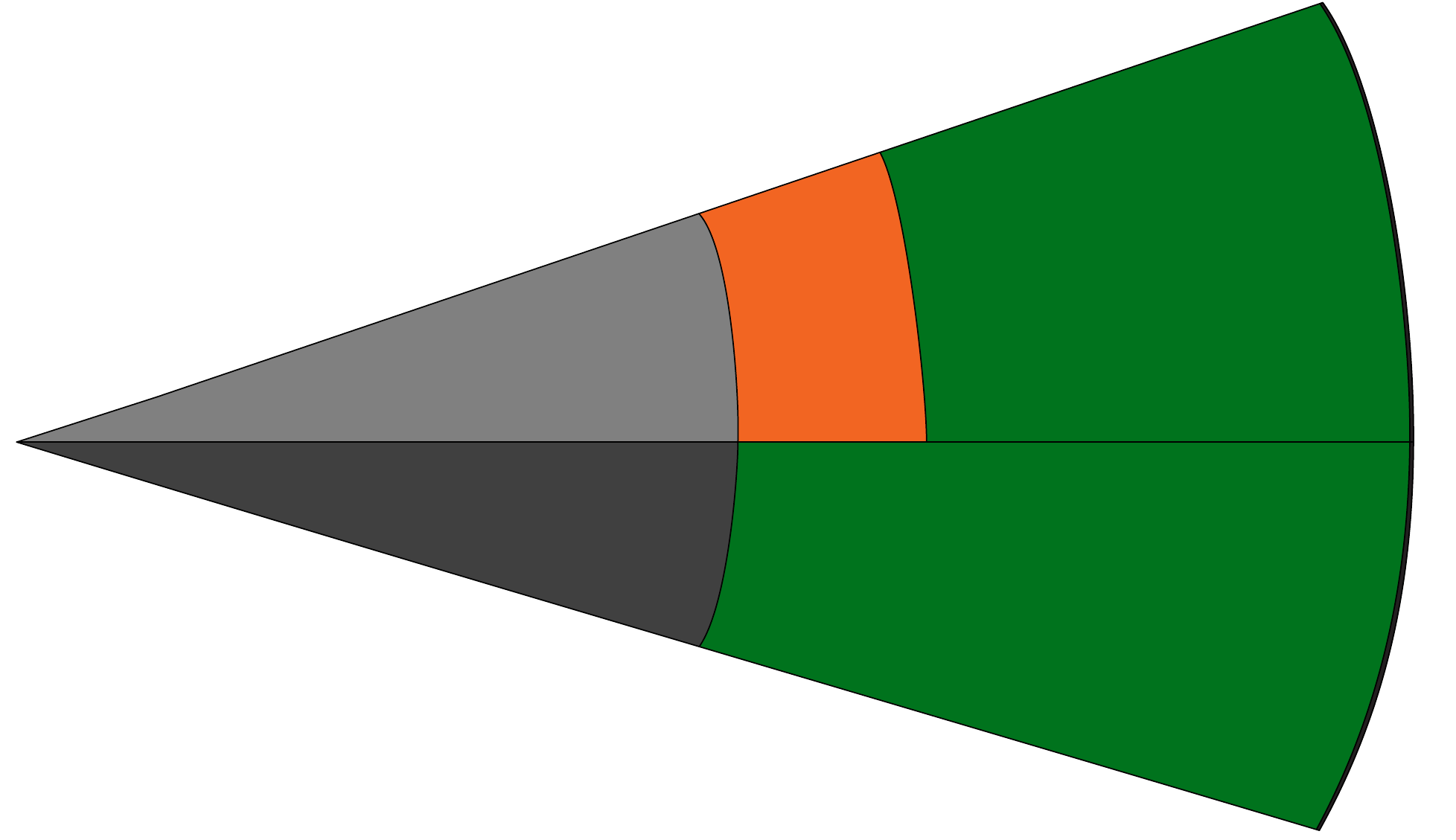}} & resurfacing/overturn & \cite{Smrekar2018}\\
 & & absence of a dynamo & \cite{orourke2020}\\
  & & & \\
 & & & \\
  & & & \\
\midrule
\multirow{4}{*}{Mars} & \multirow{4}{*}{\includegraphics[scale=0.1,angle=270]{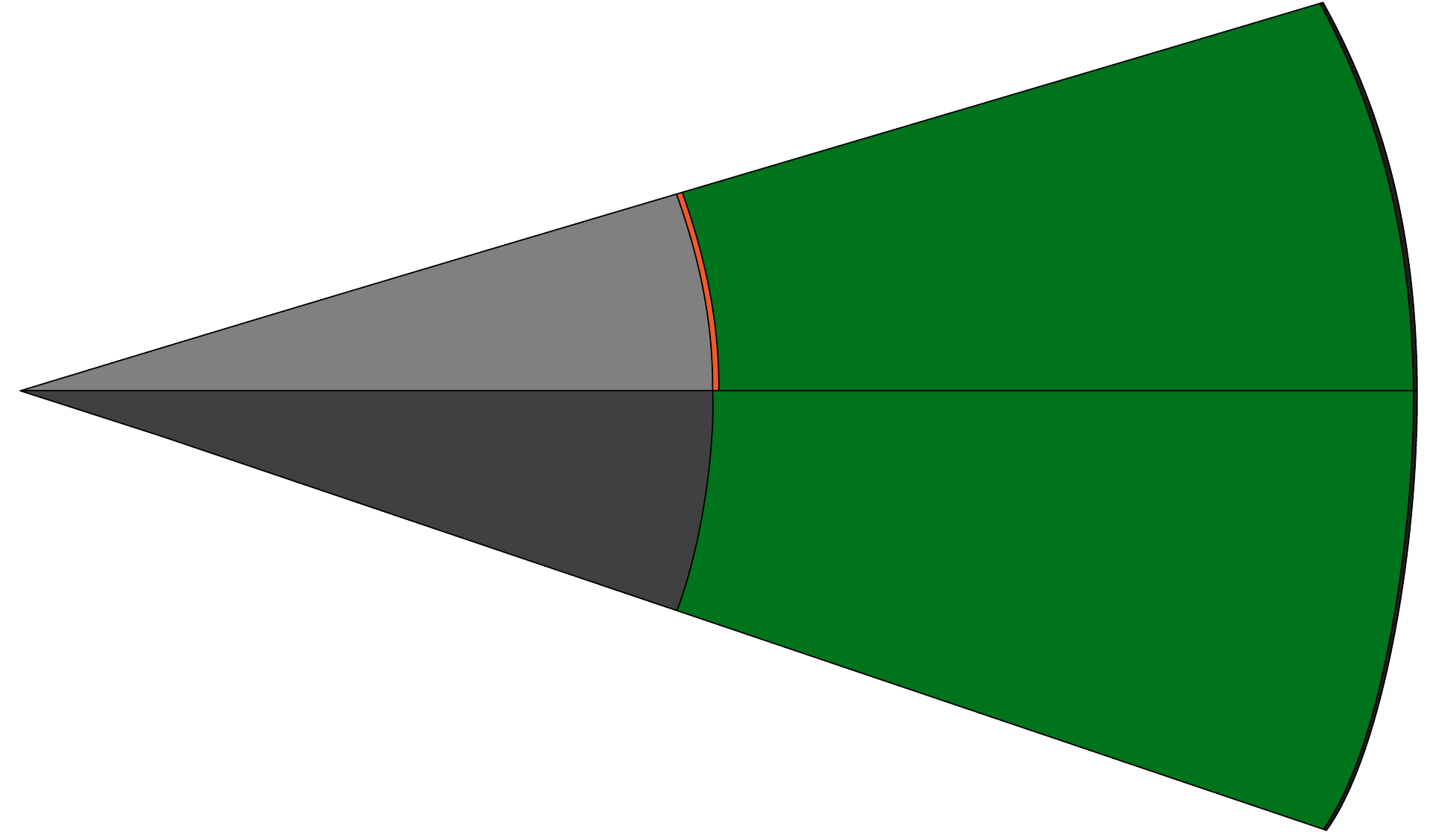}} & petrology/geochemistry & \cite{mcsween1985,wood1982}\\
 & & of \gls{SNC meteorites} & \cite{Borg1997}\\
 & & & \\
  & & & \\
\midrule
\multirow{4}{*}{Vesta} & \multirow{4}{*}{\includegraphics[scale=0.1,angle=270]{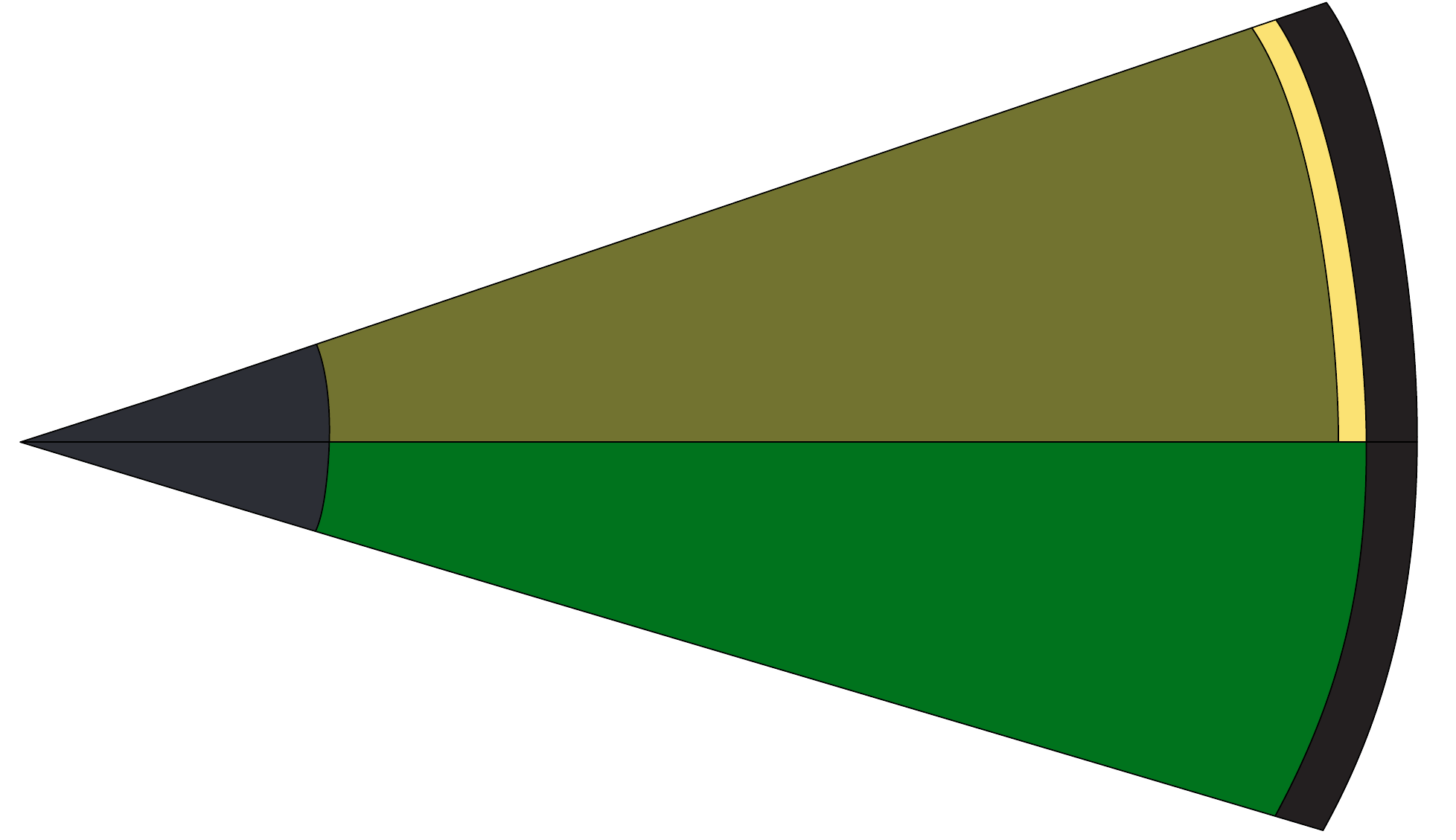}} & geochemistry of HED meteorites & \cite{righter1997}\\
 & & & \\
 & & & \\
  & & & \\
\bottomrule
\end{tabular*}
\begin{flushleft}
$^a$Representative/proposed interior with lithologies shown approximately to scale.  Key to colors: Yellow: late-stage, evolved-melt.  Green: silicate mantle.  Dark gray: solid metallic core. Light gray: liquid metallic core. Red/green gradient (on Moon): chemical gradient in mantle representative of cumulate overturn. Orange: existing melt. Black: crust.   Top half-slices represent either an existing basal magma ocean (for Venus and Mars) or a past shallow magma ocean (for Vesta).
\end{flushleft}
\end{table*}

\subsection{An Inner Solar System Synthesis}
\label{sec:synthesis}

\begin{figure}
	\centering
	\includegraphics[width=0.5\textwidth]{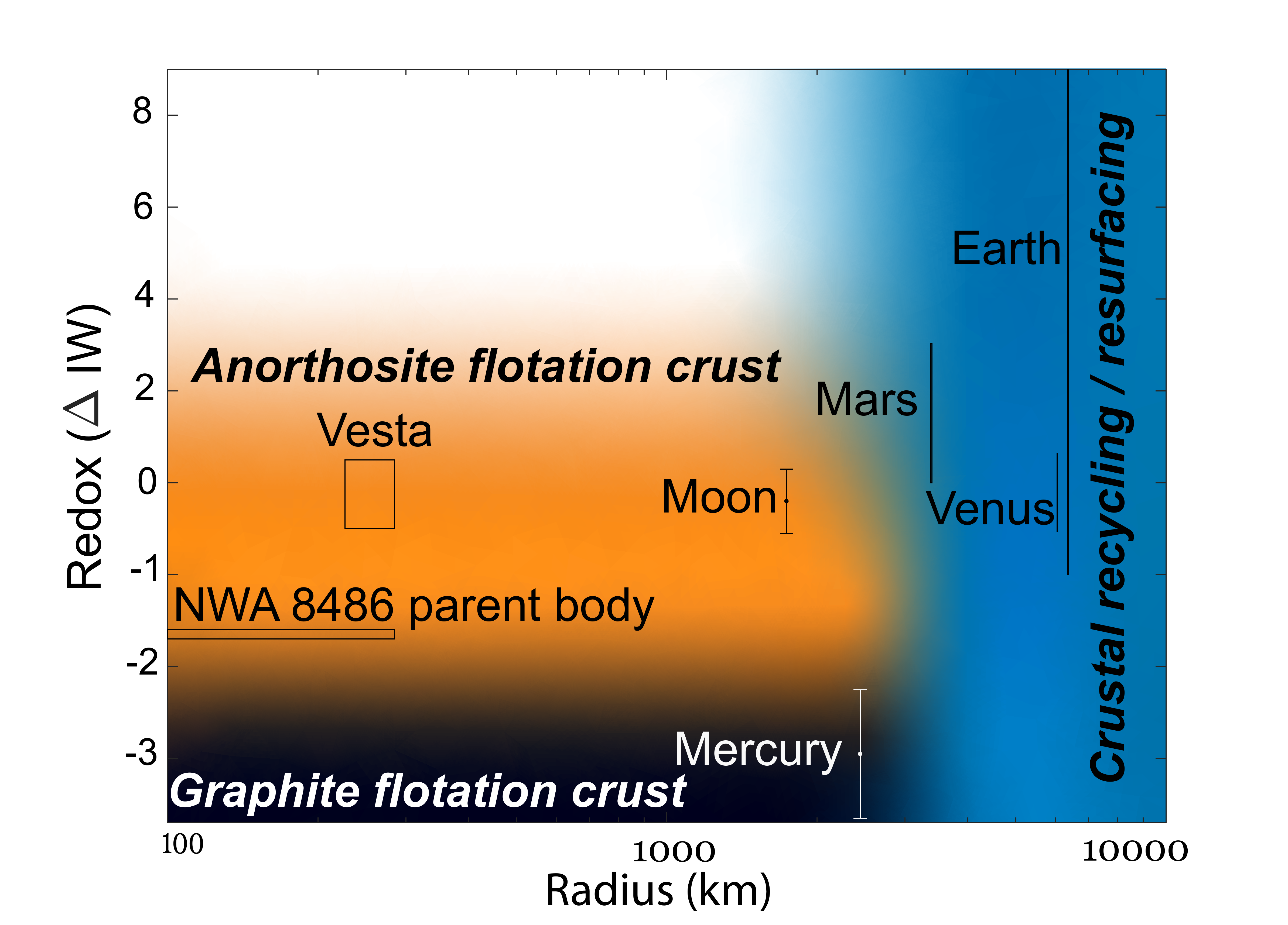}
	\caption{Plot of radius of vs. oxygen fugacity of different objects in the inner Solar System with suspected past or present magma oceans.	Regions of parameter space for different magma ocean outcomes are labeled. Crustal recycling or resurfacing has only been observed on larger bodies.  Graphite flotation crusts form only in mantles that are comparatively reducing with low H$_2$O abundance. Vesta is in red as it does not posses a flotation crust, despite an oxygen fugacity similar to that of the Moon. 
	}
	\label{FIG:solarsystem}
\end{figure}

There is circumstantial evidence that magma oceans were widespread in the inner Solar System.  Because the physical and chemical state of a magma ocean can profoundly influence the evolution of the interior, e.g. via overturn of the cumulate crust \citep{Elkins-Tanton2003} and core formation \citep{Rubie2007}, as well as the composition of the atmosphere (Sec. \ref{sec:atmospheres}), variation in those properties could underpin the diversity among inner Solar System bodies \citep{Schaefer2018}.  Arguably the two most important parameters controlling magma ocean outcomes are (i) the size of the body, which regulates the supply of energy in accreting planetesimals (Eqn. \ref{eqn:accretion_energy}) and governs the adiabatic temperature profile and hence the depth of the magma ocean (Eqn. \ref{eqn:adiabat}) \citep{Albarede2007}; and (ii) the oxygen fugacity of the mantle, as that is related to the abundance of iron (as FeO) and the composition of gases that are released to the surface during \gls{partial melting} \citep{Kasting1993}.   Figure \ref{FIG:solarsystem}) plots the approximate locations of inner Solar System bodies in this two-parameter space.     

Assuming that the inner rocky planets all accreted more or less from the same reservoir, the oxygen fugacity of a mantle was set by (a) the abundance of incorporated water (as a variable source of oxygen); and (b) the charge disproportionation of 3Fe$^{2+} \rightarrow\ $ 2Fe$^{3+}$ + Fe$^0$ in \gls{silicate perovskite} at high pressure \citep{Bindi2020}, and the sequestration of Fe$^0$ (metallic iron) into the core, leaving the mantle more oxidized \citep{Armstrong2019}.  The latter is predicted to have occurred in Earth's deep mantle but not in that of Mars or the Moon due to lower pressures there, meaning that the mantles of these smaller bodies are more reduced \citep{Deng2020}.  Sr and Eu abundances may be evidence for these reduced mantle, at least for the Moon.  Under reducing conditions, i.e. one log unit below the iron-w{\"u}stite (IW) \gls{redox buffer}, Eu$^{3+}$ reduces to Eu$^{2+}$.   Sr and Eu$^{2+}$ are incompatible in major minerals common on the Moon.  Depletion of Sr and Eu in lunar basalts suggests that the oxygen fugacity of formation was one log unit below the IW buffer \citep{Papike1996}.  Such reducing conditions would have meant any carbon would have been primarily out-gassed as CO or COS rather than CO$_2$ \citep{Renggli2017}.  In contrast, planets with comparatively oxidized mantles would be more likely to have CO$_2$-containing atmospheres.    

Thus, large planets which formed further from the Sun and which (it is assumed) accreted more water-rich material would have the most oxidized mantles, while close-in small planet would have the most reduced.  Earth has the most oxidized mantle, with Mars as a close second, followed by Venus and tiny, highly reduced Mercury (Fig. \ref{FIG:solarsystem}).  The amount of Fe left in the mantle would also influence the density of the residual melt as magma oceans began to crystallize, with mantles with high mantle Fe contents like Mars being more like to evolve residual liquids with densities that cause them to sink to the core-mantle boundary, forming a basal magma ocean (Sec. \ref{sec:dynamics}).   

\section{The Exoplanet Perspective}
\label{sec:exoplanets}

While magma oceans in the Solar System may not have persisted to the present day, with Venus \citep{orourke2020} and Io \citep{Khurana2011} as possible exceptions, they almost certainly are present on some planets around other stars.  In the past two decades more than 4000 candidate exoplanets have been discovered\footnote{https://exoplanetarchive.ipac.caltech.edu/}.  Among this diverse population are objects that are small enough to be primarily rocky and that exist in environments where they could host magma seas or oceans and could experience high rates of melting (Sec. \ref{sec:heat_sources}). As instrumentation improves, methods of detection are refined, and more space- and ground-based facilities are developed (Table \ref{tbl:methods}), the number of detected and characterized exoplanets continues to increase.

\subsection{Methods of Detection and Characterization}
\label{sec:methods}

\begin{table*}[width=2.0\linewidth,cols=3,pos=h]
\caption{Exoplanet Observation Methods}\label{tbl:methods}
\begin{tabular*}{\tblwidth}{l|l|l|l}
\toprule
\bf{method} & \bf{parameters} & \bf{notable surveys/missions} & \bf{references}\\
\toprule
\multirow{2}{*}{transit photometry} & \multirow{2}{*}{$P$, $R_p$} & \multirow{2}{*}{\emph{CoRot}, \emph{Kepler}, \emph{TESS}} & \cite{Deeg_2018,Baglin_2006};\\
& & & \cite{Ricker2014,Koch_2010}\\
\midrule
\multirow{3}{*}{Doppler RV} & \multirow{3}{*}{$P$, $M_p \sin i$, $e$} & HIRES, HARPS, & \cite{Fischer_2016,Wright_2018};\\
& & ESPRESSO, & \cite{Gonzalez2018};\\
& & EXPRES & \cite{Petersburg2020} \\
\midrule
\multirow{2}{*}{transmission spectra} & \multirow{2}{*}{atmosphere} & \emph{Hubble}, \emph{Spitzer},  & \cite{Crossfield_2015,Kreidberg2018b};\\
& & \emph{JWST} & \cite{Deming2019} \\
\midrule
\multirow{2}{*}{secondary eclipse} & \multirow{2}{*}{$T_{\rm eq}$, $A$} & \emph{Kepler},\emph{Spitzer},  & \cite{Alonso2018,Deming2019};\\
& & \emph{JWST} & \cite{Beichman2018} \\
\midrule
\multirow{2}{*}{phase curves} & \multirow{2}{*}{}day/nightside  & \emph{Kepler},\emph{Spitzer}, & \cite{Koll2015,Deming2019};\\
& variations &  \emph{JWST} & \cite{Parmentier_2018} \\
\midrule
\multirow{2}{*}{direct detection} & \multirow{2}{*}{}emission, & GPI, SPHERE, ELTs, & \cite{Bonati_2019, Birkby2018};\\
& reflection & \emph{LUVOIR}, \emph{HabEx} & \cite{Luvoir2019,Gaudi2020};\\
\bottomrule
\end{tabular*}
\end{table*}

\subsubsection{Transit Photometry}
\label{sec:transits}

Of the many methods used to detect and characterize exoplanets (Table \ref{tbl:methods}), transit photometry has proven to be the most successful and most widely employed, with over three-quarters of currently known exoplanets discovered via this method, mostly by space-based telescopes and especially by \emph{Kepler} (Table \ref{sec:methods}).  The successor to \emph{Kepler}, the Transiting Exoplanet Survey Satellite (\emph{TESS}), is particularly well-suited to discovering Earth-size planets on short-period orbits around bright, nearby stars; planets such as these are candidate hosts of magma oceans.  The transit method detects the small, repeated change in the apparent brightness of a star as an unresolved planet on a fortuitously highly-inclined orbit crosses the stellar disk \citep{Deeg_2018}. From these data, the planet's orbital period $P_K$ can be estimated as well as planet radius:
\begin{equation}
\label{eqn:transit_radius}
R_p \approx  R_* \sqrt{\delta} = 1 R_{\oplus} \frac{R_*}{R_{\odot}} \sqrt{\frac{\delta}{8.4 \times 10^{-5}}}
\end{equation}
where $R_*$ is the stellar radius and $\delta$ (the ``transit depth") is the fractional change in stellar brightness during the transit.  Extreme photometric precision, only routinely achieved from space, is required to detect Earth-size planets around Sun-like stars (Eqn. \ref{eqn:transit_radius}).  Although such planets can be detected around smaller M-type dwarf stars \citep{Gillon2016,DIttmann2017}, for a given $P_K$ they will have lower $T_{\rm eq}$ and be less likely to host magma oceans (Eqn. \ref{eqn:stellar_energy}).  Even where detection and precise determination of $\delta$ is feasible, often it is limited knowledge of the host star properties that impact the precision/accuracy of determinations of planet radius and irradiance or $T_{\rm eq}$ \citep{Berger2020}.   

\subsubsection{Doppler Radial Velocity}
\label{sec:rvs}

The Doppler radial velocity (RV) method, while initially employed as a method to detect new planets, is now often used to validate planet candidates found by the transit method and determine their mass.  A star's RV (motion along the line of sight) is measured by the Doppler shift in its spectrum; if the star hosts one or more planets, those can be revealed by detecting the periodic motion of the star around the systemic center of mass or barycenter.  In the case of a single planet, the planet mass $M_p$ is related to the amplitude $K$ of the RV signal by:
\begin{equation}
\label{eqn:rv_mass}
\begin{split}
M_p \sin i =  K \left(1-e^2\right)^{1/2}\left(\frac{P_KM_*^2}{2\pi G}\right)^{1/3} \\
\approx \frac{K}{0.64\,{\rm m\,s}^{-1}}\left(\frac{P_K}{1 {\rm \,day}}\right)^{-1/3}\left(\frac{M_*}{M_{\odot}}\right)^{2/3}
\end{split}
\end{equation}
where $i$ is the orbital inclination to the plane of the sky, $M_*$ is stellar mass (assumed much larger than $M_p$), and $e$ is the eccentricity (taken to be $\ll1$).   Since $\sin i < 1$ only the minimum planet mass can be measured with RVs alone, but $\sin i \approx 1$ for transiting planets.  For these planets, $P$ and orbital phase are also precisely known, allowing more exact measurement of the remaining Keplerian orbital parameters; for circular orbits this is the single parameter $K$.  Current RV precision is generally limited to about 1 m\,sec$^{-1}$.  The goal of the rising generation of instrument projects is to achieve $\sim$0.1 m\,sec$^{-1}$, but stellar sources of noise (``jitter") must be overcome \citep{Fischer_2016, Wright_2018}.     

An RV-determined mass (Eqn. \ref{eqn:rv_mass}), combined with a transit-determined radius (Eqn. \ref{eqn:transit_radius}), yields a mean density and can be compared to models of planet interiors \citep{Lopez_2017,Dorn2017}.  Mass and radius can also be combined to calculate a surface gravity $g = GM_p/R_p^2$, a key parameter for models of planetary atmospheres \citep{Ciardi_2019}. Radial velocities can also constrain eccentricity, which through tidal dissipation can drive melting on close-in planets (Sec. \ref{sec:heat_sources}) .

\subsubsection{Transit Spectroscopy}
\label{sec:spectroscopy}

If a transiting planet has an atmosphere, a very small fraction of the star light that reaches a distant observer will pass through the upper layers of that atmosphere.  Condensates and gases in the atmosphere will absorb and/or scatter that light in a wavelength-dependent manner (Fig. \ref{FIG:lavaworldcartoon}), causing the planet’s transit-derived apparent radius to vary slightly with wavelength.  Observations at multiple wavelengths, i.e. spectra, can be used to detect and constrain the structure of the atmosphere  and identify constituents \citep{Kreidberg2018b}.  For lava worlds, this could include sulfur compounds such as SO$_2$ as evidence of widespread volcanism \citep{Kaltenegger2009}.

Transiting planets on near-circular orbits will also be periodically occulted by their host stars, and the light that is reflected and/or emitted from the spatially unresolved planet can be isolated by differencing observations during and outside of this ``secondary eclipse" (Fig. \ref{FIG:lavaworldcartoon}).  Since $R_p \ll a$ the planet's reflected signal is very small and, with the exception of Jupiter-size planets on ultra-short period orbits \citep{Angerhausen2015,Mallonn2019}, usually undetected.  Due to the lower temperature of a planet compared to its star,  smaller planets are only detected at infrared wavelengths where the planets radiates significantly and a signal could be detected, and if the observations are obtained from space.  The infrared emission from the planet can be used to estimate its emitting temperature, a value expected to be close to $T_{\rm eq}$ \citep{MacDonald2020}, and by energy balance, albedo and heat re-distribution \citep{Garhart2020}.  Lava worlds with high surface temperatures (not necessarily equal to $T_{\rm eq}$) therefore lend themselves to this method.  The infrared regime also contains many broad features produced by molecular absorption lines and thus the spectrum of this emission is a diagnostic of the composition and structure of any atmosphere.  Sufficiently high-cadence observations during the ingress and egress portions of a secondary eclipse offer the opportunity to partly resolve spatial variation in emission over the planet, or ``eclipse mapping" \citep{Williams2006,Rauscher2018}.  

\subsubsection{Phase curves}
\label{sec:phasecurves}

As a planet orbits its star, the \gls{phase angle} for a distant observer varies periodically (Fig. \ref{FIG:lavaworldcartoon}). Measurement of the change in flux from the unresolved planet with phase provides information about the scattering properties of the surface or atmosphere (at optical wavelengths), or longitudinal variation in the temperature and atmospheric structure between the illuminated and dark sides of the planet (in the infrared).  Infrared phase curves, in particular those from \textit{Spitzer}, have been useful in measuring the day- and night-side temperature contrast on the planet, which is  the efficiency of heat transport, i.e. by an atmosphere and rotation \citep[][Sec. \ref{sec:heat_transfer}]{Koll2015}, which models suggest depends on the rotation and composition of the atmosphere \citep{Zhang_2017}.  determining if the emission is strongly peaked at the substellar point, as expected for an airless, tidally locked planet, or if there is a westward or eastward shift of this peak or ``hotspot" due to heat transport and the thermal inertia of the atmosphere \citep{Rauscher2018}.  Observations at multiple wavelengths probe longitudinal variations in albedo and cloud structure \citep{Parmentier_2018}.  To date, most studies to date have been on giant planets on short-period orbits  \citep[e.g.,][]{Zellem2014,Dang2018} with analyses attempted only a handful of rocky planet on $\sim$1-day orbits \citep{Batalha_2011, Sanchis_Ojeda_2013, Hu_2015, Demory_2016a,Angelo_Hu_2017}.

\subsubsection{Direct Detection}
\label{sec:direct}

Direct detection can identify and characterize planets on non-transiting orbits, separating their signal from the central star either spatially (in images) or by velocity (in spectra).  This is usually considered most feasible at infrared wavelengths where the planet-to-star contrast is the most favorable.  Direct imaging offers an unambiguous extraction of the planet's signal, e.g. to obtain a spectrum.  To date, this method has has lent itself to the detection of young, super-giant planets in orbits of tens of AU (see review by \citealp{Bowler2016}).  Direct imaging might eventually detect planets on wide orbits heated by a giant impact, or their tidally-heated satellites (see below).  In contrast, direct spectroscopic detection identifies the Doppler-shifted signal of a spatially unresolved planet and its Keplerian variation via cross-correlation \citep{Birkby2018}.  This could, in principle, be used to detect the emission from hot planets on close-in orbits.  Some directly imaged planets might have satellites: \citet{Agol_2015} have proposed a \textit{spectroastrometric} separation of signals from a planet and its satellite using the wavelength-dependent shift of the image centroid.

\subsubsection{Sensitivity and Biases}
\label{sec:biases}

All methods of exoplanet detection and characterization have limited sensitivity and intrinsic biases which influence the observed distribution and characteristics of exoplanets.  The transit method detects only those planets on highly-inclined orbits (within $R_*/a$ of edge-on for circular orbits, or$\sim$1-10\% of a randomly-oriented population).  As a consequence of this geometric requirement and Kepler's Third Law, the transit method is biased towards planets on close-in planets which are more likely to transit and transit more often.  Thus, it is well-suited to detecting highly-irradiated planets, as long as the cadence is sufficient to resolve the transit, which will be less than an hour.   The Doppler RV method is also biased towards short period planets, since, for a given planet mass, the Doppler amplitude increases with decreasing orbital period (Eqn. \ref{eqn:rv_mass}).  Since the mass of rocky planets scales roughly as $R_p^4$ while the transit depth scales as $R_p^2$ (Eqn. \ref{eqn:transit_radius}), the Doppler RV method is less sensitive to smaller planets. While transit photometry may be able to readily detect Earth-size planets on short-period orbits, their masses might remain uncertain until more precise instruments and better mitigation of stellar noise become available.  Direct detection is highly biased towards giant planets, particularly in young systems \citep{crossfield2015} and it may contribute only in rare cases such as post-impact planets (Sec. \ref{sec:post-impact}).  

\subsection{The Diversity of Lava Worlds around Other Stars}
\label{sec:diversity}

\subsubsection{Ultra-Short Period Planets}
\label{sec:usps}

Highly-irradiated planets on ultra-short period orbits (USPs) include the most likely lava world candidates. The period cut-off for USPs is arbitrary and varies, but the convention is one day \citep[e.g.,][]{Sahu_2006, Sanchis_Ojeda_2014, Winn_2018, Dai_2019}, with Kepler-974c as the current record-holder at $P = 4.25$ hr \citep{Ofir2013,Rappaport2013}.  Approximately 0.5\% of Sun-like G-type stars host USPs and they appear slightly more common around cooler K-type stars \citep{Sanchis_Ojeda_2014}.  To date, nearly 80 USPs have been discovered by transit photometry; nearly all have radii less than 2\rearth\ \citep{Winn_2018} and are thus likely to be primarily rocky \citep{Weiss2014}.  Indeed, most of those USPs with masses established by Doppler RV observations, are consistent with rocky composition with a few measurements (or limits) that suggest a thick atmosphere (Fig. \ref{FIG:mass-radius}) \citep{Demory_2011, Winn_2011, Dai_2017}. 

USPs orbit inside the observed inner edge of protoplanetary disks around \gls{T Tauri star}s, thus these objects could not have arrived at their present location by migration through a disk.  USPs tend to be the innermost planet in multi-planet systems, and the larger mutual inclinations and wider orbital spacing of these particular systems suggest that that USPs were scattered inwards by gravitatinal interaction with companion planets \citep{Sanchis_Ojeda_2013,Lee2017,Dai2018,Winn_2018,Petrovich2019}.  Some USPs could have possessed thicker atmospheres which were then removed by the intense radiation and particle fluxes from the host stars \citep{Owen_2013, Dai_2017, Lopez_2017} (see Sec. \ref{sec:atmos_evol}).   While few USPs have measured masses or mean densities, lower imits for those of Kepler-947c have been estimated using the Roche criterion for disruption by the tides from the host star \citep{Rappaport2013,Price2020}; these suggest an iron-rich interior. Three of the best-studied USPs are described and discussed in Sec. \ref{sec:examples}. 

\subsubsection{Post-Giant Impact Planets}
\label{sec:post-impact}

The planets of the early inner Solar System probably hosted magma ocean phases sustained by heat from accretion of planetesimals and giant impactors (Secs. \ref{sec:heat_sources} and \ref{sec:solarsystem}).  Around very young stars, rocky planets are presumably in the same magma ocean phase; in principle these could be directly detected, especially soon after a giant impact \citep{Stern_1994, Miller_Ricci_2009, Bonati_2019, Lupu_2014}.   \citet{Meng2014} have reported infrared variability in a young star consistent with the debris of such a collision.  Identifying such planets and studying their emission will enhance our understanding of the corresponding era in the Solar System's history.  Because of the heat capacity of a deep magma ocean produced by such an event, and the continuous foundering of any crust, planets will remain super-luminous in the infrared and possibly detectable for several million years \citep{Bonati_2019, Lupu_2014}.  The potential for direct detection is strongly influenced by the insulating effects of the atmosphere, which throttles the loss of energy to space, increasing the magma ocean lifetime and making this phase statistically more likely to be observed, but at the same time rendering the planet less detectable in the infrared \citep{Bonati_2019}. As the magma ocean cools, its emission spectrum is expected to evolve \citep{Hamano_2015}.  Any planet heated by a recent giant impact would presumably be accompanied by a tell-tale debris disk or dust cloud that could distinguish it from other scenarios\citep{Meng2014}.

\subsubsection{Tidally-Heated Planets and Satellites}
\label{sec:tidal-planets}

The Jovian satellite Io owes its intense volcanism to the tidal deformation it experiences on its eccentric orbit close to Jupiter (see Sec. \ref{sec:io_lakes}).  Planets on close orbits around their host stars (or satellites around Jupiter-like planets) could experience significant internal heating, either due to the eccentricity of the orbit or non-synchronous rotation (Sec. \ref{sec:heat_sources}).  These ``exo-Ios" would experience large-scaling melting and widespread volcanism \citep[e.g.,][]{Jackson_2008, Henning_2009, Cassidy_2009, Peters_2013, Driscoll_2015, Oza_2019}). In systems of multiple planets (or satellites), mutual interactions could maintain eccentric orbits, interior heating, and thus lava oceans or volcanic activity for much of the stellar lifetime.  These, in principle, may be observable \citep{Henning_2009,Peters_2013, Oza_2019}.  The persistence of measurable orbital eccentricity against orbital circularization by tidal dissipation in a system with known age, or the degree to which the orbit is excited by companion planets could constrain tidal dissipation in the planet and hence the modified quality factor $Q' = 3Q/(2k_2)$ (appearing in Eqn. \ref{eqn:tides}).  A low value compared to that typical of solid rocky planets ($Q' \sim 100$) could indicate a  magma ocean \citep{Barr2018}.  \citet{Henning_2009} predict significant tidal heating for $P_K = 10-30$ days and $e \lesssim 0.1$; partial melting on a global scale occurs for $P_K<2$ days from tidal heating alone.  While tidal effects may increase equilibrium temperatures by only a few K \citep{Henning_2009}, other manifestations such as volcanic hotspots or volcanic gases in the atmosphere could be detected \citep{Henning_2009,Kaltenegger2009}.  Tidally-heated ``exomoons", i.e. Io-like satellites of giant planets, might be detected by the orbital phase variation of their contribution to the combined infrared emission of the unresolved planet plus satellite \citep{Peters_2013,Forgan_2017}.   However, detection  will only be feasible in cases of large (i.e., Earth-size), hot ($\approx$850\,K) satellites around planets that are on wide orbits around relatively dim host stars, i.e., M dwarf stars \citep{Peters_2013}. 

\begin{figure}
	\centering
	\includegraphics[scale=.6]{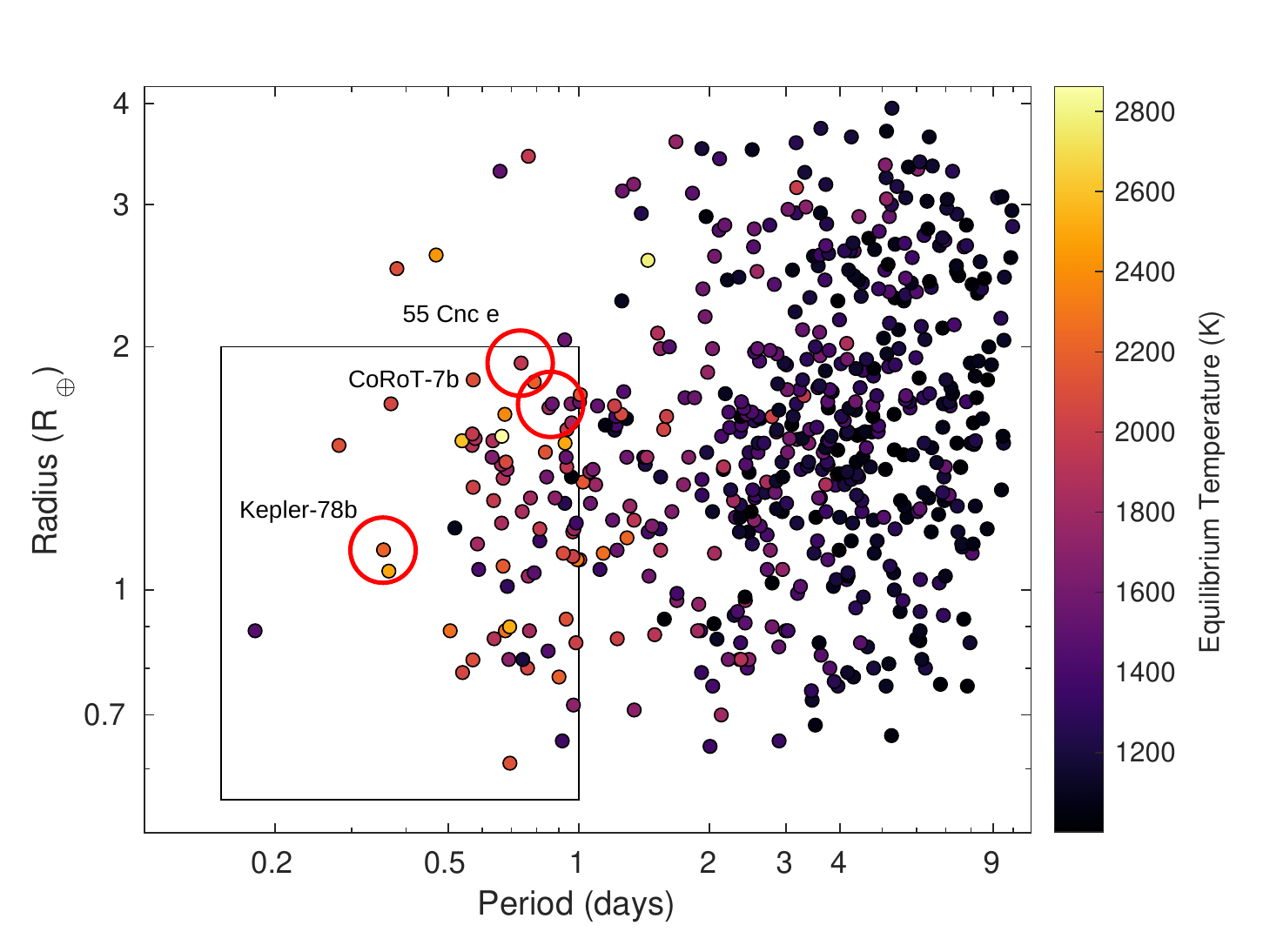}
	\caption{Orbital period versus radius of planets with periods less than 10 days and with equilibrium temperatures greater than 1000K. A box marks the domain of Ultra Short Period planets ($P_K < 1$\,day) that are plausibly rocky.  Some USPs around M dwarf stars may not be hot enough for melting ($T_{rm eq} < 1300$K), while some planets on wider orbits around more luminous stars will be at least partially covered by a magma ocean. \cnce, Kepler-78b, and CoRoT-7b, discussed in detail in Sec. \ref{sec:examples}, are circled. Stellar and planetary data were downloaded from the NASA Exoplanet Archive December 2 2020. Equilibrium temperatures for some planets were not reported and were derived from stellar and orbital parameters as in, for instance, \citet{Sheets_2014}, assuming a planet with albedo $\ll$1 and efficient redistribution of heat.}
	\label{fig:rad_equil}
\end{figure}

\begin{figure}
    \centering
    \includegraphics[scale=.6]{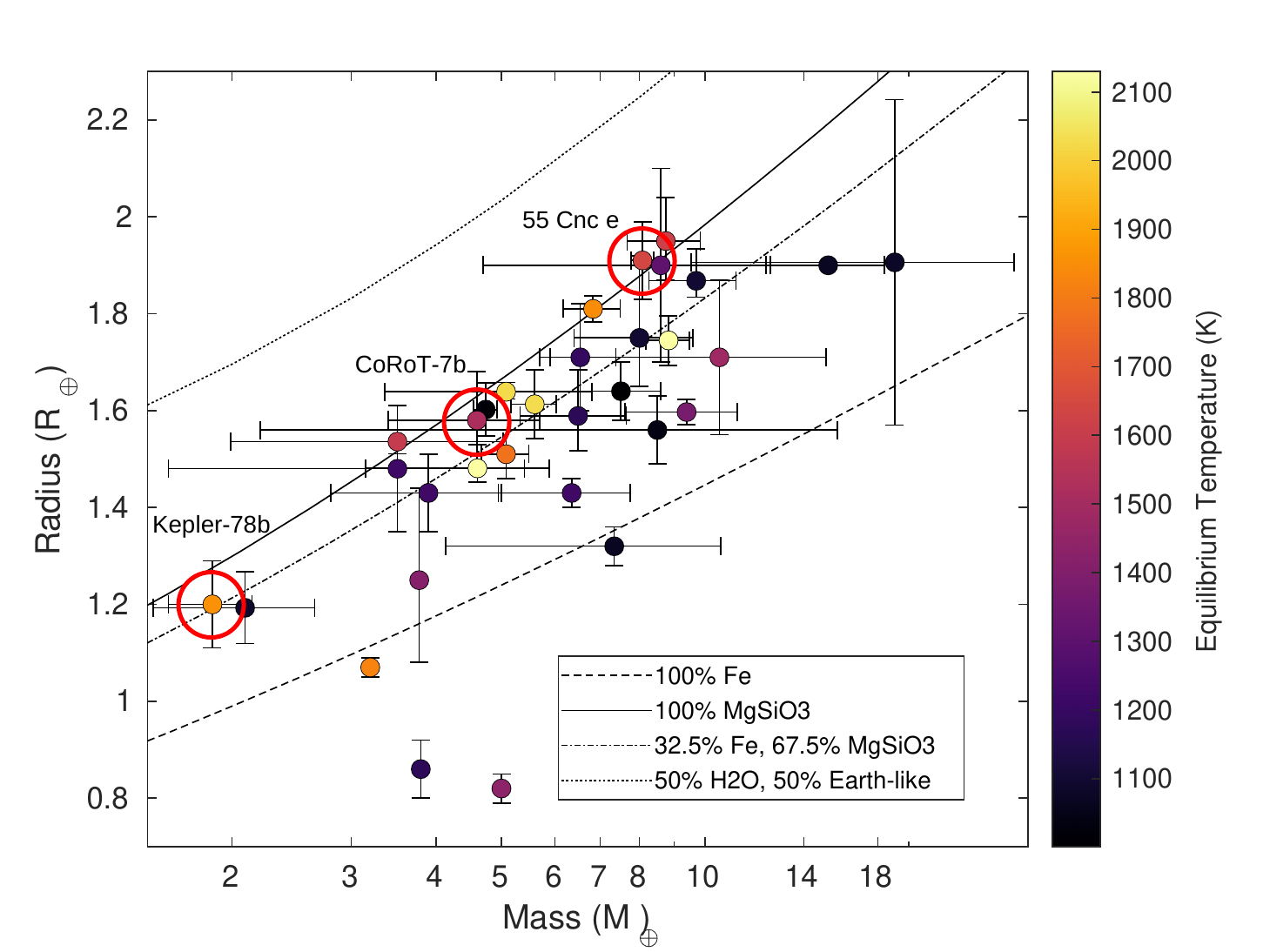}
	\caption{Earth- and super-Earth-size exoplanets with equilibrium temperatures $>1000$\,K for which both mass and radius have been measured. \cnce, Kepler-78b, and CoRoT-7b are circled in red. The masses and radii are from \citet{Dai_2019} and the NASA Exoplanet Archive, with the latter up-to-date as of August 2 2020. Models of planet composition from \citet{Zeng_2019} are plotted for pure iron, Earth-like (silicate mantle and iron core), pure silicate, and Earth-like planets overlain by a water envelope (50\% by mass).}
    \label{FIG:mass-radius}
\end{figure}

\subsection{Three Prominent Lava Worlds}
\label{sec:examples}

\subsubsection{CoRoT-7b}
\label{sec:corot7b}

CoRoT-7b, a $1.68\pm0.09$\rearth ``super-Earth" on a 20-hour orbit around a G-type star, was the first exoplanet identified as plausibly rocky \citep{Leger_2009}.  High stellar RV noise compared to the expected planet signal have hindered an accurate mass estimate, with values from $2.3\pm1.8$\mearth to $8\pm1.2$\mearth reported \citep{Queloz_2009,Hatzes_2010,Ferraz-Mello_2011,Pont_2011,Boisse_2011}.  This translates into uncertainty in bulk density, complicating models of interior composition.  \citet{Leger_2011} find that an Earth-like proportion of silicate mantle and metallic iron core is consistent with the median reported RV mass range ($6.9\pm1.2$\mearth).  A mass at the upper end of the range of measurements \citep[e.g.,][]{Queloz_2009,Hatzes_2011,Ferraz-Mello_2011} would require a comparatively larger Mercury-like core \citep{Valencia_2010,Wagner_2012}. 

If CoRoT-7b is tidally locked and lacks an atmosphere, day-side temperatures could be as high as 2600\,K, and the planet would possess a hemispherical dayside magma ocean \citep{Leger_2009}. Night-side temperatures will be far lower; assuming inefficient heat transport, no greenhouse effect, and a geothermal heat flux of 300 \mwpersqm, \citet{Leger_2009} predict night-side temperatures as low as 50\,K.  An atmosphere would serve to move heat from the day-side to the night-side and would serve to significantly diminish this dichotomy.   Dissipation of tides raised by the host star in th planet may also induce a significant amount of heating on the nightside \citep{Barnes_2010}. The depth of the magma ocean on CoRoT-7b depends inversely on its surface gravity, which controls the pressure at a given depth \citep{Leger_2011}, as well as transport of heat by mantle subsolidus convection to the night side \citep{Kite2016}. The planet's proximity to the host star is usually assumed to have led to the removal of any substantial atmosphere, but a larger planet radius based on a more accurate stellar radius from a parallax by the \emph{Gaia} mission suggests a volatile-containing planet \citep{Stassun_2017, Dai_2019}.   Regardless, CoRoT-7b could retain a thin atmosphere composed of out-gassed volatile compounds or more \gls{refractory} elements sputtered from the surface.  An \gls{exosphere} composed of refractory elements might be expected to be trailing away from the parent star \citep{Mura_2011}. To date, no absorption lines originating from an exosphere have been detected in CoRoT-7b’s transmission spectrum \citep{Guenther_2011}.  
    
\subsubsection{Kepler-78b}
\label{sec:kepler78b}

Kepler-78b is on an 8.5-hour orbit around a K-dwarf; with a radius of $1.2\pm0.09$\rearth and a mass of $1.69\pm0.41$\mearth, it is likely to have an Earth-like interior \citep{Howard_2013, Pepe_2013, Dai_2019}.  Like CoRoT-7b, stellar activity has limited accurate Doppler RV measurement of its mass, but in this case detrending by Gaussian process regression has partially ameliorated these systematics \citep{Grunblatt_2015}. Possible compositions range from Moon-like and iron-deficient to Mercury-like and iron-rich \citep{Hatzes_2014, Stassun_2017}. 

Using \kepler\ photometry of the phase curve and secondary eclipse, \citet{Sanchis_Ojeda_2013} has constrained the dayside $T_{\rm eq}$ and the reflected light (albedo) of Kepler-78b to 2300--3100\,K and 0.4--0.6, respectively, although they note that the constraint on albedo are loose and allow for values outside this range.  This is invariably hot enough for a surface lava-ocean \citep{Pepe_2013} and, while crude, is in agreement with \citet{Rouan_2011}, \citet{Batalha_2011}, and \citet{Sheets_2014}, who have argued for a similarly high albedo for the magma ocean surface of another USP, Kepler-10b.

\subsubsection{55 Cancri e}
\label{sec:fiftyfivecnce}

\cnce\ (also designated $\rho$ Cnc e) is a $1.91\pm0.08$\rearth planet \cite{Demory_2016a} that is the innermost ($P_K = 18$\,hr) of five known planets orbiting a bright, nearby K-type dwarf star.  The star also has an M-dwarf companion at a projected separation of 1060 AU \citep{Mugrauer_2006}.  Perturbations by this stellar companion could have misaligned the orbit of \cnce\ with respect to the host star's spin axis \citep{Kaib_2011}, but a detection of this misalignment using the \gls{Rossiter-McLaughlin effect} \citep{Bourrier_2014} has not been confirmed due to the predicted weak signal \citep{Lopez-Morales2014}.  It is possible that the entire planetary system was misaligned by the gravitational influence of the M dwarf companion; interactions among the planets and with the secondary star in the 55 Cnc system have potentially had consequences for the system’s evolution \citep{Hansen_2015, Nelson_2014, McArthur_2004}.   

RV observations with multiple instruments point to a mass at or near $8.08\pm0.31$\mearth \citep[e.g.,][]{Demory_2011, Endl_2012, Nelson_2014, Demory_2016a}. The bulk composition of the planet is controversial, since the nominal mass-radius is marginally ($\approx2\sigma$) inconsistent with a purely silicate mantle-iron core composition (Fig. \ref{FIG:mass-radius}) and allows for a substantial atmosphere of volatiles \citep{Winn_2011, Demory_2011, Gillon_2012}.  More exotic compositions have been proposed, i.e., a rocky world enriched in refractory Ca and Al and lacking an iron core \citep{Dorn_2018} or one that is carbon-rich, with carbides replacing silicates, but with a metallic core \citep{Madhusudhan_2012}.  The latter scenario was motivated by a report that the parent star has a high C/O ratio \citep{Delgado_Mena_2010}, a claim that has since been disputed \citep{Teske_2013}.

Observations of secondary eclipses in the infrared, i.e., by \emph{Spitzer}, are an independent probe of any atmosphere.  On the one hand, \emph{Spitzer} found the night side of the planet to be around 1600K; if the planet is tidally locked, as expected, this suggests appreciable but relatively inefficient heat transport from the day side (at 2700\,K) to the night side hemisphere by an atmosphere \citep{Demory_2016a}. Moreover, there is a $\approx$40\,deg eastward shift in the location of peak emission on the dayside - the expected effect of a super-rotating atmosphere.  \citet{Demory_2016a} propose two explanations: a thick atmosphere that transports heat from the dayside across the terminator but which condenses out on the cooler nightside, or an airless world with a low-viscosity magma ocean that convects heat to the nightside. A re-analysis of \textit{Spitzer} phase curves  supports heat transport by a thick atmosphere \citep{Angelo_2017}.

Ideas on the most plausible composition of \cnce’s atmosphere have evolved from H-, He-, or supercritical water-dominated compositions \citep{Demory_2011,Winn_2011,Gillon_2012} to those with higher molecular weight volatiles or even refractories that are evaporated or sputtering of refractory-rich compounds analogous to the exosphere of Mercury.  The mixing ratio of hydrogen is controversial since it is expected to rapidly escape from a planet this close to its host star. \citet{Ehrenreich_2012} did not detect a H exosphere while \citet{Tsiaras_2016} report the detection of features in an infrared transmission spectrum consistent with HCN.    \citet{Ridden_Harper_2016} and \citet{Bourrier_2018a} report tentative detections of Ca$^+$ and Na in the exosphere.

Especially intriguing is potential orbit-to-orbit variability in the depth of both the secondary eclipse and primary transit, in different spectral bands.  Variability in the secondary eclipse depth at 4.5$\mu$m as measured by \textit{Spitzer} suggests a transient source of infrared opacity in the atmosphere.  \citet{Demory_2016b} propose volcanic plumes as one possible source; such plumes would raise the effective emitting altitude (unity optical depth) of the planet's atmosphere to where the temperature and infrared emission are lower, thus decreasing the depth of the secondary eclipse.  Likewise, \citet{Tamburo_2018} propose volcanic clouds of alumina, \gls{olivine}, and pyroxene grains to explain variations in the scattered light from the planet detected in the optical.  \citet{Sulis_2019} limit the planet's geometric albedo to 0.47 ($2\sigma$), but this does not rule out atmospheres with clouds of mineral grains \citep[e.g.,][]{Mayorga2020}.  A circumstellar dust torus similar to the one in the Io-Jupiter system could be an alternate source of the observed variable IR opacity \citep{Demory_2016b}.  Observations of the primary transit in both the optical \citep{Sulis_2019} and infrared \citep[4.5$\mu$m band of \textit{Spitzer},][]{Demory_2016b,Tamburo_2018}. show no significant variability, e.g. by volcanic plumes or a dust torus.  Modulations in the host star's brightness in the far-ultraviolet have been detected \citep{Bourrier_2018b}. \cnce\ may be close enough to its host star for magnetic interactions to occur, exerting influence on the stellar wind and causing a starspot that rotates with the planet over its 18-hr period \citep{Winn_2011,Sulis_2019,Folsom2020}. \citet{Bourrier_2018a} propose that this interaction could result in a coronal rain of cooler gas, similar to what is observed in the Sun, and could explain the observed UV flux modulation.

\section{Dynamics and Stability of Magma Oceans}
\label{sec:dynamics}

\subsection{Vertical Stability}
\label{sec:vertical}

Magma oceans that are cool globally or locally will eventually crystallize, and the (negative) buoyancy of the mineral grains with respect to the surrounding melt and the melt with respect to the underlying mantle at ambient conditions are important but poorly understood factors controlling the evolution and stability of magma oceans  \citep{caracas2019melt}.  Buoyant minerals will tend to float to the surface of a magma ocean on a timescale governed by the melt viscosity and vigor of convection \citep{Solomatov2015}.  Denser minerals will tend to sink and accumulate at the base of the ocean, where they could penetrate the solid mantle if there is sufficient density contrast.   Where surface temperatures are sufficiently high, more volatile elements in the melt will evaporate, and can be either lost to space or condense at cooler regions of the planet's surface.  These processes of physical segregation will drive chemical evolution in the melt, which in turn will influence any further crystallization and the fate of those grains.   Independently, the effect of pressure at depth on the density of solids and liquids, described by equations of state (EOS), mean that both the mass of the planet and the depth of the ocean play important roles in the dynamics. 

At low pressures, most solids are denser than their parent liquids (with a few exceptions such as water ice, plagioclase in a silicate melt, and graphite in a reducing melt) and thus crystals will tend to sink.  However, liquids tend to be more compressible than solid phases because the atoms in the former are mobile.  This means at a sufficiently high pressure (or depth) the solids that are crystallizing from a liquid will be less dense, and will rise to a level where they attain neutral buoyancy (the \emph{cross-over depth}).  The difference in compressibility between liquids and solids arises from the difference in the \gls{coordination number}s of cations in liquids vs. crystalline solids.  In silicates, this coordination number quantifies the number of \gls{cation-anion polyhedra} in silicates and the degree of packing of the atoms; for a given composition, the higher the average coordination number, the greater the density.  In liquids the average coordination number of cations (a continuous distribution) is typically slightly lower than its solid counterpart, which is why the liquid is less dense.  The average coordination number increases continuously with pressure in a liquid \citep{solomatova2019pressure}, whereas the cations in solid counterparts do not change their coordination numbers unless a phase transition occurs, and thus the liquid eventually becomes denser.

\begin{figure*}
\label{fig:eos}
	\centering
		\includegraphics[width=0.9\textwidth]{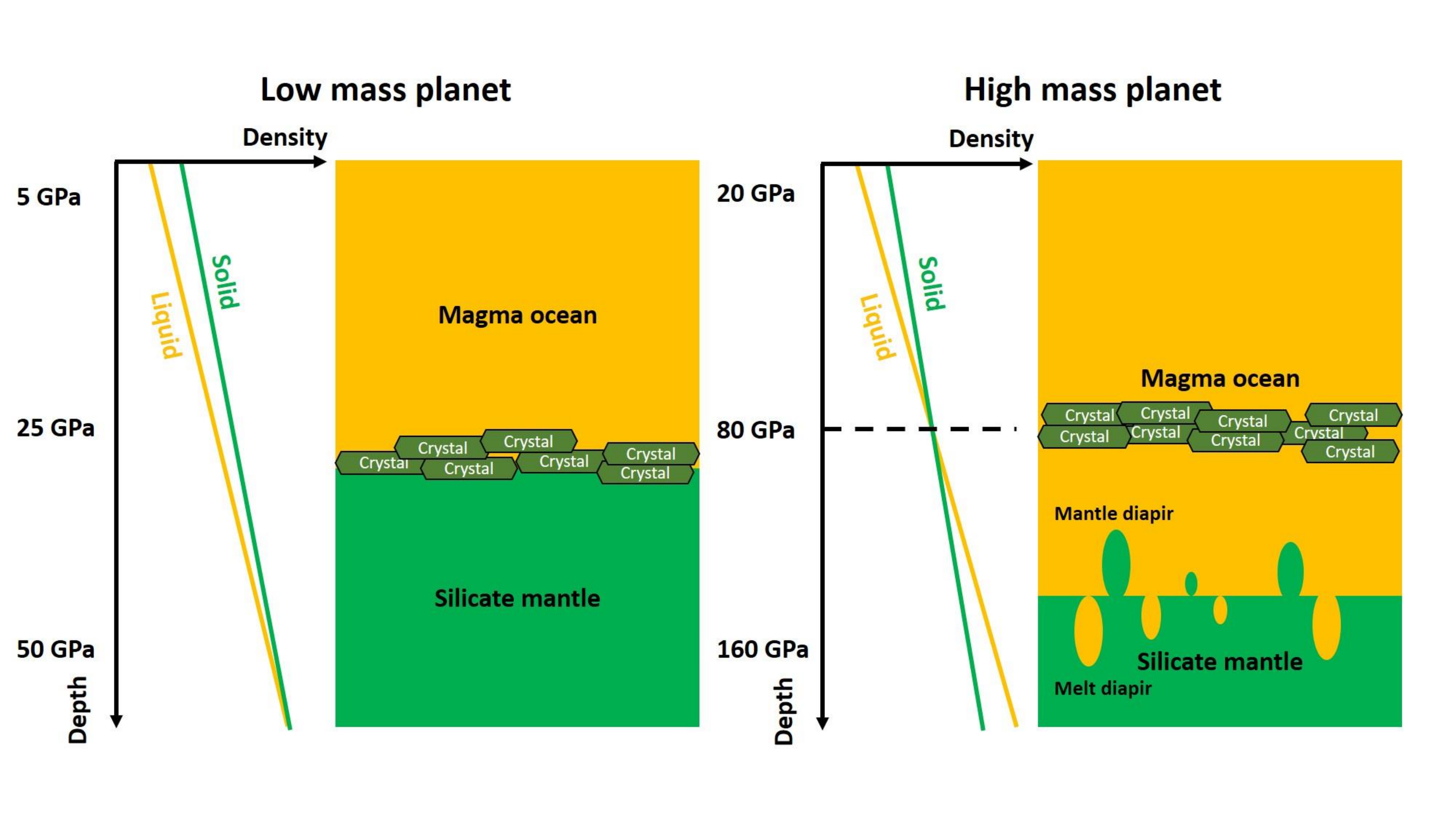}
	\caption{Difference in magma ocean behavior on a low mass planet with low gravity (left) and a high mass planet with high gravity (right).  In the former, the magma ocean is shallower than the cross-over depth where the liquid phase has a density equal to the crystallizing solids, and crystals accumulate at the base of the ocean.  In the latter, the cross-over depth is within the magma ocean column, the ocean bifurcates, and the lower ocean can become denser than the underlying mantle and destabilize.}
	\label{FIG:magmaocean_mass}
\end{figure*}

For example, in the single giant impact scenario of lunar formation, the Earth's mantle could have been completely molten to the core-mantle boundary at a pressure of 130 GPa \citep{solomatova2019pressure}.  Below a cross-over pressure depth of 115 GPa , the liquid melt is denser than the first crystallizing solids, i.e. \gls{bridgmanite}, and crystals formed at that depth would be buoyant.  All else being equal, on lower mass planets crystallization will produce a cumulate "pile" at the base of the magma ocean.  In a deep enough magma ocean on a sufficiently massive planet, the cross-over depth occurs within the magma column, crystals accumulate in a layer at that depth, and physically separated and chemically distinct upper and lower magma oceans can form (Fig. \ref{FIG:magmaocean_mass}).  The lower ocean could become denser than any underlying mantle and leading to turnover and migration to the core-mantle boundary.  This phenomenon motivated the hypothesis of a basal magma ocean in Earth \citep{Labrosse2007}  

As crystallization proceeds, the composition of the residual melt and hence the density evolves.   In magma oceans with a Mg-Fe-silicate (mafic) composition, cooling and crystallization will first produce Mg-rich minerals with higher melting points, e.g. forsterite and enstatite.  Thus the iron content of the residual melt increases and the density also increases due to the higher atomic weight of Fe compared to Mg.  In the case of Earth, as bridgmanite forms, the increasing density of the melt causes the cross-over depth to occur at lower pressure (shallower), reaching 50 GPa when the magma ocean is 50\% crystallized, at which point viscosity rises and convection slows dramatically \citep{caracas2019melt}.  Even in the absence of pressure effects, the late-crystallizing, iron-rich cumulates themselves may be denser than the underlying mantle, triggering overturn \citep{Miyazaki2019}.  Overturn of ilmenite (FeTiO$_3$-bearing cumulates) is a well-studied scenario to explain mare volcanism on the Moon \citep{Li2019,Yu2019,Zhao2019}.      

Magma ocean chemistry can also evolve due to evaporation \citep{Schaefer2009,Perez-Becker2013,kite2016atmosphere}.  This is the case when surface temperature exceed 2000\,K, e.g., for exoplanets like CoRoT 7-b or 55 Cnc e on very close orbits around solar-type host stars (Sec. \ref{sec:examples}).  In the case of planets with tenuous atmospheres, evaporates can condense at cooler regions, i.e. the poles or, in the case of tidally-locked planets, near the day-night boundary or terminator (Fig. \ref{FIG:lavaworldcartoon}), influencing the surface chemistry there.  The vapor can also escape to space from small planets with low gravity, causing permanent changes in the composition of the magma ocean.  ``Evaporating planets" -- highly periodic occultation of stars by dust tails from small, undetected planets on ultra-short period orbits -- are evidence for a process of simultaneous evaporation and dust condensation \citep{Rappaport2012,Sanchis-Ojeda2015}.  

\citet{Kite2016} explored the consequences of evaporation on the physical and chemical evolution of a magma ocean or ``lava pond" centered at the sub-stellar point on a tidally-locked planet. The evaporation from a silicate melt proceeds from most to least volatile:  Na > K > Fe > (Si, Mg) > Ca > Al.  Since Na and K have a comparatively low atomic weight, their loss will increase the density of the residual melt, at least in a chemical boundary layer at the surface, which will eventually mix into the magma column.   In steady-state, there is compensatory melting of the silicate mantle at the base of the magma ocean which dilutes this enrichment.  As evaporation proceeds, the elimination of Fe lowers the density of the boundary layer and could stabilize it as a ``lag".  The appearance of lags depends on the substellar temperature and  Fe content of the mantle and is a potential observable that could be used to constrain planet composition \citep{Kite2016}.  Composition might also be inferred by probing the mineralogy of the dust grains that condense in an escaping silicate vapor atmosphere \citep{Bodman2018}.  
 
\begin{figure*}
	\centering
		\includegraphics[scale=.50]{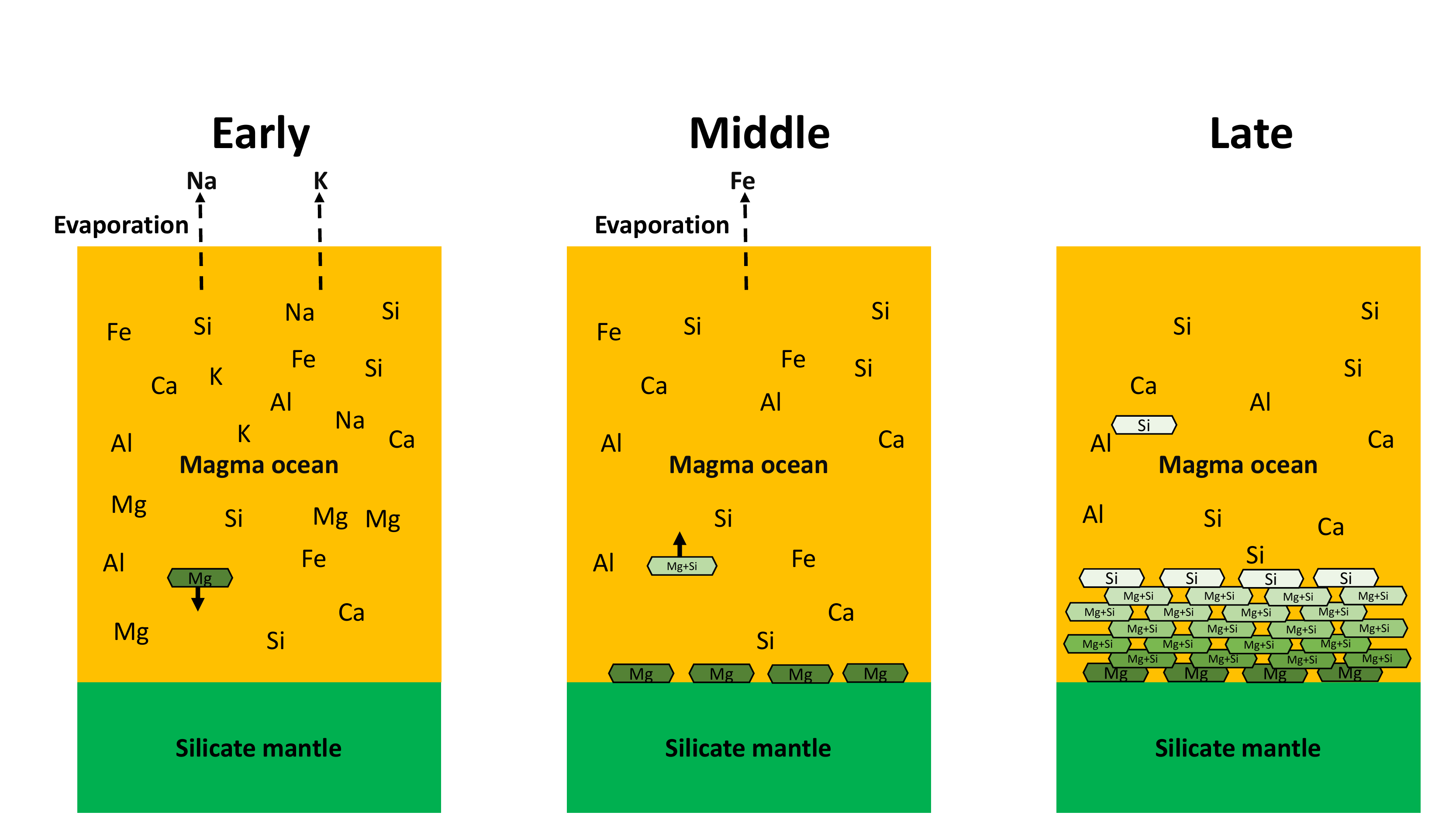}
	\caption{Possible stages in the evolution of a magma ocean experiencing chemical evolution due to both crystallization and evaporation.  Left: At lower temperatures or early stages of magma ocean evaporation/crystallization, volatile Na and K evaporate and refractory Mg-rich minerals crystallize and sink, both  increasing the relative Fe content and density of the melt.  Middle: The progressive evolution of the melt eventually causes later crystallizing to be buoyant.  At higher temperatures Fe is volatile and evaporates, and the density of the melt decreases.  Right: At late stages, crystallizing Si-rich minerals tend to be neutrally or slightly negatively buoyant in the Fe-poor but Ca- and Al-rich melt.  The timescales for crystallization and evaporation need not be comparable and the order of stages could be reversed.}
	\label{fig:stages}
\end{figure*}

More complex situations could include both crystallization, which increases the Fe content and density of the melt, and evaporation, which could either increase or decrease the density, with the relative importance of these two effects depending on the mass (surface gravity) and equilibrium temperature of the planet.  Evaporation will dominate on smaller, hotter planets and crystallization will dominate on cooler, more massive planets (Fig. \ref{fig:stages}).    

\subsection{Effect of Rotation}

The rotation of the planet can also affect the behavior of a magma ocean via its effect on circulation and overturn \citep{maas2019dynamics}. Vigorous convection of a crystallizing magma ocean stirs the minerals grains and distributes them throughout the magma ocean, homogenizing the crystallization process.  On a rotating planet, the Coriolis acceleration tends to suppress meridional (north-south) flow (Sec. \ref{sec:heat_transfer}) and three-dimensional flow patterns tend to become more two-dimensional (vertical and equatorial) and there is less mixing between latitudes.  The effect is larger on planets with a smaller global Rossby number $Ro$ (faster rotation).  Since the \emph{local} Rossby number $\propto 1/\sin \theta$ (Eqn. \ref{eqn:rossby}), the effects of rotation are negligible at the equator but can become important towards the poles.

Suppression of circulation will have two major effects.  First, on synchronously-rotating planets with thin atmospheres, circulation from the substellar point towards the terminator will be suppressed towards the poles.  This will not affect the extent of the magma ocean, since heat transport by a circulating magma ocean is small compared to the radiation terms in the surface energy balance equation (Eqn. \ref{eqn:rad_balance}) \citep{Kite2016}, but will allow different latitudes to become chemically segregated.  Second, at higher latitude the vigor of convection will decrease and settling of mineral grains will increase, leading to meridional gradients in the chemistry of the melt \citep{maas2019dynamics}.  

\subsection{Interaction with the Mantle}

In most scenarios the silicate portion of a planet is not completely melted and the magma ocean is in contact with a solid silicate mantle, usually below, or in the case of a basal magma ocean, above.   Both the magma ocean and the mantle will be convecting -- the latter by solid-state (subsolidus) convection.  Nearly all previous work has assumed that convection in these layers is self-contained and independent, but in this case the boundary between the magma ocean and the mantle is a phase-change boundary, with a comparatively small change in density, rather than a compositional boundary with a large density contrast like the core-mantle boundary.  Material can cross this boundary provided the necessary latent heat of crystallization is released or absorbed \citep{Labrosse2018}.    \citet{Agrusta2020} showed that the presence of a magma ocean can significantly enhanced convective heat transport in the underlying solid mantle, especially in case where a solid layer is sandwiched between upper and lower (basal) magma oceans.  This essentially means the entire mantle contributes to the thermal budget of a magma ocean, which could greatly increase its longevity against complete crystallization \citep{Agrusta2020}.  Such a scenario for delayed crystallization has been used to estimate the lifetime of the lunar magma ocean and to infer the Moon's formation time \citep{Maurice2020}.

In the extreme case a density inversion could develop as a result of, e.g., partial crystallization and enrichment of Fe in the residual liquid, leading to a \gls{Rayleigh-Taylor instability} and, eventually, overturn and the development of downward-propagating diapirs.  Overturn could be continuous and gradual \citep{Boukare2018} or large-scale and catastrophic \citep{Elkins-Tanton2003}.  The process could become self-sustaining as a result of the gravitational energy released during the overturn event.  This can lead to the formation of a basal magma ocean.

A basal magma ocean could also affect heat transport in the metallic core of a planet, particularly as it crystallizes.  The latent heat released during crystallization of a basal magma ocean would suppress heat loss from the core.  \citet{zeff2019fractional}, proposed that a impact-generated, crystallizing magma ocean in the Fe-rich mantle of Mars would be susceptible to sinking to the core-mantle boundary as a result of Fe enrichment.  This chemically distinct layer with its supply of latent heat and relatively low thermal conductivity would insulate the core, hindering heat flow, and could explain the cessation of the Martian dynamo early ($\approx$4.3 Gyr ago) in the planet's history (see Sec. \ref{sec:mars}).

\section{Atmospheres of Lava Worlds}
\label{sec:atmospheres}

\subsection{The Role of an Atmosphere}
\label{sec:atmos_role}

Any atmosphere on a lava world can significantly influence the stability, extent and lifetime of the magma ocean \citep{elkinstanton2008,Lebrun2013,Nikolaou2019}, as well as its detectability (Fig. \ref{FIG:lavaworldcartoon}).   Atmospheres can greatly elevate surface temperatures by the ``greenhouse" effect and will transport heat to the night side and/or poles of a planet (Sec. \ref{sec:heat_transfer}), causing surface temperatures to be more uniform and enlarging any magma ocean.  Atmospheres also provide a reservoir and transport for volatiles that are exchanged with the magma ocean, and since these gaseous species also affect the radiative properties and structure of the atmosphere, e.g., via clouds, complex feedbacks can develop \citep{Olson_2019}.  These atmospheres will be dynamic (short residence times), and this exchange with the interior could be crucial for their evolution, since often the proximity of the planet to the host star, which causes elevated surface temperatures permissive of a magma ocean, also drives the erosion of an atmosphere by elevated X-ray, UV, and particle fluxes.  Finally, an atmosphere can prevent direct observations of the surface and any magma ocean; thus it is through an interpretation of observations of the intermediary atmosphere that the existence and properties of a magma ocean might have to be established.   

Most of our understanding of planetary atmospheres is based on studies of the present Solar System or, by chemical inference and modeling, the early Solar System, with Mercury (possessing only a sputter-produced, low-density exosphere) and Venus (with a thick CO$_2$-dominated atmosphere) serving as two useful end-member analogs (Fig. \ref{FIG:lavaworldcartoon}).  The theory for many of the underlying physical and chemical processes, including the greenhouse effect \citep{Ekholm1901,Ingersoll1969,Komabayasi1967} have been derived from studies of Earth's atmosphere and oceans as applied to the rest of the Solar System.  However, the diversity of rocky planet atmospheres is undoubtedly far greater than is represented by the inner Solar System, with H/He rich atmospheres \citep{Tian2005,Pierrehumbert2011,Ramirez2017}, and more reducing CO and CH$_4$-rich atmospheres as plausible scenarios \citep{Rimmer2019,Zilinskas2020}.   Also unrepresented are more ``exotic" phenomena such electromagnetic interactions \citep{Castan2011,Pu2017,Kislyakova2018} for which we have only theory and laboratory experiments.  Here, we restrict our consideration to plausible primordial or proto-atmospheres that could have co-existed with early magma-oceans, including those proposed for the early Solar System \citep{Massol2016}, the processes that drove evolution of those atmospheres, and two possible end-states: the complete loss of an atmospheres as on Mercury, and a thick high-molecular weight greenhouse atmosphere such as on Venus.  

\subsection{Primordial Atmospheres}
\label{sec:atmos_primordial}

Due to the release of gravitational energy (Sec. \ref{sec:heat_sources}), the latter stages of the accretion of an Earth-sized planet will be highly energetic (Eqn. \ref{eqn:accretion_energy}), and accompanied by melting and intense shocking which will drive volatiles such as \water\ and carbon-bearing molecules (e.g., CO, CO$_2$, CH$_4$) into an atmosphere \citep{Abe1988,Schaefer2007,Schaefer2017}.  In the canonical picture of rocky planet formation, accretion proceeds with successively larger bodies \citep{Morbidelli2012}; for Earth these included the Moon-forming impactor or impactors  (see Sec. \ref{sec:moon}).  For Earth, at least, the abundance of highly siderophilic elements in the mantle are evidence for the accretion of a ``late veneer" consisting of a comparatively small amount of chondritic-like material after the giant impact and magma ocean phase.  This may have been responsible for a significant fraction of the volatile inventory, including water, at Earth's surface  \citep{Albarede2009,Peron2017}.  The composition of the atmosphere will depend greatly on the oxidation state of mantle silicates and a magma ocean; this in turn will depend on the composition of the progenitor \gls{planetesimals} but also on pressure effects in the mantle: in the deep mantle of Earth-sized planets, charge disproportionation of Fe$^{2+}$ into (metallic) Fe$^{0}$ and Fe$^{3+}$ and migration of the former to the core will lead to a more oxidized mantle (see Sec. \ref{sec:synthesis}) and thus a more oxidized primordial atmosphere rich in H$_2$O, CO$_2$, and N$_2$.  A late veneer of unaltered planetesimals, however, could drive atmospheric chemistry towards a more reducing state \citep{Zahnle2020}.  

Some volatiles, particularly H$_2$O, are highly soluble in silicate magma and thus the surface pressure of a steam atmosphere will be set by the total H$_2$O inventory and its solubility in the magma.  With an optically thick atmosphere the surface temperature is related to the surface pressure via the adiabatic gradient; since a magma ocean will exist only if the surface temperature exceeds $\approx$1300\,K, this produces a stabilizing feedback that will tend to maintain the surface inventory of water (as steam) at the level required to maintain a magma ocean, e.g. about 300 bars (or one terrestrial ocean) in the case of Earth  \citep{Matsui1986,Zahnle2007}.  A substantial CO$_2$ atmosphere suppresses magma crystallization and outgassing of H$_2$O, allowing a planet to retain more water, long term, in its interior \citep{Bower2019}.     

Earth-mass planets can also accrete H/He-rich gas directly from a planet-forming circumstellar disk \citep[e.g.,][]{Lee2015}, and the mass of the envelope will increase with planet mass.  The radius distribution of \emph{Kepler} planets is indirect evidence that rocky planets often capture (and sometimes lose) a H/He envelope that contributes substantially to their radius but only a few percent to their mass \citep{Fulton2017}.  A molecular hydrogen-rich atmosphere can produce significant surface warming by collision-induced absorption in the infrared \citep{Pierrehumbert2011} and will buffer the magma ocean to reducing conditions \citep{Olson_2019}.  However, the light elements H and He are expected to readily escape as the upper atmosphere is heated by X-ray and UV radiation from the central star \citep{Owen2019}.   The less massive the body, the less H/He accreted and the shorter the lifetime of such an atmosphere \citep{Ikoma2012}.  On the other hand, \citet{Kite2019} showed that a sufficiently thick H/He atmosphere can maintain a magma ocean on a rocky planet, into which that atmosphere can dissolve: this could limit the radii of ``sub-Neptune" planets discovered by \emph{Kepler}.  

\begin{figure*}
\centering
\includegraphics[scale=0.58]{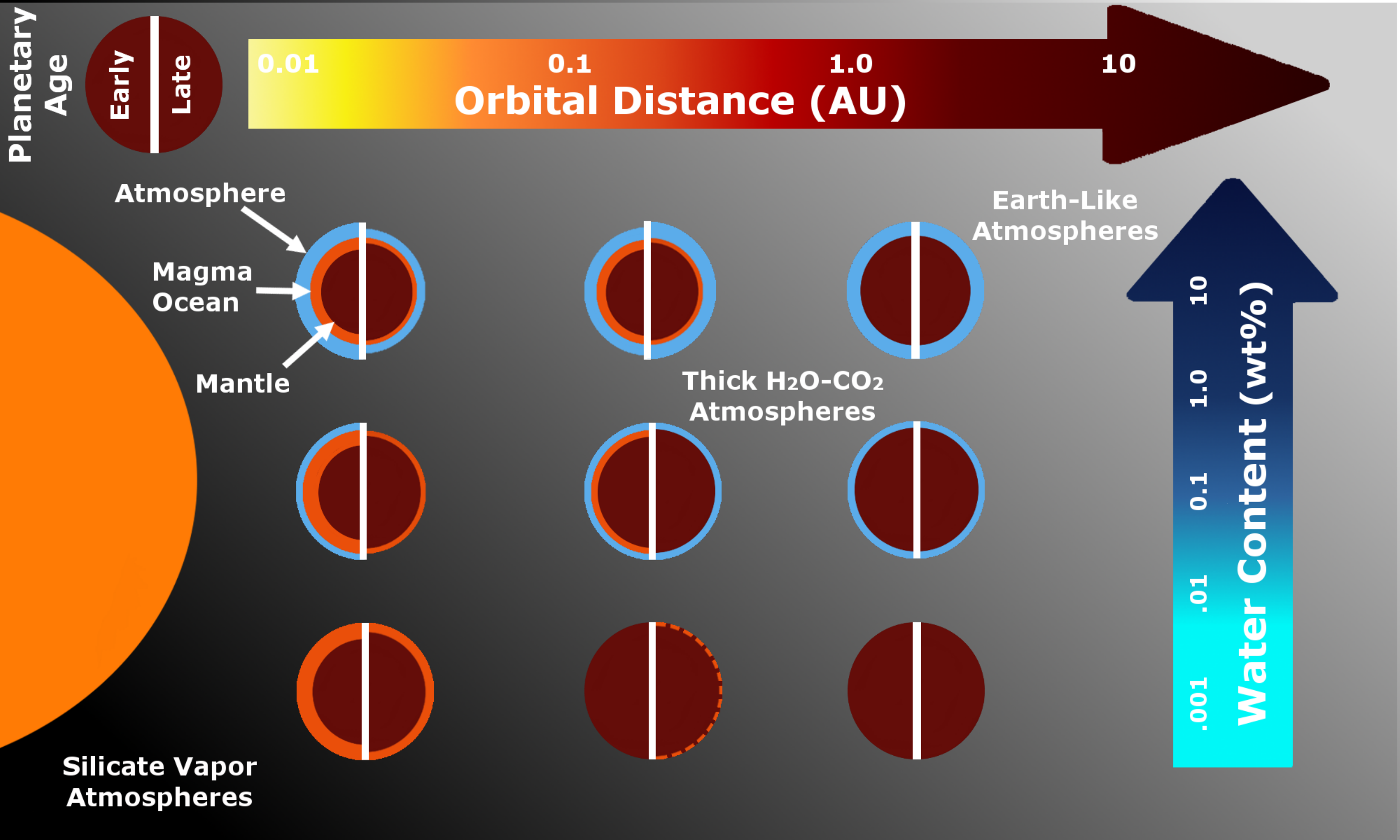}
\caption{Different scenarios for the appearance and evolution of magma oceans on rocky planets depending on orbital distance and initial water content (which also is a proxy for other volatiles).   Each circle represents a planet in its early (proto-atmosphere but post-accretionary magma ocean) and late (Gyr) phases (left and right, respectively).  The orange component represents the magma ocean and the blue component represents the atmosphere.  The background represents zones where the outcomes are silicate vapor atmospheres (lower left), Venus-like steam-CO$_2$ atmospheres (middle) and Earth-like atmospheres (upper right).  Ultra-short period planets (left column) will experience massive loss of volatiles via atmospheric escape; only very volatile-rich planets might retain any atmosphere. Magma oceans will be widespread on such planets, depending on the luminosity of the host star.  Planets on short-period orbits (middle column) will experience a runaway greenhouse and moderate loss of volatiles; magma oceans will appear only if the greenhouse effect is sufficiently strong, or late when the star is evolving into a luminous giant, the planet is tidally locked, and has no atmosphere (bottom scenario).  Gradual crystallization of such magma oceans will expel volatiles into atmospheres.  Temperate planets (right column) will have no long-term magma oceans nor experience significant atmospheric escape or a runaway greenhouse.  A moderate volatile content can form an ``Earth-like" atmosphere.
}
\label{FIG:atmomaglifetime}
\end{figure*}

A primordial atmosphere and an underlying magma ocean can be sustained by continued impact input of energy and volatiles, depending on the irradiance from the host star.  Planets sufficiently close to the star can maintain their magma oceans almost indefinitely \citep{Hamano_2015}, even if they lose their primordial H/He, as long as it retain a sufficient inventory of volatiles (chiefly \water\ and CO$_2$) to constitute a thick greenhouse atmosphere, i.e. the planet is sufficiently massive and/or far enough away from the star.  Planets that are somewhat closer will lose this atmosphere (see below) and the magma ocean will crystallize.  Planets that are very close-in, e.g. the USPs (Sec. \ref{sec:usps}) can maintain a magma ocean without the benefit of a greenhouse atmosphere.  Figure \ref{FIG:atmomaglifetime} illustrates the dependence of magma ocean evolution based on stellar irradiance and water content.

\subsection{Evolution of Atmospheres on Lava Worlds}
\label{sec:atmos_evol}

Atmospheres will evolve by any continued accretion of residual planetesimals, escape of light elements to space, exchange with the interior, including any surface or near-surface magma ocean, and irradiance, including photochemically active UV irradiance, by the central star (Fig. \ref{FIG:oceancartoon}).  This evolution may be regulated or destabilized by negative or positive feedbacks, e.g. the solubility of a steam greenhouse atmosphere in molten silicates \citep{Matsui1986,Abe1988}. 

The presence of highly siderophilic elements in Earth's mantle indubitably supports accretion of a ``late veneer" of chondritic-like planetesimals after crystallization of the magma ocean effectively segregated Earth's core and mantle but the exact timing of this input, its relation to both to the ``Late Heavy Bombardment" recorded by large lunar impact basins and to the handful of known early Archean impact structures, and the causative dynamic, e.g., migration of the outer giant planets, are areas of active research \citep{Bottke2017,Lowe2018,Mojzsis2019}.  Presumably many planetary systems experience such episodes, depending sensitively on their specific orbital geometry.  While impacts can erode an Earth-like atmosphere, recent three-dimensional calculations by \citet{Kegerreis2020} suggest that, for Earth-mass planets, atmospheric erosion is efficient only for those impacts approaching the scale of the Moon-forming event.  This is also predicted by new scaling laws derived by \citet{Denman2020} from numerical \gls{smooth particle hydrodynamics} experiments.   Nonetheless, this process could be much more important for H/He-rich atmospheres \citep{Biersteker2019}.  

Light elements can escape to space from Earth-mass planets with upper atmospheres heated by X-ray and UV radiation from the host star \citep{Catling2017,Owen2019}.  An important figure of merit for such a mode of escape is the Jean's parameter, which is the ratio of the gravitational to thermal energy of atoms or molecules at an altitude $h$ where the mean free path between collisions becomes comparable to the atmospheric scale height (the exobase),
\begin{equation}
    \label{eqn:jeans}
    \Lambda = \frac{GM_p \mu}{(R_p + h) k_B T},
\end{equation}
where $k_B$ is the Boltzmann constant, and $\mu$ the atomic weight of the atom or molecule considered.  Due to gravitational settling and condensation, only light atomic species are present at these altitudes and only ionizing X-ray and UV (XUV) radiation with wavelength $\lambda < 91.2$\,nm from the star can be absorbed.  Nevertheless, inefficient cooling allows temperatures to reach many 1000s of K.  If $\Lambda > 3$ molecular escape occurs. In hot, extended atmospheres of light elements, i.e. H and He,  $\Lambda \lesssim 2$ and escape occurs as a hydrodynamic flow \citep{Volkov2011}.  Such a flow can drag heavier C, N, and O atoms into space \citep{Catling_2009}.

Hydrogen contained in molecular form, principally the \water\ of a steam atmosphere and oceans, but also less abundant constituents such as CH$_4$ and HCN can be lost by UV photodissociation and diffusive and turbulent transport to the exobase where it can escape.  This is the standard explanation for the loss of any initial water inventory on Venus, and similar processes may be widespread among rocky planets close to young stars \citep{Tian2018}.  Due to the intense greenhouse effect, a thick \water-dominated atmosphere on a planet that is irradiated at a level exceeding modern Venus can maintain a magma ocean for significantly longer than on a planet further out.  One consequence of this long-lived atmosphere is the loss of water by photodissociation and H escape, which can ultimately lead to lower surface temperatures and crystallization of the magma ocean \citep{Hamano2013}.    

A major question concerns the fate of the oxygen produced by the photodissociation of \water\ and escape of H and whether O$_2$  will accumulate in the atmosphere \citep{Tian2015,Luger2015}.  In isolation from the interior, an atmosphere will become more oxidizing, e.g. any CO or CH$_4$ replaced by CO$_2$, as has been proposed for the Archean Earth \citep{Catling2001,Zahnle2013}. This process could ultimately lead to the accumulation of significant O$_2$ -- considered an important biosignature --  and potential ``false positives" for biosignatures \citep{Luger2015} \citep[see discussion by Meadows ][]{Meadows2017}.  This is distinct from photodissociation of CO$_2$ (see Sec. \ref{sec:venus-like}).  Free oxygen will react with any reducing elements like ferrous Fe on the surface, and especially throughout a rapidly convecting magma ocean \citep{Schaefer2016,Wordsworth2018}.  This oxygen can also react with impact ejecta during any tail-end of the accretionary phase of rocky planets \citep{Kurosawa2015}.  Massive water loss may occur on highly irradiated planets close to active stars \citep{Johnstone2020}; at such high XUV irradiance, ionization of both H and O strongly couples the species and leads to efficient loss of O along with H \citep{Guo2019,Johnstone2020}.  Quantifying the evolution of XUV radiation from stellar hosts over timescales of Myr to Gyr is crucial to understanding these processes \citep{Tu2015}.  

Non-thermal escape mechanisms can play an equal or even larger role in the removal of atmospheres from planets on close-in orbits, particularly in removing heavier atoms of C, N, and O.  The rates of escape depend on the behavior of the star and the interaction with the planet and any magnetic field, as well as XUV heating of the upper atmosphere, which determines its vertical structure.  The kinetic energy imparted to the atom is variously derived from recombination after photoionization or even photodissociation \citep{Shematovich_2018,Howe2020}, collisions with ions from  star or planet planetary ions accelerated by the magnetic field in the stellar wind \citep{Lundin_2007}, or electromagnetic interactions with the stellar wind \citep{Catling2017}.  The foundations of a better understanding of these phenomena is being laid with statistical determinations of flaring rates of stars with a range of spectral type and rotation rates/ages.  However, in the case of solar-type host stars of magma ocean planets, flares have low contrast with respect to the stellar photosphere and even the most sensitive space-based surveys can only detect the largest events.  For example, the average energy of flares detected in \emph{Kepler} data is $\approx$400 times the \emph{maximum} observed on the Sun.   On dimmer M dwarf stars the higher contrast allows detection by ground- and space-based surveys \citep{Rodriguez2020,Gunther2020}.  Detection of coronal mass ejections, e.g. by spectroscopy, is even more challenging and remains frontier work \citep{Leitzinger2020,Odert2020}.  

It is also important to place this activity within a temporal context, since elevated activity when stars are younger and more rapidly rotating could mean that most non-thermal mass escape occurs then as well.  Here the all-sky photometry obtained by the  \tess\ mission of stars with a range of established ages will contribute \citep{Feinstein2020}.  Also just as important but uncertain is how the planet's magnetic field, if any, will influence this escape, since it can have opposing roles of concentrating stellar wind particles onto an atmosphere while simultaneously reducing the kinetic energy per particle \citep{Blackman2018}.  Nonetheless, extensive modeling efforts have been made in order to quantify atmospheric escape as a function of different parameters such XUV emission \citep[e.g.,][]{Rodriguez-Mozos2019}. 

Broadly speaking, it is expected that less massive planets on closer orbits will be less likely to retain a significant atmosphere, but the boundaries are unclear and the extent to which the interior will out-gas is important \citep{Kite2020}.  \citet{Olson_2019} predict that a planet less massive than 0.5 M$_{\oplus}$ will lose most of its atmosphere in $\sim 100$\,Myr.   

\subsection{End-State Atmospheres}
\label{sec:end-state}

\subsubsection{Thick H$_2$O-CO$_2$ Atmospheres}
\label{sec:venus-like}

Thick, Venus-like atmospheres of CO$_2$ and variable amounts of H$_2$O are the outcome on planets with significant volatile content and that are sufficiently massive, distant, and/or have a magnetic field which can protect against significant erosion such that the atmosphere is largely retained over Gyr (Fig. \ref{FIG:atmomaglifetime}.  Surface temperatures on Venus (740K) are well below the solidus of an Earth-like mantle, but Venus may have maintained a magma ocean for a significant amount of time \citep{Chassefiere2012}.  If Venus was significantly closer to the Sun, had a thicker or wetter atmosphere, and/or a lower albedo it could sustain a global magma ocean indefinitely \citep{Hamano_2015}.   Assuming a modified adiabatic profile that accounts for changes in the gas properties with temperature, the surface temperature $T_s$ for a given surface pressure $P_s$ will be
\begin{equation}
    T_s \approx T_{\rm eq} \left(P_s/P_{\rm rc}\right)^{\zeta \frac{\gamma-1}{\gamma}},
\end{equation}
where for a Venus-like CO$_2$-dominated atmosphere, the ratio of specific heats $\gamma = 1.3$, $\zeta = 0.78$ \citep{Robinson2012}, the pressure at the radiative-convective boundary $P_{\rm rc} \sim 0.1$\,bar \citep{Robinson2014}, and the equilibrium heat flux is $230 (0.72/a)^2$\,\wpersqm.  A Venus twin orbiting interior to 0.23\,au, or out to about 0.3\,au if the atmosphere was thicker or the planet more massive, or had a lower albedo, would have $T_s > 1300$\,K.  Water would be highly soluble in a magma ocean; CO$_2$ much less so:  at 0.57 wt ppm bar$^{-1}$ \citep{Ni2013}.  Only 1.5\% of the mass of Venus' CO$_2$ atmosphere would dissolve into a 100 km deep ocean; less if H$_2$O were also dissolved, displacing CO$_2$.     

CO$_2$ absorption of UV ($\lambda = 120-210$\,nm) photons can lead to dissociation:  CO$_2 \rightarrow $CO + O or possibly CO$_2 \rightarrow $C + O$_2$ \citep{Lu2014}.  The extent to which this could lead to the buildup of molecular O$_2$ is debated, since molecular oxygen can react with ferrous iron in the surface (in this case, a magma ocean) and lightning in the atmosphere can catalyze the reverse reaction \citep{Harman2018}.  However, rates of electrical discharge on Venus -- and more broadly in dry, hot atmospheres -- are uncertain \citep{Lorenz2018,LOrenz2019}.  It is possible that while the accumulation of O$_2$ may be limited, CO can gradually accumulate as the magma ocean and interior are oxidized.  (CO is only a trace component in the present Venusian atmosphere).   A Venus-like atmosphere enriched in CO, as well as SO$_2$, given the expected destabilization of sulfide-bearing minerals known to exist on the surface of Venus higher temperatures \citep{Zolotov2018}, could be a model for many of the Earth and super-Earth exoplanets discovered on short-period orbits (Sec. \ref{sec:exoplanets}).  

\subsubsection{Silicate Vapor Atmospheres}
\label{sec:mercury-like}

\begin{figure}
	\centering
		\includegraphics[width=0.48\textwidth]{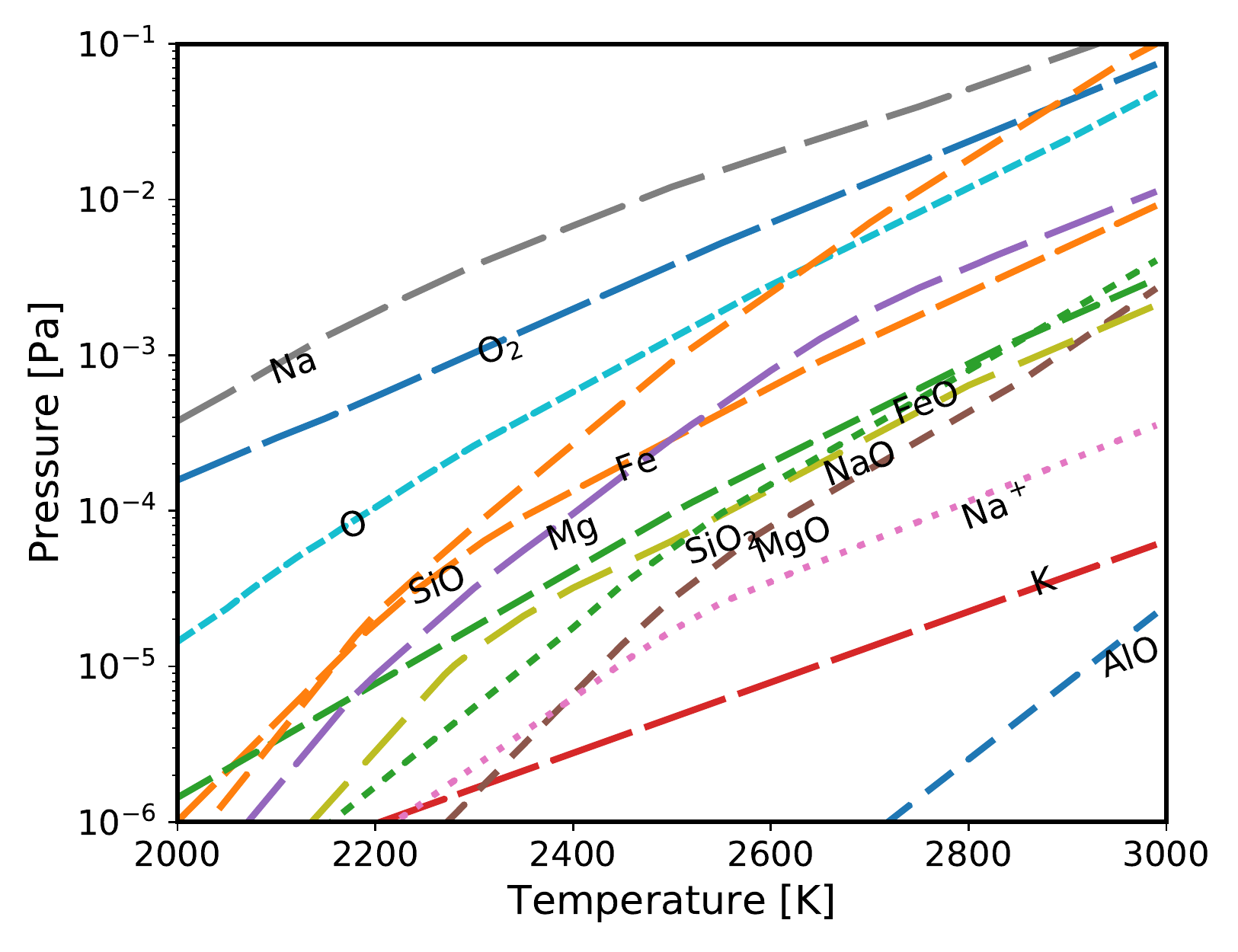}
	\caption{Approximate vapor pressure of constituents of a rock vapor atmosphere at 2000-3000\,K in equilibrium with a surface with the bulk composition of Earth's mantle.  Based on \citet{Schaefer_2009} with some extrapolation.  The dominant species in such an atmosphere are Na, O$_2$, and, at the highest temperature, Fe.}
	\label{FIG:silicatevapor}
\end{figure}

Depending on their mass, planets on very short period orbits could be depleted in volatiles due to their formation close to the parent star \citep{Lissauer2007,Lopez2017} and/or lose virtually their entire atmosphere through XUV- and particle-driven loss.  They would become more Mercury-like, albeit with significantly higher surface temperatures and stellar particle fluxes, depending on the planet's magnetic field.  Unlike Mercury, tidal locking is expected to lead to 1:1 synchronous rotation, extreme day-night temperature contrasts, and the presence of a hemispheric magma ocean at the surface (Fig. \ref{FIG:lavaworldcartoon}).   As temperatures exceed 2000\,K, the vapor pressure of more volatile constituents of silicate mantles / magma oceans, i.e. Na and K, along with O for valence balance, become appreciable, forming a tenuous atmosphere \citep{Schaefer_2009} (Fig. \ref{FIG:silicatevapor}).  With increasing temperature, more refractory elements contribute to this atmosphere: Fe, Si, Mg, and lastly Al and Ca when the temperature approaches 3000\,K (Fig. \ref{FIG:silicatevapor}).  The gradient in vapor pressure from the sub-stellar point towards the terminator will drive a thermal wind, with atoms being deposited at the margins of the magma ocean or possibly near the terminator \citep{Schaefer2009,Kite2016} (Fig. \ref{FIG:lavaworldcartoon}.  This wind can be partially ionized and to the extent it is collisional, will be affected by any planetary magnetic field \citep{Castan2011} or potential removal by the electric and magnetic fields of the stellar wind.  If the removal of elements is more efficient from mixing of the magma ocean due to the gradient in temperature then a chemical residual layer or ``lag" can develop, which can be less or more dense depending on the stage of evaporation, i.e. the atomic mass of the evaporated constituents.  Depending on the FeO content of the mantle, evaporation of lighter Na and K produces a heavier, unstable lag, while evaporation of heavier Fe (at a later stage) produces a stable lag \citep{Kite2016}.

In the extreme cases, vaporization of a highly irradiated planet and loss of the vapor or dust condensates to space could significantly erode the planet.  The dusty cometary winds of such evaporating planets will obscure the star if the planet is on a transiting orbit and candidate systems have been identified, including Kepler-1520b \citep{Rappaport2012}, K2-22b \citep{Sanchis-Ojeda2015}, and HD 240779 \citep{Gaidos2019b}.  Vaporization reduces the mass and the surface gravity, which in turn allows vaporization to proceed more quickly, leading to the potential runaway evaporation of the entire planet \citep{Perez_2013}.  However, the process could potentially self-arrest if a  buoyant, refractory (i.e. Mg, Ca, Al-rich) magma ocean and crust were to develop that does not mix with the interior.  These dust tails, back-lit by the star, provide a means of probing the interior composition of such objects that would otherwise not be accessible \citep{Budaj2015,vanLieshout2016,Gaidos2019a}.

\section{Summary and Future Directions}
\label{sec:summary}

\subsection{Significant Advances}
\label{sec:advances}

While evidence for past magma oceans in the Solar System has been accumulating for the past half century, beginning with the analysis of Apollo 11 samples, it is only in the last decade that a magma ocean as a near-universal phase of early rocky planet evolution has become both widely appreciated and investigated. With the detection of rocky planets on short-period orbits around other stars it has also become more likely that we can study \emph{extant} magma oceans directly, albeit from a great distance. Arguably the most significant advance in the realm of Solar System exploration was the \emph{Dawn} mission to Vesta, which greatly strengthened the connection between the asteroid and the HED meteorite suite which are products of magma ocean crystallization \citep{McSween2013}.  A preliminary understanding of Vesta's interior structure was provided by measurements of its gravity field; more might be learned once any remnant magnetic field, like that suggested by the HED meteorites \citep{Fu2012}, is measured.  Measurements of Mercury and Mars have provided intriguing but indirect evidence for magma oceans.  For the time being, evidence for Earth's possible magma ocean phase will be restricted to improving measurements of the isotopic composition of the mantle relative to the hypothetical Earth-forming reservoir.   Venus will remain an enigma until surface exploration is resumed.

Space-based photometry has ushered in an era of detection via transits of Earth-size planets around other stars; the limited Doppler RV measurements of mass that have been possible thus far indicate rocky-metal compositions similar to Earth.  Due to detection bias, known planets tend to be on short-period orbits of a few days or less, and equilibrium temperatures exceeding the solidus raise the prospect of observable, stable magma oceans.  Transits of three well-characterized examples (CoRoT-7b, 55 Cnc e, and Kepler-78b) were detected by three different missions: \emph{CoRoT}, \emph{MOST}, and \emph{Kepler} and have been the most extensively characterized in terms of mass, equilibrium temperature, and the presence or absence of an atmosphere.  Nevertheless, the limited accuracy and degeneracy of interpretation of some measurements, particularly mass, leave room for substantial controversy such as composition of the interior and the presence or absence of an atmosphere, \citep[e.g.,][]{Dai_2019}.

\subsection{Major Outstanding Questions and Key Needs}

 Our understanding of the behavior of lavas/magma on Earth, elsewhere in the Solar System, or on ultra-short period planets around other stars is only as good as our knowledge of the rheologic and thermodynamic properties of those materials.   Properties can be determined by laboratory measurements \citep[e.g.,][]{hofmeister2016,heap2020thermal}, empirical fits to data collected in the field \citep[e.g.,][]{castruccio2013}, or, more rarely, direct measurement in the field \citep[e.g.,][]{Chevrel2018}. Volatiles play a key role in determining lava lake behavior, and our knowledge of volatile solubility, speciation, and diffusivity have come in large part from experiments \citep[e.g.,][]{wallace2015,ni2018,zhang2010}. 
More data are needed, particularly for compositions that radically deviate from the calc-alkaline compositions of the terrestrial trend between basalt and rhyolite.  For example, \citet{Leger2011} predicted an eventual magma ocean composition of mixed CaO-Al\textsubscript{2}O\textsubscript{3} on CoRoT-7b. Recent studies by \citet{morrison2019} and \citet{Sehlke2020} used laboratory measurements to determine the rheology of both primary and impact-induced predicted lunar melts.  More data on carbon- and sulfur-rich melts are needed to understand a possible magma ocean and volcanism on Mercury and perhaps planets with analogous, reduced compositions.  The exploration of lakes of liquid sulfur in undersea volcanoes \citep{deRonde2015} may offer insight into phenomena on Io.

Almost certainly Vesta was \emph{not} the only small body to host a magma ocean in the Main Asteroid Belt, simply perhaps the largest intact survivor.  The energetics of planetary accretion seem to have been sufficient to melt and even vaporize a significant mass fraction \citep{Davies2020}.  Enrichment of heavy isotopes of Si and Mg in planets with respect to chondritic meteorites has been interpreted as evidence for isotopic fractionation during evaporation from molten planetesimals \citep{Young2019}.  There have been several surveys that have identified a few basaltic asteroids that, unlike the ``Vestoid" asteroid family, cannot be dynamically linked to Vesta \citep{Lazzaro2000,Moskovitz2008,Roig2008,Oszkiewicz2017}.  Finds and falls of unique basaltic as well as the anorthosite-rich meteorites \citep{Bland2009,Frossard2019} also hint at additional diversity in similar bodies, obscured, in part, by the effects of space weathering \citep{Yamamoto2018}.  Expanded surveys of the asteroid belt, as well as systematic analysis of current and future meteorites collected, e.g. Antarctica, may provide a fuller picture of the diversity of parent bodies that hosted or are fragments of crystallized magma oceans.    

\subsection{Future Solar System Missions}

Several planned or proposed missions to planets and satellites should yield more definitive and detailed information on magma oceans or lava lakes they hosted or host.  First in line is \emph{Bepi-Columbo}, due at Mercury (i.e., its first flyby) in October 2021 \citep{Spohn2001,Schulz2006,Milillo2010}.  \emph{Bepi-Columbo} will further the investigation of the composition of the Hermian crust and the plant's magnetic field \citep{Rothery2020}.   This includes potentially confirming \emph{Messenger}'s tentative detection of carbon by the MIXS X-ray and MGNS $\gamma$-ray and neutron spectrometers \citep{Rothery2020} as a test of the hypothesis that a Hermian magma ocean had a graphite flotation crust \citep{peplowski2016}.  Ultimately, a sample return mission could provide definitive answers to the history of the planet and its putative magma ocean \citep{VanderKaaden2019}.

Ever since volcanism was detected by Voyager 1, Io has remained a compelling but hazardous target for spacecraft missions, with only limited data from \emph{Galileo} and now \emph{Juno} \citep{Mura2020}.  The Io Volcano Observer (IVO) is a proposed mission that would orbit Jupiter and study the satellite via a series of flybys to determine how and where heat is produced by tidal dissipation, and whether Io hosts a subsurface magma ocean via measurements of the induced magnetic field produced by eddy currents, the distribution of heat and volcanism, and its gravity field \citep{McEwen2020}.

Future missions to the Moon include the Chang'e 5 sample return from Mons R\"{u}mker, scheduled at the end of 2020 \citep{Zhao2017}, and, possibly, a continuation of the Soviet/Russian \emph{Luna} series in 2021 with the \emph{Luna 25} lander and continuing with planned rovers and sample return from the south pole \citep{Tretayakov2020}.   Mons R\"{u}mker is a 3.5 Gyr-old basaltic volcanic dome complex which could be informative about late-stage volcanism on the Moon \citep{Zhao2017}.  Of course, human return to the Moon would bring with it the opportunity for high volume sampling, although perhaps from a limited region of the surface.  And there may be surprises in store ``for free" from the curated Apollo samples that are being released for analysis for the first time\footnote{https://sservi.nasa.gov/articles/apollo-next-generation-sample-analysis-program/}.

Sample return from Mars is a long-term goal of the planetary science community \citep{Muirhead2020}, and the first concrete step will be taken with sample caching by the \emph{Perseverance} (Mars 2020) mission to Jezero Crater \citep{Grady2020}.  Although the goals of the mission are to search for evidence of past life in the sedimentary deposits around the landing site, any returned samples would ultimately provide counterparts to the Martian meteorites (SNCs) and/or indirect evidence for or constraints on any past magma ocean (e.g., the composition of the mantle).  A potential alternative to collecting Martian material would be to return samples from one of Mars' satellites Phobos and Deimos.  The \emph{Martian Moons eXploration} (MMX) mission will return samples from Phobos at the end of the 2020s \citep{Kuramato2018}. If these satellites accreted from an impact generated disk, as has been proposed, they will consist of an admixture of Martian and impactor material \citep{Hyodo2017,Pignatale2018} that is far easier to retrieve (but probably much more difficult to interpret) than the crust of Mars itself.

Venus will continue to remain an enigma for some time.  There are no approved missions to return to the planet, although many have been proposed or are under current study.  Missions to investigate the structure and composition of the atmosphere \citep{Garvin2020} could reveal whether volcanism still persists, as suspected, while high-resolution radar and gravity surveys will better characterize the planet's interior structure and response to tidal deformation \citep{Wideman2020} and thus describe the crust which could rule out or possibly detect a basal magma ocean.  Sample return from Venus, while obviously desirable, is an enormous technical challenge and is not on the horizon \citep{Sweetser2003}. 

\subsection{Beyond the Solar System}

While we await these Solar System missions, some of the most fruitful observations and ground-breaking discoveries may occur for much more distant objects, i.e. exoplanets.  The first of what could be multiple extended missions for the \tess\ satellite has now started.  This promises to expand and refine our sample of close-in planets around bright stars \citep{Barclay2018}.  The longer baseline and faster 10-minute cadence for the full dataset, which will better resolve short-duration transits, will allow \tess\ to improve the statistics of Earth- to super-Earth-size ultra-short period planets and identify many that are suitable for follow-up.

New RV spectrographs such as ESPRESSO should allow more USPs to be ``weighed" with greater precision.  Spectrographs such as SPIRou \citep{Donati2018} and NIRPS \citep{Wildi2017} which observe at infrared wavelengths are, in principle, less affected by stellar photosphere noise \citep[``jitter"][]{Marchwinski2015}, likely to be the limiting source of noise in many systems.  SPIRou and NIRPS also simultaneously measure polarization, which can be used to constrain the strength and configuration of the stellar magnetic fields and the positions of active regions, information useful for mitigating jitter.  Of particular interest is whether continued population of the mass-radius diagram of USPs (Fig. \ref{FIG:mass-radius}) will uncover denser planets with Mercury-like interiors and massive iron cores like Kepler-974c produced by collisions or segregation of metal and silicates \citep{Scora2020}.\footnote{The composition of former short-period planets can be investigated \emph{post mortem} as elemental debris in the atmosphere of the white dwarfs that disrupted them \citep{Doyle2019}.}  This decade will also see ``first light" at one or more $\ge$30\,m-aperture Extremely Large Telescopes, eventually with instrumentation that can obtain informative spectra of the atmospheres of Earth-size planets during transits \citep{Marconi2018,Szentgyorgyi2018,Mawet2019}.

Perhaps the biggest leap forward will be provided by the infrared \jwst, scheduled for launch in late 2021, which will be able to detect the secondary eclipses and/or measure the phase curve of many hot Earth-size planets and by doing so estimate their equilibrium temperature, albedo and efficacy of heat transport from the day to night sides \citep{Beichman2018}.  A smaller optical-infrared space telescope (\emph{ARIEL}) dedicated to exoplanet observations should follow several years later \citep{Edwarfs2019}.      Time variability and eclipse mapping of the offset of any hot spot should provide more insight into the circulation of any atmosphere; in the absence of an atmosphere, peak temperatures would be indicative of the presence or absence of a magma ocean.  Infrared spectra of the largest/brightest examples could also be used to constrain the mineralogy of any solid surface \citep{Hu2012}.  Even with a thick atmosphere obscuring the surface, a magma ocean might reveal itself via the relative solubility of important atmospheric gases and hence their abundance in the atmosphere, e.g. CO, CO$_2$, and H$_2$O  \citep{Pawley1992,Wetzel2013,Armstrong2015,Yoshioka2019} as well as H/He \citep{Kite2019}.   Over the horizon are the next generation of proposed space observatories like \emph{Origins Space Telescop} \citep{Cooray2019}, \emph{HabEx} \citep{Gaudi2020}, and \emph{LUVOIR} \citep{Luvoir2019}; these offer the greater sensitivity and angular resolution, enhancing our ability to explore lava worlds from afar. 

\section*{Acknowledgements}

R. D. acknowledges fruitful discussions with J. Birnbaum and T. Keller.  F. Nimmo , E. Kite, and two anonymous reviewers gave expert and constructive comments on the manuscript.

\section{Bibliography}

\printglossaries



\printcredits

\typeout{get arXiv to do 4 passes: Label(s) may have changed. Rerun}
\end{document}